\documentclass[letterpaper, 12pt, onehalfspacing, leqno]{article}

\usepackage{amssymb}
\usepackage{amsmath}
\usepackage{amsfonts}
\usepackage{amsthm}
\usepackage{graphicx}
\usepackage{enumerate}
\usepackage[margin=1in]{geometry}
\usepackage{setspace}
\usepackage{dsfont}
\usepackage{natbib}
\usepackage{mathrsfs}
\usepackage{color}
\usepackage{comment}
\usepackage{latexsym}
\usepackage{cancel}
\usepackage{subcaption}
\usepackage{booktabs} 
\usepackage{multirow}
\usepackage{float}
\usepackage[colorlinks, citecolor=blue]{hyperref}

\usepackage{xcolor} 
\usepackage{pgfplots}
\pgfplotsset{compat=1.18}

\theoremstyle{definition}

\title{Learning from an Unknown DGP: Experimental Evidence on Belief Updating with AI Recommendations\footnote{Kovach: Department of Economics, Purdue University, \emph{mlkovach@purdue.edu}. Martin: Department of Economics, UC Santa Barbara, \emph{danielmartin@ucsb.edu}. Tserenjigmid: Department of Economics, UC Santa Cruz, \emph{gtserenj@ucsc.edu}. We thank Anastasiia Morozova and Leshan Xu for their excellent research assistance. Martin thanks the Sloan Foundation for supporting this research under the ``Cognitive Economics at Work'' grant. IRB approval was obtained at University of California, Santa Barbara (11-22-0691) and Purdue University (2025-349). Preregistration can be found at https://aspredicted.org/fz8s-535m.pdf, and Appendix~\ref{app:preregistration} summarizes how our analyses relate to the preregistration.}}
\author{\centering Matthew Kovach  \and Daniel Martin \and Gerelt Tserenjigmid}

\date{July 10, 2026}

\begin{document}

\maketitle

\begin{abstract} We use a controlled experiment to study how beliefs are updated after receiving qualitative information (AI recommendations) from an unknown data-generating process (DGP). Across 60,252 pairs of prior and posterior beliefs, we document three behavioral patterns: updates close to zero when recommendations confirm extreme priors, larger updates when recommendations contradict extreme priors, and smaller updates for intermediate priors. These three behavioral patterns suggest four testable properties of belief updating, which we assess at the aggregate and individual levels. Finally, we examine how well updates are captured by three models of belief updating.
\end{abstract}

\newpage
\section{Introduction}\label{sec:intro}

Understanding how a decision maker (DM) updates their beliefs after receiving information is a foundational issue in economics and is of both theoretical and practical interest. The standard approach to modeling belief updating is to assume that the DM knows the information structure or data-generating process (DGP) of the received information and updates their beliefs accordingly. As a result, experiments that study belief updating often make the DGP explicit through transparent information structures, including colored marbles or balls drawn with replacement from cups or urns with known compositions \citep*{holt_smith_2009, danz_vesterlund_wilson_2022}, binary signals with known reliability \citep*{coutts_2019}, biased computerized information sources with explicitly specified conditional signal distributions \citep*{charness_oprea_yuksel_2021}, or canonical base-rate updating problems in which base rates and signal reliabilities are provided \citep*{esponda_vespa_yuksel_2024}. In these treatments with a transparent DGP, the experimenter communicates the relevant mapping from states to signals, so the Bayesian posterior is well defined and deviations from the Bayesian benchmark can be interpreted as errors, biases, or frictions in processing information.\footnote{See \cite*{benjamin2019errors} for a review of the experimental literature on belief updating.} However, in many settings in the world, the process by which information is generated is neither transparent nor described in probabilistic terms, and there is less evidence on how beliefs are updated in such settings.\footnote{A notable exception is the experiment by \cite*{afrouzi2023overreaction}, where participants observe a series of 40 past realizations from a simple AR(1) process and are then asked to make forecasts for future values without knowing the DGP. They find overreaction to recent observations.} 

To help fill this gap in the literature, we implemented a controlled experiment where the DGP for information about the state of the world was unknown to participants, but they received messages with qualitative content. We say that a DGP is \textit{unknown} when the DM is not endowed with a likelihood function or a set of possible likelihood functions indicating how messages are generated from states. By \emph{qualitative content}, we mean that a message restricts or summarizes the underlying state without directly reporting the likelihood of the state.\footnote{\cite*{hoong2025improving} show that providing coarse qualitative information is the optimal form of recommendation in the face of belief updating biases, which could potentially explain why we see such recommendations so often in practice.} We study this type of information because received information often takes this form in real-world settings, whether it be a suggested action, a most likely state, or an alert. For instance, a transaction may be flagged as fraudulent, a patient may be recommended for additional medical screening, or an applicant could be classified as high risk. These messages convey information about an underlying state, but do not indicate the likelihood of the state or the mapping from states to messages.

In the world, unknown DGPs could come in many forms: human advice, expert judgment, news, analyst forecasts, and institutional alerts. As a result, there are potentially many ways to generate an unknown DGP in an experiment. For instance, a qualitative message could be sourced from a human novice or expert, a crowd forecast, or a statistical model. Additionally, experimental participants could be told more or less about the source of those messages. In our experiment, the qualitative content comes in the form of artificial intelligence (AI) recommendations, which are particularly well suited for studying how beliefs are updated when the DGP is unknown. First, AI is often described as being a ``black box'' in news media, so there is reason to believe participants will treat the DGP for AI recommendations as unknown. Second, we can reliably estimate the AI's DGP because we can evaluate its recommendations for a large number of instances. Third, the DGP of the AI is stable, reducing unmodeled issues arising from dynamic changes in recommendations.\footnote{Our experiment is not designed to show how belief updating with AI recommendations is different from belief updating for other types of unknown DGPs. However, in Section~\ref{sec:humansvsAI} we discuss how the response to human recommendations might differ from the response to AI recommendations.}

Our online experiment adapts the ``Bouncer'' task introduced by \citet*{caplin2024abc}. In each of 160 rounds, participants viewed pictures of a face and reported their belief that the person was over 21 when the picture was taken. After the elicitation of their initial belief that the person was over 21 (henceforth, \emph{prior} belief),\footnote{We use this description for ease of use and acknowledge that this is a prior only relative to the AI recommendation, not relative to all information.} an on-screen ``AI Assistant'' announced one of two recommendations regarding the person's age: ``More likely Over 21'' or ``More likely Under 21'' (for simplicity, we will refer to these as the \emph{Over 21} or \emph{Under 21} recommendations, respectively). Participants could then revise their prior before reporting a final belief that the person was over 21 (henceforth, \emph{posterior} belief). Our main goal is to document and understand belief updating (i.e., how prior and posterior beliefs are related), so we incentivized truthful elicitation by having the prior and posterior beliefs be equally likely to determine payoffs and using the binarized quadratic scoring rule \citep*{hossain2013binarized} to determine the bonus payment based on a randomly selected belief. 

Participants in our experiment were randomized into one of two treatments. In the NOINFO treatment, participants received no information about the machine learning (ML) model used to generate the AI recommendation, the accuracy of the AI recommendation, or the mapping from ML outputs to the AI recommendation. In the INFO treatment, participants were provided details about how the ML model was trained, the AI's overall accuracy, and the AI's state-contingent recommendation rates. We varied the information provided to participants across these treatments for several complementary reasons. First, we wanted to investigate whether providing likelihood information would cause participants to update more like they do in experiments where the DGP is explicitly stated. Second, this variation allows us to examine how updating changes when participants have more or less uncertainty about an unknown DGP's accuracy. In addition, knowledge about an AI's accuracy can vary through experience or direct information, so this treatment variation lets us examine robustness along that dimension.

Our experiment generated a panel of 60,252 pairs of prior and posterior beliefs from 377 U.S. residents that are roughly balanced across the two treatments. From these pairs, we document three behavioral patterns in participant updating:
\begin{enumerate}
    \item \textbf{Updates close to zero when recommendations confirm extreme priors:} when priors are extreme (close to zero or one) and the recommendation confirms them, average updates are close to zero.  
    \item \textbf{Larger updates when recommendations contradict extreme priors:} when priors are extreme (close to zero or one), and the recommendation contradicts them, average updates move more toward the recommendation.
    \item \textbf{Smaller updates for intermediate priors:} when priors are not extreme, average updates are smaller than when the AI recommendation contradicts extreme priors.  
\end{enumerate} 

\begin{figure}
\centering
\begin{subfigure}[t]{0.48\textwidth}
\includegraphics[width=\linewidth]{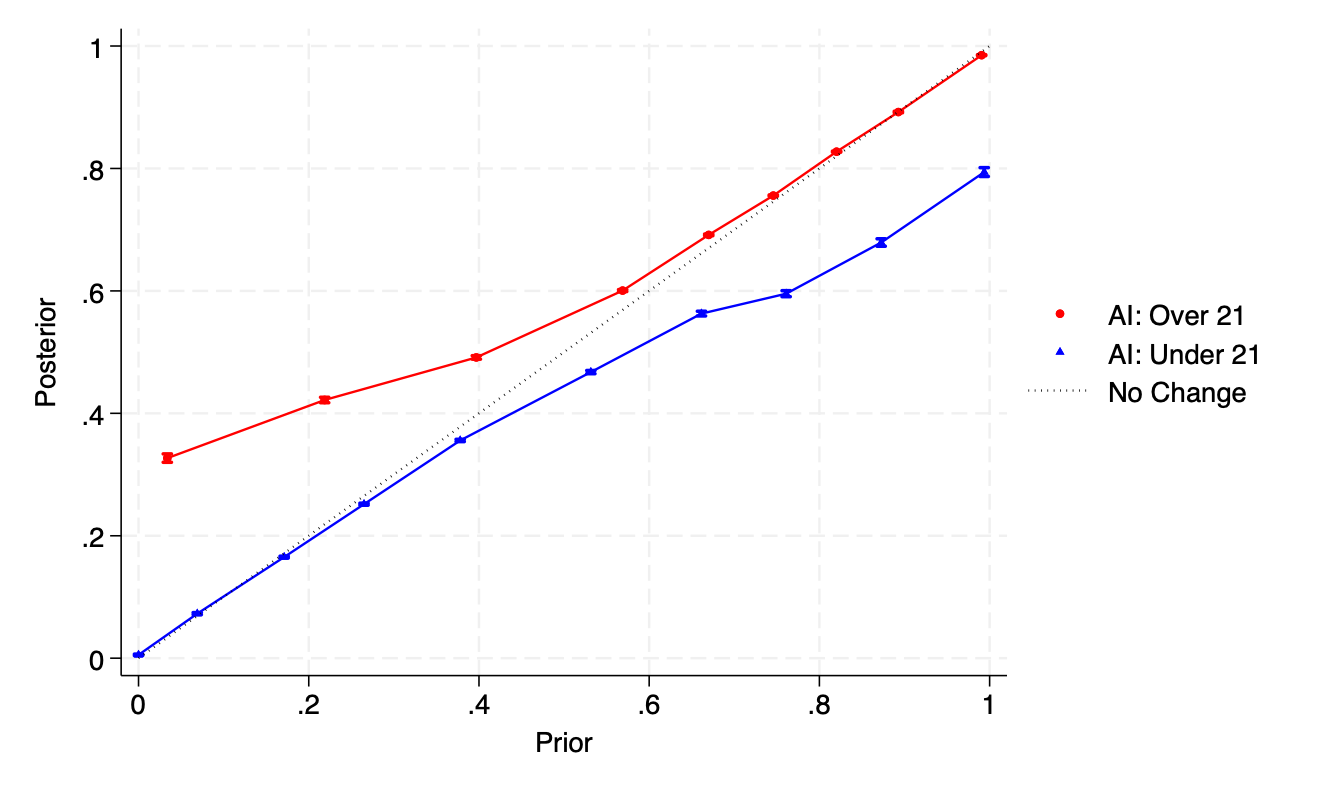}
\caption{Data (Average Beliefs)}
\label{fig:main-data}
\end{subfigure}
\begin{subfigure}[t]{0.49\textwidth}
\includegraphics[width=\linewidth]{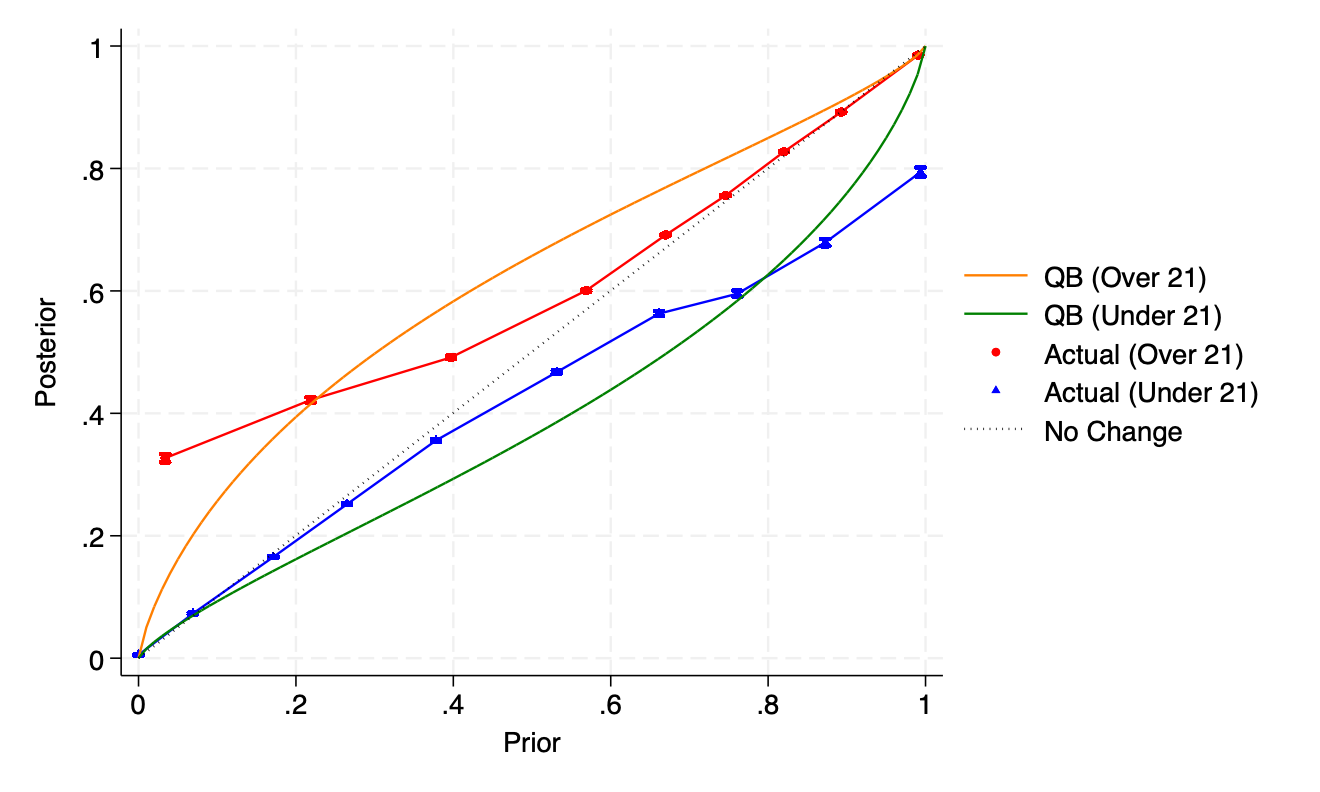}
\caption{Quasi-Bayesian (qB) Prediction}
\label{fig:main-qb}
\end{subfigure}
\begin{subfigure}[t]{0.49\textwidth}
\includegraphics[width=\linewidth]{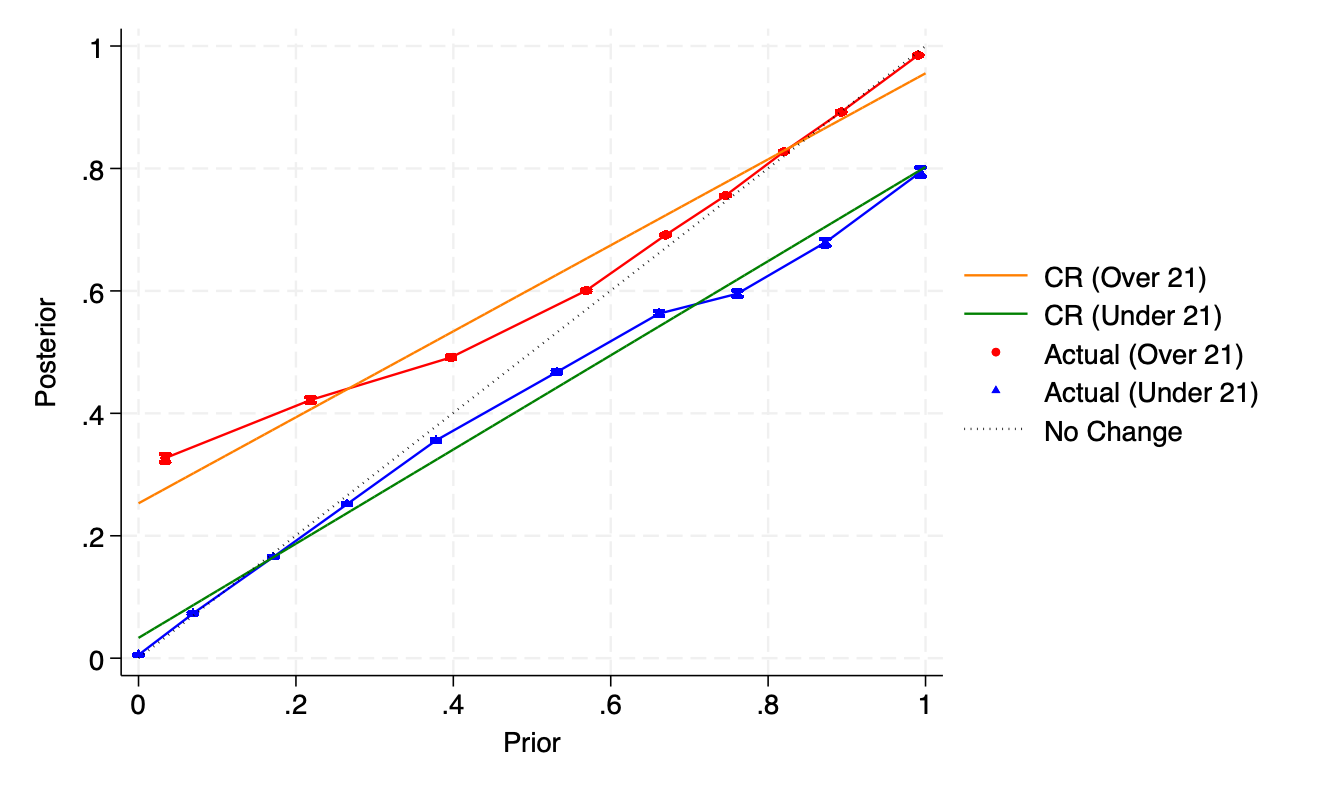}
\caption{Contraction Rule (CR) Prediction}
\label{fig:main-cr}
\end{subfigure}
\begin{subfigure}[t]{0.49\textwidth}
\includegraphics[width=\linewidth]{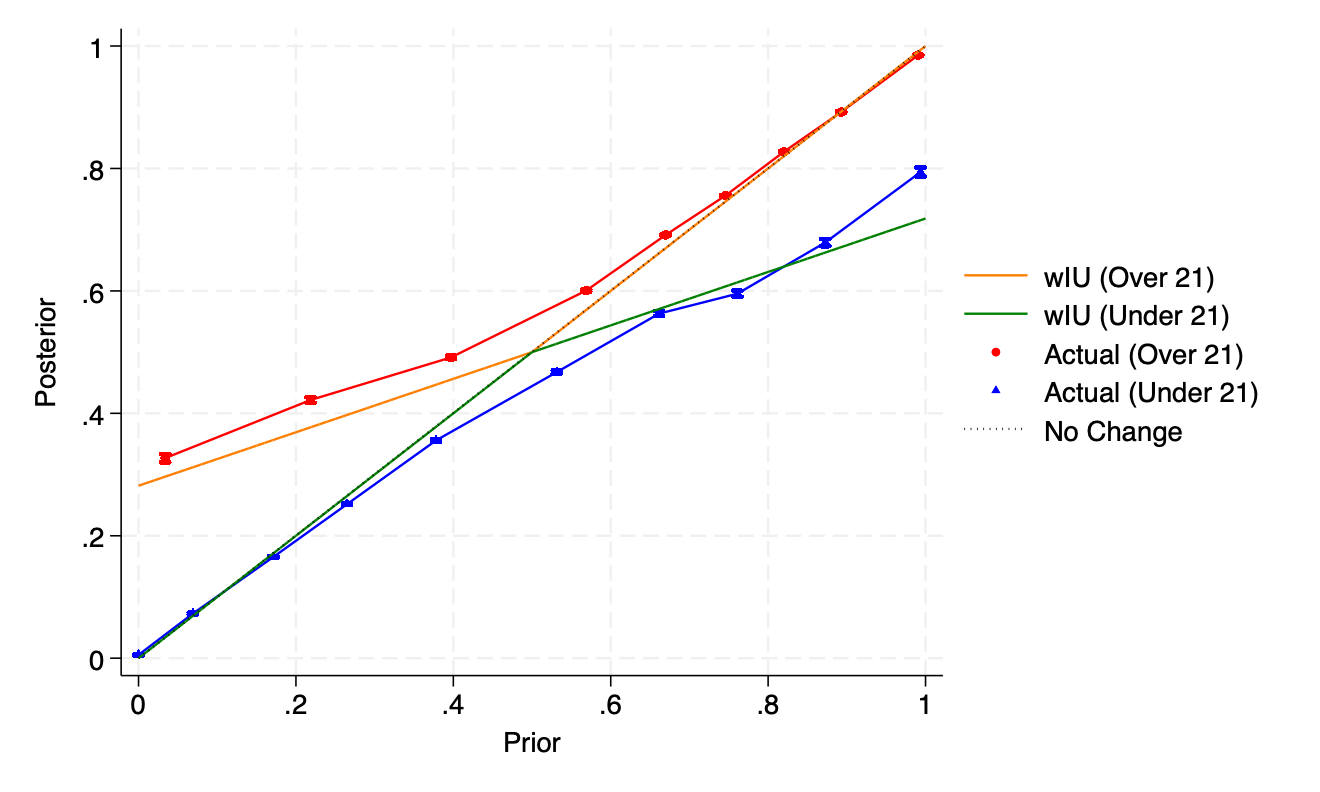}
\caption{weighted Inertial Updating (wIU) prediction}
\label{fig:main-wiu}
\end{subfigure}
\caption{Aggregate average beliefs and estimated predictions for all models by AI recommendation. For each recommendation, prior beliefs are binned by decile and then the average prior and posterior beliefs for that bin are plotted. For average beliefs, the curve above the 45-degree line (i.e., red curve) is for the Over 21 recommendation, and the curve below the 45-degree line (i.e., blue curve) is for the Under 21 recommendation. Vertical bars show $\pm 1$ standard error of the posterior mean for the empirical average belief series.}
\label{fig:main}
\end{figure}

These three behavioral patterns are illustrated in Figure~\ref{fig:main-data}, which shows the average prior and posterior for each decile of prior belief by recommendation in the aggregate data (pooled across participants and treatments). The red dots above the 45-degree line (connected by solid red line segments) are for the Over 21 recommendation, and the blue triangles below the 45-degree line (connected by solid blue line segments) are for the Under 21 recommendation. Consistent with the first pattern, average posteriors are close to the prior (close to the 45-degree line) when the AI recommendation is Over 21 and the prior is in the upper part of the belief distribution. Symmetrically, when the AI recommendation is Under 21 and the prior is in the lower part of the belief distribution, average posteriors are also close to the 45-degree line. On the other hand, average posterior beliefs are farther from the 45-degree line when the prior is extreme and contradicts the AI recommendation. That is, when the prior is close to zero and the recommendation is Over 21 or when the prior is close to one and the recommendation is Under 21. In addition, when priors are intermediate (when they are closer to one half), average posterior beliefs move away from the 45-degree line, but not to the same extent as for extreme priors and contradictory recommendations. We see these three behavioral patterns in both treatments (see Figure~\ref{fig:treatment}) and in splits of our data by AI correctness, gender, age, response time, beliefs about AI accuracy, confidence in own prior, task difficulty, round, and ability (see Appendix~\ref{app:splits}).

These three behavioral patterns suggest four candidate properties of belief updating. The first is \emph{consistency} with the AI recommendation: average posterior beliefs move weakly toward the recommendation, rising after an Over 21 recommendation and falling after an Under 21 recommendation. The second is \emph{monotonicity}: for each recommendation, average posterior beliefs are weakly increasing in prior beliefs. The third is \emph{reactionary updating}: average movement toward the recommendation is weakly larger when the recommendation more strongly contradicts the prior, so updating after Over 21 is weakly larger at lower priors and updating after Under 21 is weakly larger at higher priors. The fourth is \emph{threshold updating}: for sufficiently confirming priors, average updating becomes indistinguishable from zero. We call this point the \emph{non-updating threshold}, and it is a behavioral parameter that distinguishes behavior across high and low priors for each recommendation.

These properties are testable, and we evaluate all four properties (consistency, monotonicity, reactionary updating, and threshold updating) at the aggregate level and for all splits of the data, including by treatment. In the aggregate data, the first three properties are not violated. The mean signed update is positive toward the recommendation ($0.070$), the smaller posterior-prior slope within a recommendation is positive ($0.702$), and the update is increasing in the degree to which the recommendation contradicts the prior ($0.263$). The $p$-values from one-sided tests for weak violations are $>0.999$ in all three cases.\footnote{At the observation level, 98.1\% of updates are weakly consistent with the AI recommendation.} Using hinge regressions, we estimate non-updating thresholds of 0.620 after the Over 21 recommendation and 0.310 after the Under 21 recommendation. These thresholds are significantly different from the relevant degenerate boundaries ($p<0.001$). The evidence at the individual level is more mixed. Under our criteria for participants, 100.0\%, 97.3\%, 99.7\%, and 46.7\% of participants satisfy the tests of consistency, monotonicity, reactionary updating, and threshold updating, respectively, and 44.3\% satisfy all four criteria. Another 55.2\% are near misses who fail exactly one criterion. Thus, the results at the individual level show heterogeneity in which properties participants satisfy. We also find evidence of heterogeneity in the estimated non-updating thresholds across participants.

Next, we examine how well the updates in our experiment are captured by several reduced-form models of belief updating.\footnote{As is often done in the belief updating literature, we treat these updating rules as functional forms to be fit to data and not as tests that reveal a unique behavioral microfoundation. Importantly, our goal is to compare the empirical fit of these models in an environment with qualitative AI recommendations, not to identify the exact psychological process that generates updating. In Section~\ref{sec:lit}, we discuss several possible behavioral microfoundations and view separating them as a useful direction for future work.} We start by considering an approach used to study belief updating when the DGP is known. In this model, which we call the \emph{quasi-Bayesian} (qB) model, we assume the state space is binary (e.g., Over 21 or Under 21), the DM holds a single (potentially miscalibrated) prior over the state space, the DM forms a static (potentially misspecified) mental model of the DGP, and the DM updates their beliefs using a variant of Bayes' rule.\footnote{One could consider multiple-prior models, multidimensional state spaces, and dynamic mental models, but the data collected from our belief updating experiment are not designed to study those dimensions.} To accommodate a range of variants of Bayes' rule, we generalize Grether's rule \citep*{grether1980bayes}, a generalization of Bayes' rule. This specification allows for many updating distortions, including underreaction to signals, overreaction to signals, and base-rate neglect, and the parameters of the model are straightforward to estimate.\footnote{We use a functional form that also allows for wishful thinking \citep*{kovach2020twisting}.} In the setting of AI assistance, \citet*{agarwal2023combining} apply qB to radiologist beliefs about pathologies in X-rays and find that this model has lower fit error than several other models of belief updating once signal-dependence neglect (a form of correlation neglect) is additionally assumed.\footnote{When the DGP is unknown, there is no uniquely defined Bayesian benchmark without additional assumptions about the decision maker’s perceived signal structure, and sufficiently general Bayesian models may have few testable implications in such settings. We therefore focus on the qB specification used by \cite{agarwal2023combining}, which imposes signal-dependence neglect. In Section \ref{sec:bayesianism}, we show that relaxing this restriction makes the model essentially unrestricted for interior priors while preserving endpoint implications that conflict with our data. Also, in their initial NBER working paper, \citet*{agarwal2023combining} find that the qB model with signal-dependence neglect outperforms 11 other models of belief updating.} The parameters they estimate for qB suggest underreaction to AI advice, and they interpret this as \emph{automation neglect}, in which the informational content of AI advice is underweighted relative to the prior.

We also consider two non-Bayesian proposals for how DMs revise their beliefs after receiving information from a data-generating process that is not well understood, but is nevertheless informative. The first of these is the \emph{Contraction Rule} (CR) of \citet*{ke2024learning}, which captures a fixed shrinkage toward a representative belief that is consistent with the recommendation. The other is \emph{weighted Inertial Updating} (wIU) proposed and characterized for general information by \cite*{dominiak2021minimum}. This model posits that DMs choose the belief closest to their prior that satisfies an information constraint, then partially mixes that belief with the prior to form their posterior. The authors of both papers point to belief updating from AI recommendations as a particularly relevant setting for their approaches, but we are not aware of existing empirical applications of either CR or wIU.

From Figures~\ref{fig:main-qb}--\ref{fig:main-wiu}, which plot the predictions for belief updating by recommendation for each model (estimated within sample), CR and wIU appear closer to aggregate belief updating than qB. In Section~\ref{sec:comp} we assess fit more rigorously by comparing the three models using out-of-sample performance as well as the completeness and restrictiveness measures of \cite*{fudenberg2021complete} and \cite*{fudenberg2023flexible}, respectively.\footnote{We expect variation in restrictiveness because these three models vary in their parsimony. In our experiment, qB has three parameters, CR has four, and wIU has just one (the subjective version of this model has two additional parameters due to subjective thresholds).} At the aggregate level, CR and wIU have lower out-of-sample MSE than qB (paired tests across splits, both $p<0.001$) and higher completeness measures for both treatments. This ranking is unchanged when leaving out participants, leaving out images, allowing qB to use latent priors away from zero and one, and relaxing signal-dependence neglect. wIU is also the most restrictive under both of the restrictiveness benchmarks we consider, whether the eligible class imposes monotonicity or only recommendation consistency. At the participant level, a subjective version of wIU, which has a subjective threshold that behaves similarly to the representative belief in CR, has lower average out-of-sample MSE than qB and wIU (paired tests across participants, both $p<0.001$), is statistically indistinguishable from CR in average MSE ($p=0.225$), and has higher completeness than all three models. One reason the subjective version of weighted Inertial Updating has lower average MSE than the objective one is that it has three parameters, which accounts for additional heterogeneity. However, this additional flexibility reduces its restrictiveness relative to the objective version of wIU.

The four testable properties of belief updating can help to explain the relative fit of these models. Because qB updates the prior via a distorted likelihood ratio, it predicts relatively smaller changes near extreme priors and relatively larger changes near intermediate priors. As a result, this model violates reactionary updating for non-trivial parameter values. On the other hand, CR and wIU both produce larger updates at extreme contradicting priors and smaller updates at intermediate priors, so they satisfy reactionary updating. Finally, unlike qB and CR, both objective and subjective versions of wIU directly generate threshold updating because they predict that the posterior remains equal to the prior whenever the prior already satisfies the information constraint.

Our paper contributes to the experimental literature on non-Bayesian updating by providing evidence from a controlled experiment in which information is payoff relevant, but the information structure is not transparent to participants. This distinguishes our work from recent papers in which the DGP is ambiguous or uncertain, but where the DM is told a set of possible signal structures, source accuracies, or interpretive models \citep*{liang_2025,shishkin_ortoleva_2023,aina_schneider_2025,farina2026communicating} and recent papers that study belief updating when the DGP is known, but the message itself is ambiguous  \citep*{epstein2024hard, fryer2019updating} or the prior belief is ambiguous or uncertain \citep*{ngangoue2021learning, coutts2019testing, chan2024prior,cohen2000experimental}.  

In addition, we contribute to a growing literature on the interaction between humans and AI by studying how humans update their beliefs after receiving AI recommendations. AI tools appear in workplace settings and commercial interactions \citep*{bick2026rapid, bonney2026microstructure}, and these systems typically communicate through terse recommendations (``police reported ahead,'' ``engine service required'') that convey qualitative content about an underlying state without revealing the algorithm's uncertainty about the state. Despite this growth, there is limited evidence on how decision makers update their beliefs based on the general information conveyed by qualitative AI recommendations (see Section~\ref{sec:AIAdvice}).

The remainder of the paper is organized as follows. Section~\ref{sec:design} details the experimental design. Section~\ref{sec:results} presents the high-level results of the experiment.  Section~\ref{sec:pred} describes the three models and presents their estimates and predictions. Section~\ref{sec:comp} assesses the three models using our experimental data.  Section~\ref{sec:lit} situates our study within literatures on non-Bayesian updating and on human-machine interaction and discusses possible behavioral microfoundations. Section~\ref{sec:conclusion} concludes with a discussion of the possible implications and directions for future work.

\section{Experimental Design}\label{sec:design}

\subsection{Task}

In each round of our experiment, participants were shown an image of a human face and asked to make two decisions based on it. In the instructions, participants were told that all images were taken between 2010 and 2014. In half of the images, the person was over 21 years old at the time the picture was taken, and in the other half of the images, the person was under 21 years old. Participants were told that the people in the images were either over 21 or under 21 when the picture was taken, but not the fraction of people that were over 21 or under 21.\footnote{A screenshot of the exact instructions can be found in Appendix~\ref{app:screenshots}.}

The first decision participants were asked to make was to report the probability that the person in the image was over 21 at the time that the picture was taken, which we refer to as their \emph{prior belief}. Participants were provided a slider bar to indicate their belief and, to minimize anchoring effects, the slider bar had no preset value. A screenshot of the decision task can be found in Appendix~\ref{app:screenshots}.

After submitting their prior belief, participants were shown an AI recommendation as a pop-up message. This recommendation was framed as the prediction of an ``AI Assistant''. As previously stated, there were two possible recommendations: ``More likely Over 21'' and ``More likely Under 21''.
After closing the pop-up message, the recommendation was presented in a gray bar above the submit button.

Once participants closed the message box, the second decision task began. In this decision, they were allowed to revise their belief before submitting again, which we refer to as their \emph{posterior belief}. Importantly, participants were not required to change their beliefs and were reminded in the pop-up message that ``you can change your reported probability, but you are not required to do so''.

After submitting their posterior, participants proceeded to the next round and could not go back and change any of their previous choices. Altogether, participants completed 160 rounds of this task, broken into blocks of 20 rounds. To minimize fatigue effects, participants were given the opportunity to pause the experiment after each block of rounds.  For each participant, images were randomly allocated to rounds.

To incentivize truthful reporting of beliefs, a single round was selected (all equally likely) and either the prior or posterior was randomly selected from that round (both equally likely).\footnote{In the pop-up message, participants were reminded that ``your reports before and after seeing the AI Assistant's recommendation are equally likely to be selected for payment''.}  To determine payment based on the randomly selected belief and true age of the person in the image, we used the incentive-compatible binarized scoring rule \citep*{hossain2013binarized} to calculate the probability of getting a \$6 bonus payment. \citet*{danz_vesterlund_wilson_2022} show that detailed information about binarized scoring rule (BSR) incentives can increase deviations from truthful reporting, hence we initially provided no information about the payment rule aside from telling participants that they would maximize the likelihood of receiving the bonus payment by truthfully reporting their belief in each round. However, participants were given the opportunity to learn more about the payment rule after completing the experiment.

Participants were given 60 seconds to complete each round, and they were informed that if they did not complete a round within 60 seconds and that round was randomly selected for payment, then they would have no chance of receiving the bonus payment. Over 98\% of rounds were completed within the time limit, and the rounds in which the time limit was hit were excluded from our analysis sample.

\subsection{AI Recommendations}

To generate the AI recommendation that the person in the image is more likely to be over 21 or under 21, we used the outputs of the ``Caffe'' model of \citet*{rothe2018deep}. This model was trained to predict the ages of humans based on images of faces, and it outputs an age-probability score between 0 and 1 for each possible age, where the scores sum to 1 across all possible ages. To generate a recommendation for an image, we summed the age-probability scores for all ages over 21. The AI Assistant gave the ``More likely Over 21'' recommendation whenever this sum was above 50\% and otherwise gave the opposite recommendation. The AI model probability of Over 21 was never exactly 50\%.

For our primary treatment, which we refer to as the NOINFO treatment, we provided participants no information about the underlying model used to generate AI recommendations, nor any information about how its predictions were mapped into AI recommendations. For instance, we did not tell participants anything about the accuracy of the AI or whether the AI recommendations were balanced across recommendations. The reason for this was twofold. First, we wanted to maximize the extent to which the DGP was viewed as a ``black box'' by the participants, which is the focus of our research question. Second, in many situations in which individuals receive AI recommendations outside of our experiment, such as the results of an Internet search, a music recommendation, or a navigation route, they are not told about how the results are generated. 

As a complement to this treatment, we also conducted an INFO treatment, in which participants were provided with several pieces of information about the AI recommendations. First, they were told the overall accuracy of the AI (``the AI’s prediction matches whether the person was actually under or over 21 approximately 76\% of the time'') and were given qualitative information of the AI's relative performance (``based on past studies, this performance is better than most humans, but is worse than the very best humans''). Following \citet*{agarwal2023combining}, we also told participants that the AI did not have extra information beyond the picture (``it uses only the image to predict the person's age'') and who programmed the AI (``the tool is based on state-of-the-art machine learning algorithms developed by a leading team of researchers at the Computer Vision Lab at the Federal Institute of Technology in Zurich, Switzerland''). In this treatment, participants were also informed of the AI's state-conditional rates of correct recommendations alongside the AI advice in the pop-up message in each round: ``when the person is actually over 21, the AI Assistant predicts ‘More likely Over 21’ about 83\% of the time'' and ``when the person is actually under 21, the AI Assistant predicts ‘More likely Under 21’ about 70\% of the time''.

\subsection{Implementation}

Our experiment was conducted on Prolific. Participants are restricted by Prolific to be at least 18 years old, and we added a requirement that they be residents of the U.S. as well. In the posting for the experiment, participants were told that the expected time to complete the study was 40--45 minutes, that the payment for completing the experiment was $\$6$, and that they could possibly receive a bonus payment of $\$6$ based on their decisions in the experiment.

Participants who chose to complete our experiment were shown a consent form that provided additional details about our experiment. Participants who chose to participate after viewing the consent form were then asked two questions as a standard attention check. Failure to correctly answer either question on the first try resulted in participants being prevented from completing the experiment.

After completing all 160 rounds, but before being informed of which round was selected for payment and whether they received the bonus payment of $\$6$, participants were asked to complete a short survey that was not incentivized. In that survey, participants were first asked three questions about the accuracy of the AI, such as: ``How often did the AI Assistant's prediction match whether the person was actually under or over 21?'' The slider bar corresponding to that question ranged from ``0\% of the time'' to ``100\% of the time''. On the next page of the survey, participants were asked three questions about their confidence in their own prior beliefs. The slider bar for that question ranged from ``0 not confident'' to ``100 extremely confident''. The third page of the survey asked the same three questions, but now about final beliefs. On the last page of the survey, participants were asked if they had been paid to classify images before, at what point they felt fatigued, and what other feedback they had on the instructions or experiment more generally. After completing the survey, each participant was told which round and decision was randomly selected for payment and whether they received the bonus for the randomly selected decision.

\subsection{Design Decisions}

\hspace*{.5cm} \emph{Task Choice}: We chose the Bouncer task for several reasons. First, it has a binary ground truth, so performance can be assessed. Second, it requires no specialized training, which allowed us to obtain 160 reports from each participant. Third, when implemented previously by \citet{caplin2024abc}, it generated a wide range of beliefs at the participant and image level without AI recommendations.

\emph{Image Selection}: We used the same 160 images from those used in \citet{caplin2024abc}. These images were originally drawn from the IMDB-WIKI dataset, which is a large publicly available face image dataset with age labels. Importantly, none of the 160 images that we used were of people who were exactly 21 years old when the picture was taken.\footnote{Appendix Table~\ref{tab:summary-treatment-balance} reports summary statistics at the image level for this selection of images.}

\emph{Belief Elicitation}: The belief elicitation literature contains several incentive-compatible methods, including the binarized scoring rule (BSR), the multiple-price list (MPL), and the Becker-DeGroot-Marschak (BDM) method. We chose the BSR because MPL and BDM are less practical in our setting, given our desire to elicit two beliefs for 160 images.

\emph{Number of Rounds and Time Limit}: To determine the number of rounds and to test the time limit, we ran a pilot session on Amazon Mechanical Turk (MTurk) with a similar design to our final design. We did not see evidence of dynamic effects or time constraints in that pilot, so we kept those parameters the same.

\emph{Information about the AI}: 
The NOINFO treatment matches the environment that motivates CR and wIU, since these models are designed for settings in which the DGP is unknown. The INFO treatment, by contrast, provided participants with the state-contingent recommendation rates. In our pilot session, we did not see treatment differences even when we provided additional information about the AI, such as the accuracy by recommendation. However, we decided to keep this treatment variation in our final experiment to verify the robustness of our results along this dimension.

\section{Experimental Results}\label{sec:results}

In total, 377 participants completed our experiment, of which 189 were in the NOINFO treatment and 188 were in the INFO treatment. Participants were required to be U.S. residents and at least 18 years old. The median self-reported age was 35 years old, 59\% self-reported being women, and 62\% self-reported being White. We find no significant differences in these demographic characteristics across treatments: average age ($p=0.071$), fraction women ($p=0.490$), and fraction White ($p=0.485$).\footnote{Appendix Table~\ref{tab:summary-treatment-balance} reports balance across participants on these and other dimensions (e.g., survey measures, accuracy, response times, non-updating rates).}

This sample yielded a total of $60{,}320$ rounds ($377$ participants $\times$ $160$ rounds each). We made one additional analysis-sample exclusion: we excluded 68 rounds ($\approx 0.11\%$) because the 60-second time limit was hit.\footnote{The rate of hitting the time limit was not significantly different across treatments ($p=0.983$).} Hence, we analyzed 60,252 pairs of prior and posterior beliefs. From the start of the round, the median and mean response times for prior beliefs were 4 seconds and 4.8 seconds, respectively, and for the posterior beliefs were 7 seconds and 8.4 seconds, respectively.\footnote{See Figure~\ref{fig:update-time} for the incremental response time by prior and AI recommendation.} The median time to complete the experiment was 40 minutes. The average payment to participants was approximately \$10. 

\begin{figure}
    \centering

\begin{subfigure}[t]{0.49\textwidth}
\includegraphics[width=\textwidth]{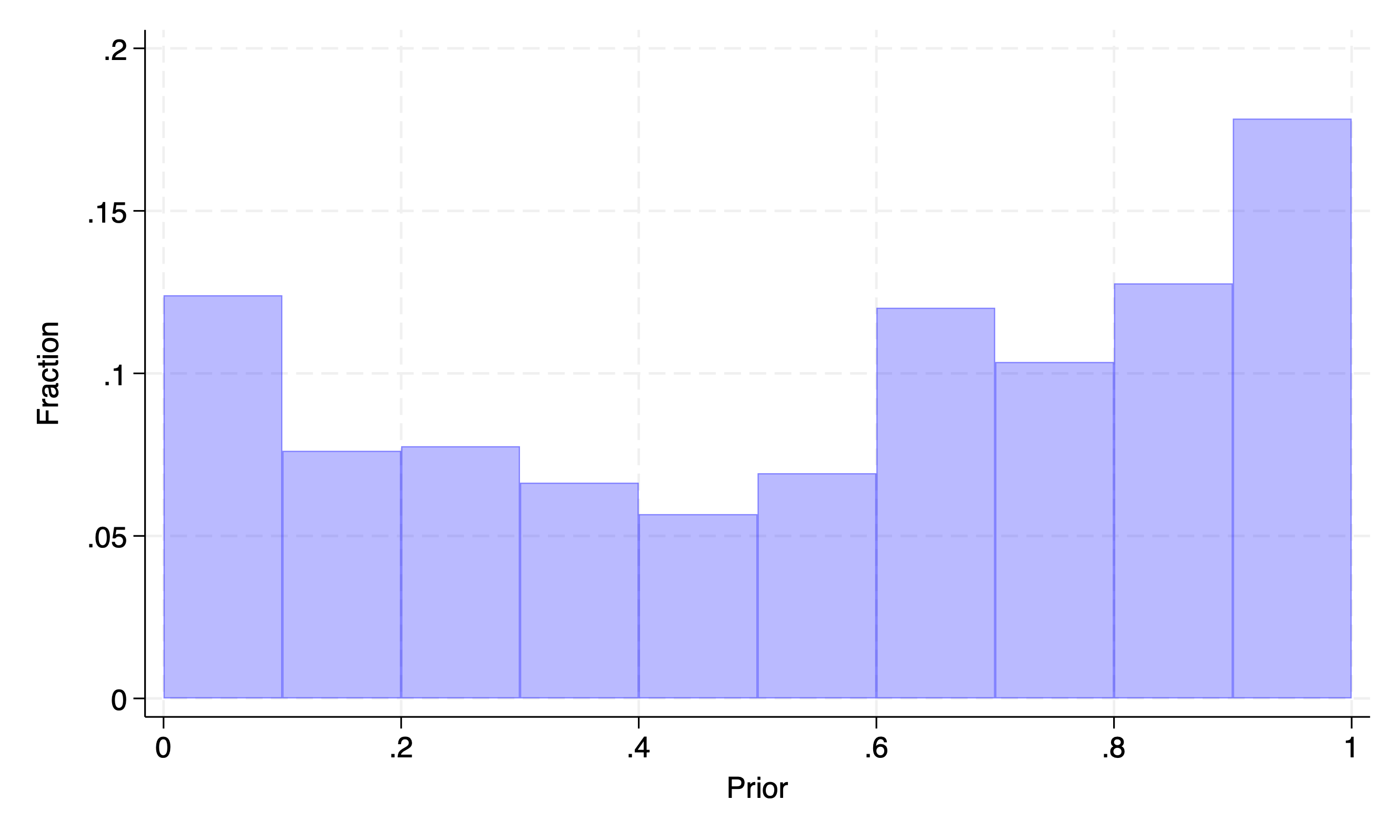}
\caption{Distribution of prior beliefs.}
\label{fig:prior-belief-distribution}
\end{subfigure}
\begin{subfigure}[t]{0.49\textwidth}
\includegraphics[width=\textwidth]{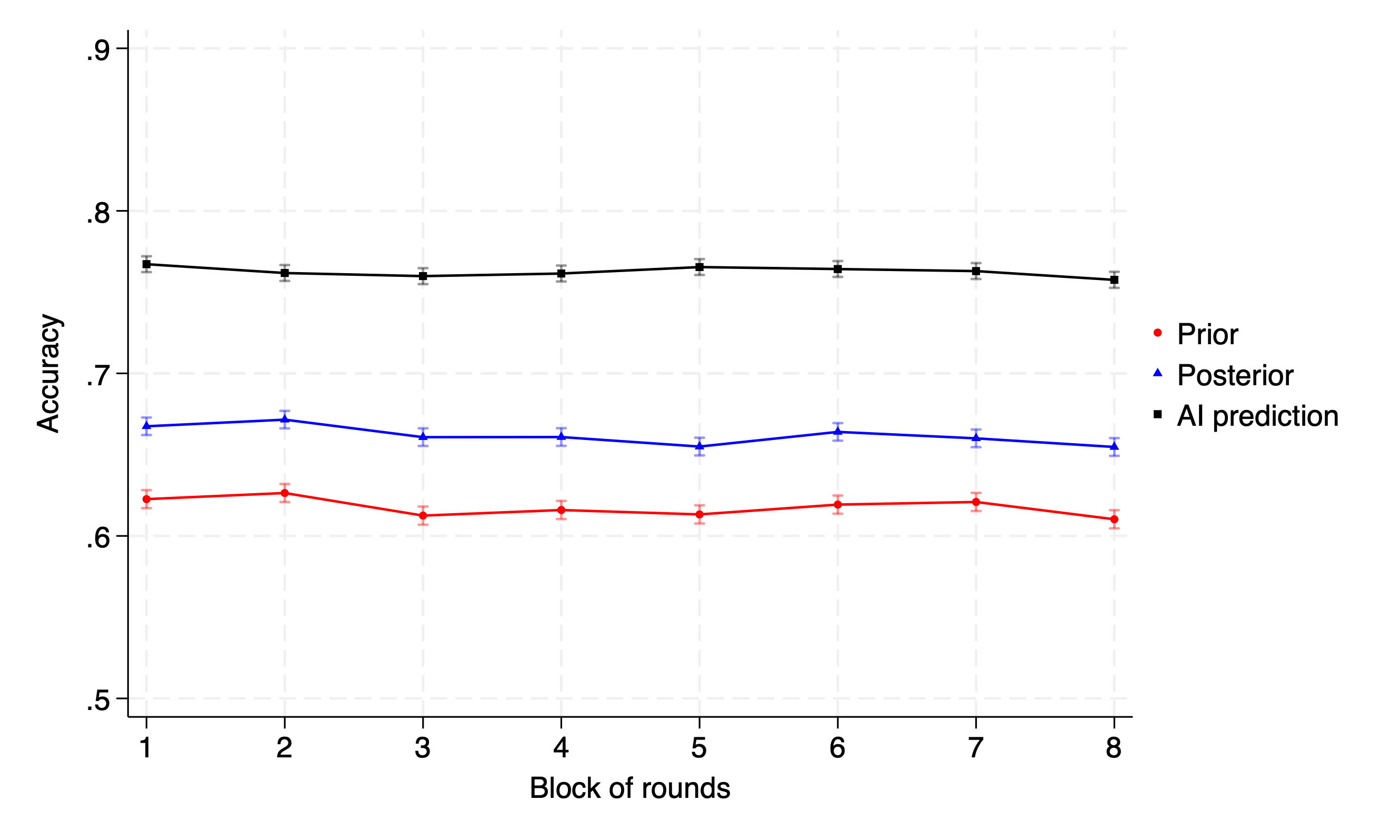}
\caption{Dynamics of binary accuracy.}
\label{fig:binary-accuracy-dynamics}
\end{subfigure}
\caption{Prior beliefs and binary accuracy in the aggregate data (pooled across participants, images, and treatments).}
\label{fig:prior-accuracy}
\end{figure}

\subsection{Prior Beliefs and Binary Accuracy} 

The distribution of prior beliefs that the person in the image was over 21 is plotted in Figure~\ref{fig:prior-belief-distribution}. We find that priors are spread across the distribution, and one measure of this spread is the largest fraction of observations in a bin. For 10 bins with a width of 10 p.p., the smallest possible fraction is 0.1 and the largest possible fraction is 1. For the aggregate data, that fraction is 17.8\%. In addition to seeing variation in priors when aggregating across participants and images, we also see variation within participants across images (mean largest bin fraction=35.4\%) and within images across participants (mean largest bin fraction=22.8\%). We do not find that prior beliefs are concentrated on the salient probabilities of 0\%, 25\%, 50\%, 75\%, and 100\%. For 25 bins with a width of 4 p.p., probabilities in the 96\%-100\% range were reported with the highest frequency (bin fraction=14.1\%) and probabilities in the 46\%-50\% range were reported with the lowest frequency (bin fraction=2.5\%).  

Based upon participants' prior and posterior beliefs, we can directly measure the accuracy of participants' beliefs by using a binary decision threshold. We follow \cite{caplin2024abc} and \cite{agarwal2023combining} in treating a participant's belief as correct in binary terms if the person in an image is older (younger) than 21 and the belief is higher (lower) than 50\%, and that the belief is 50\% correct if it is exactly 50\%. Similarly, we say an AI recommendation is correct if the person in an image is older (younger) than 21 and the recommendation is Over 21 (Under 21). The binary accuracy for the participants' priors and posteriors is 61.8\% and 66.2\%, while the binary accuracy of AI recommendations is 76.3\%. The average binary accuracy of priors and posteriors is not significantly different across treatments ($p=0.321$ and $p=0.822$, respectively). We find no significant differences in binary accuracy by the demographic characteristics, even when accounting for the gender and race of the individual in the image. Joint tests of participant gender, age, and race yield $p=0.686$ for prior accuracy and $p=0.474$ for posterior accuracy. AI recommendations have higher binary accuracy than 98.4\% of participants' prior beliefs and 91.0\% of participants' posterior beliefs, and lower binary accuracy than 1.6\% and 6.6\%, respectively. From Figure~\ref{fig:binary-accuracy-dynamics}, we do not see evidence of linear trends in the accuracy of participants' beliefs or AI recommendations (regression of round block onto prior accuracy: $p=0.192$, posterior accuracy: $p=0.057$, AI accuracy: $p=0.489$). Since the order of pictures is randomized and we do not observe evidence of dynamic changes in belief accuracy, we pool observations across rounds for subsequent analyses. In addition, Figure~\ref{fig:update-round} does not show changes in belief updating across the experiment.\footnote{In Appendix~\ref{app:dynamic-wrong-ai}, we check for evidence of dynamic changes in updating after AI advice that appears to be erroneous (e.g., AI advice contradicts an extreme prior and the participant updates little). We find suggestive evidence of lower updating on the next task after such events, but controlled one-round associations are close to zero.}

\subsection{Aggregate Tests of the Four Properties}\label{sec:threepatterns}

Figure~\ref{fig:main-data} shows three behavioral patterns that motivate the tests below. First, participants update close to zero when the recommendation confirms an extreme prior. For example, when the prior is at least 90\% and the recommendation is Over 21, the average absolute update is just 0.008. The same is similarly small when the prior is at most 10\% and the recommendation is Under 21. The hinge estimates below summarize this pattern with non-updating thresholds: average updating becomes indistinguishable from zero above 0.620 after an Over 21 recommendation and below 0.310 after an Under 21 recommendation. Second, updates are larger when the recommendation contradicts an extreme prior. When the prior is at most 10\% and the recommendation is Over 21, the average posterior increases from 0.025 to 0.318, an update of 0.292 toward the recommendation. When the prior is at least 90\% and the recommendation is Under 21, the average posterior falls from 0.976 to 0.776, an update of 0.200 toward the recommendation. Third, contradicting recommendations generate smaller updates for more moderate priors. In the nearest ten-point prior intervals on the contradicting side of 50\%, updating is about 0.080 after an Over 21 recommendation and 0.078 after an Under 21 recommendation. Thus, updating after contradicting recommendations is roughly 12--21 percentage points larger at extreme priors than at moderate priors.

These three patterns suggest four properties of belief updating. The first two are regularity properties. \emph{Consistency} requires average posterior beliefs to move weakly in the direction of the AI recommendation: after the recommendation of Over 21, average posteriors should be weakly higher than priors, and after the recommendation of Under 21, average posteriors should be weakly lower. \emph{Monotonicity} requires average posteriors to be weakly increasing in prior beliefs, separately by recommendation. The next two properties describe the shape of updating. \emph{Reactionary updating} requires weakly larger updating when the AI recommendation is more contradictory to the prior. For recommendation of Over 21, this means lower priors. For recommendation of Under 21, this means higher priors. \emph{Threshold updating} requires a non-updating threshold: after an Over 21 recommendation, there is a threshold above which average posteriors are indistinguishable from priors, and after an Under 21 recommendation, there is a threshold below which average posteriors are indistinguishable from priors.

We test each property using regressions at the observation level with two-way clustered standard errors by participant and image. The tests differ in what kind of evidence they are designed to provide. For consistency, monotonicity, and reactionary updating, the properties are weak inequalities, so the tests look for evidence of rejection. Consistency is rejected only if the mean signed update is significantly negative, where signed updating is defined as movement toward the AI recommendation. Monotonicity is rejected only if either posterior-prior slope within a recommendation is significantly negative. Reactionary updating is rejected only if the slope of signed updating on contradiction is significantly negative. Thus, for these first three properties, large one-sided $p$-values mean that we do not reject the property. Threshold updating is different: it requires affirmative evidence that the no-updating region is nondegenerate. For this property, we estimate hinge regressions for each recommendation that allow updating on the contradicting side of a threshold and no updating on the confirming side. The estimated hinge location is the regression estimate of the non-updating threshold. We require the thresholds to be significantly away from the degenerate boundary: below 1 after an Over 21 recommendation and above 0 after an Under 21 recommendation, where thresholds are reported as probabilities of Over 21.

Table~\ref{tab:aggregate-property-tests} reports the results. The aggregate evidence supports the four properties, but in different senses. For the first three properties, we fail to reject the required weak inequalities: the mean signed update is 0.070, the smaller of the two slopes of posterior beliefs on prior beliefs within a recommendation is 0.702, and the slope of signed updating on contradiction is 0.263. Because these are tests for weak violations, the corresponding one-sided p-values test for negative violations. All three are $>0.999$. For threshold updating, the test instead looks for evidence of passing. The hinge regressions place the non-updating threshold at 0.620 after the Over 21 recommendation and 0.310 after the Under 21 recommendation, and both thresholds are nondegenerate ($p<0.001$).

\begin{table}[ht!]
\centering
\caption{Regression tests of the four properties.}
\label{tab:aggregate-property-tests}
\footnotesize
\begin{tabular}{llcl}
\toprule
Property & Test & Estimate & $p$-value \\
\midrule
Consistency & Mean signed update & 0.070 & $>0.999$ \\
Monotonicity & Min. posterior-prior slope & 0.702 & $>0.999$ \\
Reactionary updating & Contradiction slope & 0.263 & $>0.999$ \\
Threshold updating & Hinge thresholds & Over 0.620; Under 0.310 & $<0.001$ \\
\bottomrule

\end{tabular}
\begin{minipage}{0.95\textwidth}
\footnotesize \emph{Notes:} All tests use regressions at the observation level with two-way clustered standard errors by participant and image. Signed updating is defined as movement toward the AI recommendation. For consistency, monotonicity, and reactionary updating, the $p$-values are one-sided tests for negative coefficients, so large values indicate no rejection. The monotonicity estimate is the smaller of the Over 21 and Under 21 slopes of posterior beliefs on prior beliefs. The threshold estimates come from hinge regressions for each recommendation. The threshold p-value tests for passing: it is the maximum one-sided p-value for thresholds away from the degenerate boundary, with the Over 21 threshold below 1 and the Under 21 threshold above 0.
\end{minipage}
\end{table}

\subsection{Treatments and Splits} 

Table~\ref{tab:ai-belief-treatment-effects} reports treatment differences in survey beliefs about AI accuracy collected after the experiment and compares those beliefs with the empirical rates.\footnote{See Appendix~\ref{app:survey} for more details on survey responses.} The overall belief about AI accuracy is not significantly different across treatments (68.2 in INFO and 67.0 in NOINFO, $p=0.544$). However, INFO increases the stated belief that the AI is accurate after an Over 21 recommendation by 5.8 percentage points ($p=0.009$), and increases the analogous belief after an Under 21 recommendation by 3.9 percentage points, though the latter difference is not statistically significant ($p=0.111$). INFO participants' beliefs are closer to the empirical rates than NOINFO participants' beliefs for all three measures, but the beliefs are still attenuated relative to the provided and empirical rates, with the largest attenuation for Under 21 recommendations. Thus, the INFO treatment moves beliefs about AI accuracy in the expected direction for Over 21 recommendations and improves recall or calibration, but those belief differences do not translate into a meaningfully different aggregate updating function.

\begin{table}[ht!]
\centering
\caption{Beliefs about AI accuracy by treatment and empirical calibration.}
\label{tab:ai-belief-treatment-effects}
\footnotesize
\setlength{\tabcolsep}{1pt}
\begin{tabular*}{\textwidth}{@{\extracolsep{\fill}}p{0.28\textwidth}ccccccc@{}}
\toprule
Belief measure & \shortstack{Empirical\\rate} & \shortstack{INFO\\belief} & \shortstack{NOINFO\\belief} & \shortstack{INFO $-$\\NOINFO} & $p$-value & \shortstack{INFO abs.\\error} & \shortstack{NOINFO abs.\\error} \\
\midrule
Overall AI accuracy & 76.3 & 68.2 (19.0) & 67.0 (22.1) & 1.3 & 0.544 & 13.8 & 18.4 \\
AI accuracy after Over 21 recommendation & 73.3 & 70.9 (18.4) & 65.1 (24.6) & 5.8 & 0.009 & 14.4 & 20.5 \\
AI accuracy after Under 21 recommendation & 80.0 & 62.2 (22.7) & 58.3 (25.3) & 3.9 & 0.111 & 21.6 & 26.6 \\
\bottomrule

\end{tabular*}
\begin{minipage}{0.95\textwidth}
\footnotesize
\emph{Notes:} The unit of observation for beliefs is a participant. Entries in the INFO and NOINFO columns are means with standard deviations in parentheses. Beliefs are measured on a 0--100 scale in the survey after the experiment. The $p$-values are two-sided Welch tests of equality across treatments. Empirical rates are calculated from the 160 experimental images. Absolute errors compare each participant's stated belief about AI accuracy with the corresponding empirical rate.
\end{minipage}
\end{table}

\begin{figure}
\begin{center}
\includegraphics[width=4.5in]{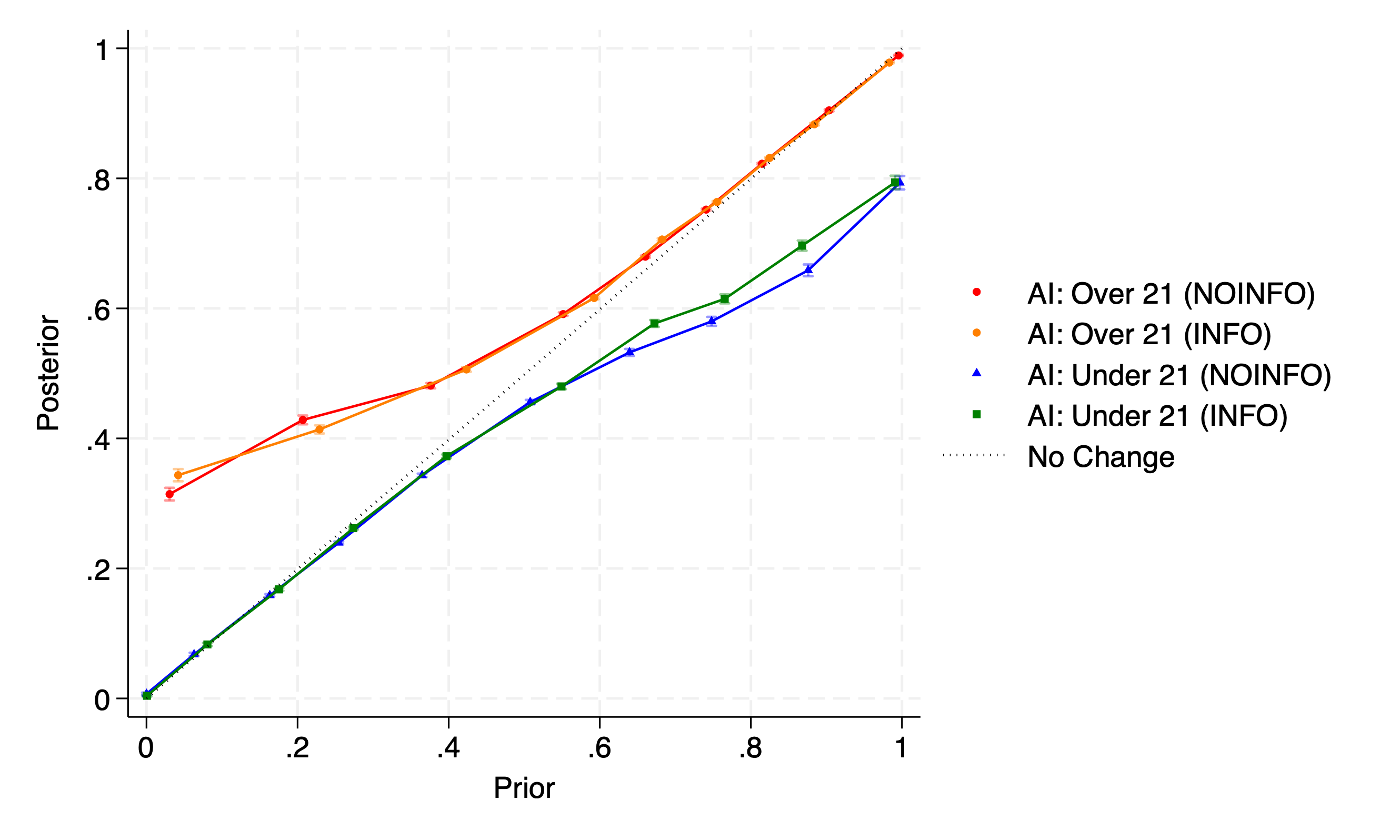}
\end{center}
\caption{Updating by treatment.}
\label{fig:treatment}
\end{figure}

However, as illustrated in Figure~\ref{fig:treatment}, updating is similar across treatments. Table~\ref{tab:property-tests-splits} re-estimates the four regression tests for the aggregate sample and for the main treatment, participant, and image splits. The treatment split separates the INFO and NOINFO treatments. The AI correctness split is at the observation level and separates cases where the AI recommendation matches the true age category from cases where it does not. The AI confidence and image difficulty splits are at the image level: an image is ``high AI confidence'' if the average value of $\max\{p,1-p\}$, where $p$ is the AI model probability of Over 21, is above the median image value of 0.772, and an image is ``hard'' if the image mean squared error of prior beliefs is above the median image value of 0.267. The round split separates rounds 81--160 from rounds 1--80. The response time split is at the participant level and classifies participants as ``fast'' if their median response time for prior beliefs is below the sample median of 3.5 seconds. The remaining splits at the participant level classify participants as women, older than the median age of 35, high ability if their MSE for prior beliefs is below the median participant value of 0.267, high belief about AI accuracy if their stated belief about overall AI accuracy is above the median of 71, and high confidence in their own prior if their stated confidence in their initial beliefs is above the median of 78. The four properties hold in the aggregate and in every split, so the qualitative shape of updating is stable across these different margins.

\begin{table}[ht!]
\centering
\caption{Behavioral property tests in aggregate and by split.}
\label{tab:property-tests-splits}
\footnotesize
\setlength{\tabcolsep}{3pt}
\begin{tabular*}{\textwidth}{@{\extracolsep{\fill}}lrrrrrrr@{}}
\toprule
Sample & $N$ & Cons. & Mono. & React. & Thresh. & Over thresh. & Under thresh. \\
\midrule
All & 60,252 & \ensuremath{\checkmark} & \ensuremath{\checkmark} & \ensuremath{\checkmark} & \ensuremath{\checkmark} & 0.62 & 0.31 \\
NOINFO & 30,206 & \ensuremath{\checkmark} & \ensuremath{\checkmark} & \ensuremath{\checkmark} & \ensuremath{\checkmark} & 0.64 & 0.31 \\
INFO & 30,046 & \ensuremath{\checkmark} & \ensuremath{\checkmark} & \ensuremath{\checkmark} & \ensuremath{\checkmark} & 0.59 & 0.30 \\
AI correct & 45,946 & \ensuremath{\checkmark} & \ensuremath{\checkmark} & \ensuremath{\checkmark} & \ensuremath{\checkmark} & 0.62 & 0.31 \\
Not AI correct & 14,306 & \ensuremath{\checkmark} & \ensuremath{\checkmark} & \ensuremath{\checkmark} & \ensuremath{\checkmark} & 0.62 & 0.29 \\
AI high confidence & 30,122 & \ensuremath{\checkmark} & \ensuremath{\checkmark} & \ensuremath{\checkmark} & \ensuremath{\checkmark} & 0.62 & 0.32 \\
Not AI high confidence & 30,130 & \ensuremath{\checkmark} & \ensuremath{\checkmark} & \ensuremath{\checkmark} & \ensuremath{\checkmark} & 0.62 & 0.29 \\
Hard image & 30,129 & \ensuremath{\checkmark} & \ensuremath{\checkmark} & \ensuremath{\checkmark} & \ensuremath{\checkmark} & 0.61 & 0.31 \\
Not Hard image & 30,123 & \ensuremath{\checkmark} & \ensuremath{\checkmark} & \ensuremath{\checkmark} & \ensuremath{\checkmark} & 0.62 & 0.29 \\
Second half & 30,133 & \ensuremath{\checkmark} & \ensuremath{\checkmark} & \ensuremath{\checkmark} & \ensuremath{\checkmark} & 0.60 & 0.34 \\
Not Second half & 30,119 & \ensuremath{\checkmark} & \ensuremath{\checkmark} & \ensuremath{\checkmark} & \ensuremath{\checkmark} & 0.63 & 0.26 \\
Fast prior report & 30,044 & \ensuremath{\checkmark} & \ensuremath{\checkmark} & \ensuremath{\checkmark} & \ensuremath{\checkmark} & 0.69 & 0.21 \\
Not Fast prior report & 30,208 & \ensuremath{\checkmark} & \ensuremath{\checkmark} & \ensuremath{\checkmark} & \ensuremath{\checkmark} & 0.57 & 0.41 \\
Woman participant & 35,800 & \ensuremath{\checkmark} & \ensuremath{\checkmark} & \ensuremath{\checkmark} & \ensuremath{\checkmark} & 0.64 & 0.24 \\
Not Woman participant & 24,452 & \ensuremath{\checkmark} & \ensuremath{\checkmark} & \ensuremath{\checkmark} & \ensuremath{\checkmark} & 0.57 & 0.40 \\
Older participant & 29,575 & \ensuremath{\checkmark} & \ensuremath{\checkmark} & \ensuremath{\checkmark} & \ensuremath{\checkmark} & 0.59 & 0.36 \\
Not Older participant & 30,677 & \ensuremath{\checkmark} & \ensuremath{\checkmark} & \ensuremath{\checkmark} & \ensuremath{\checkmark} & 0.63 & 0.26 \\
High ability & 30,218 & \ensuremath{\checkmark} & \ensuremath{\checkmark} & \ensuremath{\checkmark} & \ensuremath{\checkmark} & 0.61 & 0.28 \\
Not High ability & 30,034 & \ensuremath{\checkmark} & \ensuremath{\checkmark} & \ensuremath{\checkmark} & \ensuremath{\checkmark} & 0.62 & 0.35 \\
High belief about AI & 29,735 & \ensuremath{\checkmark} & \ensuremath{\checkmark} & \ensuremath{\checkmark} & \ensuremath{\checkmark} & 0.57 & 0.34 \\
Not High belief about AI & 30,517 & \ensuremath{\checkmark} & \ensuremath{\checkmark} & \ensuremath{\checkmark} & \ensuremath{\checkmark} & 0.71 & 0.26 \\
High confidence in prior & 29,564 & \ensuremath{\checkmark} & \ensuremath{\checkmark} & \ensuremath{\checkmark} & \ensuremath{\checkmark} & 0.69 & 0.29 \\
Not High confidence in prior & 30,688 & \ensuremath{\checkmark} & \ensuremath{\checkmark} & \ensuremath{\checkmark} & \ensuremath{\checkmark} & 0.52 & 0.34 \\
\bottomrule

\end{tabular*}
\begin{minipage}{0.95\textwidth}
\footnotesize \emph{Notes:} For consistency, monotonicity, and reactionary updating, a checkmark means that the corresponding weak inequality is not significantly violated at the 5\% level using two-way clustered standard errors by participant and image. For threshold updating, a checkmark means that the hinge-regression thresholds are significantly away from the degenerate boundary at the 5\% level. The last two columns report the hinge-regression estimates of the non-updating thresholds after the Over 21 and Under 21 recommendations. Each ``Not'' row is the complement of the corresponding split row.
\end{minipage}
\end{table}

The final two columns show that the location of the non-updating threshold varies more than the existence of threshold updating. Across splits, the Over 21 threshold ranges from 0.52 to 0.71, while the Under 21 threshold ranges from 0.21 to 0.41. The treatment comparison is similar across treatments: the Over 21 threshold is 0.64 in NOINFO and 0.59 in INFO, and the Under 21 threshold is approximately 0.31 in both treatments. This variation is consistent with heterogeneity across participants and images in how AI advice is used, but the conclusion is unchanged: for sufficiently confirming priors, average updating is close to zero, while contradicting recommendations generate larger belief movements.

Table~\ref{tab:exclusion-robustness} in Appendix~\ref{app:property-robustness} reports robustness checks that exclude non-updaters, participants with extreme no-updating rates, bottom-tail response times for prior beliefs, large INFO recall errors on the survey question about AI accuracy, or degenerate prior distributions. The property tests remain supported in each restricted sample.

\subsection{Non-Updaters and Partial Non-Updaters} 

Participants were explicitly told that they could leave their belief unchanged, and in our analysis sample, the prior and posterior beliefs are the same for 78.6\% of our observations. A natural question is whether this non-updating represents a deliberate decision by participants not to update their beliefs. The property tests above show that non-updating is not random: it is concentrated where priors already satisfy the qualitative content of the recommendation. The response-time evidence in Appendix~\ref{app:accRT} is consistent with participants spending more time on decisions when the AI recommendation is contradictory, but also with spending some time on all decisions. Also, Figure~\ref{fig:update-value} shows that for extreme priors, the value to updating is low under the assumption of signal-dependence neglect, but other priors that conflict with the AI recommendation have the highest value for updating. 

We also find there is heterogeneity at the participant level in the amount of non-updating. 8.5\% of participants (32 of 377) never update their priors. We keep these participants in our analysis, but our results are qualitatively the same if we exclude them. Conceptually, there is no reason to exclude them, as putting full weight on the prior will lead to no updating in all three models of belief updating we consider. Non-updaters have higher confidence in own prior (78.5 vs. 73.5 for others) and lower belief about AI accuracy (62.3\% vs. 68.1\% for others), but neither difference is statistically significant ($p=0.199$ and $p=0.127$, respectively). They are also not significantly different from the rest of the participants in prior accuracy (61.1\% vs. 61.8\% for others, $p=0.594$).

Also, 80.1\% of participants (302 of 377) are \textit{partial non-updaters}: they update sometimes, but less than 50\% of the time. In Section~\ref{sec:pred}, we show that both partial non-updating and the aggregate level of non-updating (78.6\%) are consistent with both the objective and subjective versions of wIU, but not qB or CR.

\subsection{Individual-Level Property Tests}\label{sec:ind-properties}

The aggregate tests can mask heterogeneity across participants, so we also estimate versions of the regression tests for each participant. Consistency, monotonicity, and reactionary updating are weak properties: no updating or a flat relationship does not by itself violate them. We therefore classify a participant as violating consistency only if the mean signed update is significantly negative at the 5\% level, as violating monotonicity only if either posterior-prior slope within a recommendation is significantly negative, and as violating reactionary updating only if signed updating is significantly decreasing in contradiction. A participant satisfies threshold updating if the hinge regressions for that participant imply nondegenerate non-updating thresholds for both recommendations.

At the individual level, the satisfaction rates are 100.0\% for consistency, 97.3\% for monotonicity, 99.7\% for reactionary updating, and 46.7\% for threshold updating. Overall, 44.3\% of participants satisfy all four criteria, and another 55.2\% are near misses who fail exactly one criterion. We therefore interpret the evidence at the individual level as showing satisfaction rates above 97\% for the three weak properties, while threshold updating varies across participants.

\section{Three Models: Estimates and Predictions}\label{sec:pred}

We now examine how well the behavioral properties documented in Section~\ref{sec:threepatterns} are captured by three different models of belief updating. The first is a model from the literature on departures from Bayesian updating, which is based on the approach proposed by \cite{grether1980bayes}. We refer to this model as \emph{quasi-Bayesian} (hence, qB) because it nests Bayesian updating and many variants (see \cite{benjamin2019errors}).\footnote{The idea of quasi-Bayesian updating, in which a DM uses Bayesian machinery with behavioral biases, is influential in the behavioral literature on belief updating (e.g., \cite{rabin2002inference,rabin2010gambler}).}

We also consider two models of belief updating that assume no knowledge of the information structure. This matters when information comes from a ``black box'' because Bayes' rule is undefined when the DGP is unknown. These models, \emph{weighted Inertial Updating} (wIU) and the \emph{Contraction Rule} (CR), are also motivated by AI-generated information. These models replace signals with \emph{general information}, or a set of potential distributions, and the DM uses this set to form a posterior belief.

To formalize these ideas, we will require a bit of notation. Let $S$ be a finite set of states and $\mu\in\Delta(S)$ be a prior belief. In our experiment, the images that individuals consider have many features, but we will only consider the two payoff-relevant states: whether the individual is over or under 21. Thus, we model updating in our experiment using a binary state space where $s_1=\text{Over 21}$ and $s_2=\text{Under 21}$. 

We do not take a stand on how participants form their prior beliefs from the images. A participant may form an approximately Bayesian perceptual belief from visual information and then use a non-Bayesian rule to incorporate a black box recommendation. As emphasized by \cite{ortoleva2022alternatives}, different updating environments may call for different behavioral models. There is therefore no contradiction between non-Bayesian updating over AI recommendations and approximately Bayesian visual perception, as in Bayesian brain accounts or economic models of perceptual inference such as \cite{caplin2015testable}.\footnote{See \cite{agarwal2023combining} for a model in which humans jointly updated over perceptual signals and AI signals.}

The DM receives information in the form of a recommendation $R$ from a set of possible recommendations $\mathcal R$. For the qB model, $R$ is a signal. For wIU and CR, $R$ corresponds to an information set $I(R)\subset\Delta(S)$, which is a convex set of probability distributions that are consistent with the recommendation. Hence, the qualitative content of a recommendation $R$ is captured by $I(R)$. In our experiment, $R_1$ and $R_2$ correspond to the Over 21 and Under 21 recommendations, respectively. 

Finally, let $\mu_R\in\Delta(S)$ be the posterior after receiving a recommendation $R$. We want to model how the DM revises their prior $\mu$ and obtains a new belief (posterior) $\mu_R$ after receiving a signal $R$ or the qualitative content generated by $R$. 

\subsection{Bayesian and Quasi-Bayesian Updating}

The Bayesian benchmark starts from a perceived recommendation process. If the DM believes the AI recommendation is generated by a signal structure $\delta:S\to\Delta(\mathcal R)$, where $\delta(R|s)$ is the probability of recommendation $R$ in state $s$, then Bayes' rule implies
\begin{eqnarray}\label{eq:bayes}
\mu_R(s_1)=\frac{\delta(R|s_1)\mu(s_1) }{ \delta(R|s_1)\,\mu(s_1) + \delta(R|s_2)\,\mu(s_2)}.
\end{eqnarray}

Equation~\eqref{eq:bayes} is the correct Bayesian benchmark only if $\delta(R|s)$ is the recommendation process that is relevant for the DM after seeing the image. This is a nontrivial restriction in our setting because the elicited prior is formed after the participant observes private visual information. Let $X$ denote that visual information, so that $\mu(s_1)=\Pr(s_1|X)$. A fully Bayesian updater should use the conditional recommendation probabilities $\Pr(R|s_1,X)$ and $\Pr(R|s_2,X)$:
\[
\Pr(s_1|X,R)=\frac{\Pr(R|s_1,X)\Pr(s_1|X)}
{\Pr(R|s_1,X)\Pr(s_1|X)+\Pr(R|s_2,X)\Pr(s_2|X)}.
\]
The INFO treatment gives participants state-contingent recommendation rates, $\Pr(R|s)$, but not the conditional recommendation process $\Pr(R|s,X)$. Thus, the INFO treatment provides enough information to compute the Bayesian posterior in Equation~\eqref{eq:bayes} only under \emph{signal-dependence neglect}: the assumption that the AI recommendation is treated as conditionally independent of the participant's private visual information given the state. This restriction is also the closest analogue to the model with the lowest error in the initial NBER working paper of \citet*{agarwal2023combining}. We therefore use signal-dependence neglect as the baseline qB specification, while also reporting qB variants based on prior-conditioned Bayesian benchmarks.

\begin{figure}
    \centering
\begin{subfigure}[t]{0.49\textwidth}
\includegraphics[width=3.15in]{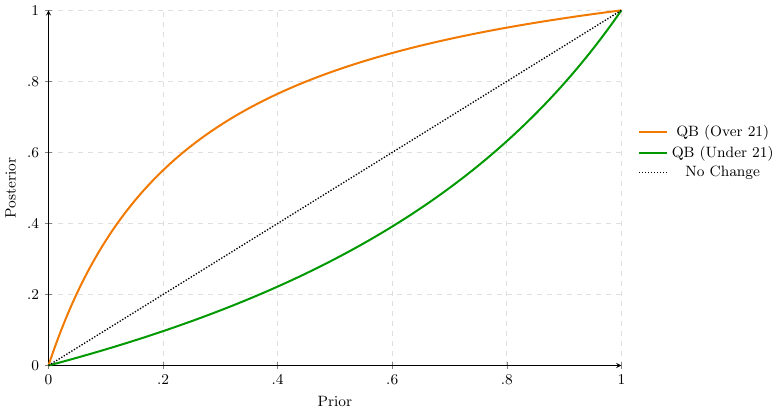}
\caption{$\beta_1=1, \beta_2=1, \alpha=0$}
\end{subfigure}
\begin{subfigure}[t]{0.49\textwidth}
\includegraphics[width=3.15in]{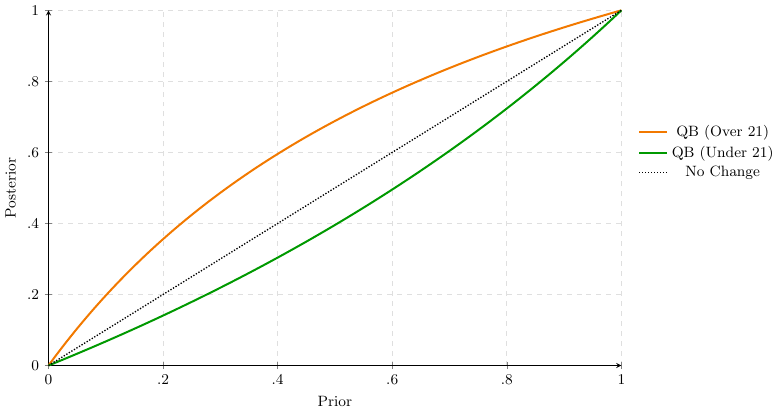}
\caption{$\beta_1=0.5,\beta_2=1, \alpha=0$}
\end{subfigure}
\caption{Predictions of qB assuming $\delta(R_1|s_1)=0.825$ and $\delta(R_2|s_2)=0.700$, as in our experiment.}
\label{fig:qb-predictions}
\end{figure}

The qB model generalizes Equation~\eqref{eq:bayes} by allowing the DM to distort the information in the recommendation, distort the prior, and have a reduced-form bias toward one state:
\begin{eqnarray}\label{eq:qb}
\mu_R(s_1)=\frac{\delta(R|s_1)^{\beta_1}\mu(s_1)^{\beta_2}e^\alpha}{ \delta(R|s_1)^{\beta_1}\mu(s_1)^{\beta_2}e^\alpha + \delta(R|s_2)^{\beta_1}\,\mu(s_2)^{\beta_2}},
\end{eqnarray}
where $\beta_1\geq 0$ controls how strongly the recommendation process enters the posterior, $\beta_2\geq 0$ controls how strongly the prior enters the posterior, and $\alpha\in\mathbb R$ captures a state-direction bias toward $s_1$. When $\beta_1=\beta_2=1$ and $\alpha=0$, qB reduces to Bayes' rule. Figure~\ref{fig:qb-predictions} illustrates the predictions of qB for different values of $\beta_1$ using the empirical frequencies in our experiment of each recommendation in each state. The left panel shows the nested case of Bayesian updating.

The baseline qB specification uses the empirical state-contingent recommendation rates: $\delta(R_1|s_1)=0.825$ and $\delta(R_1|s_2)=0.300$ after the Over 21 recommendation, and $\delta(R_2|s_1)=0.175$ and $\delta(R_2|s_2)=0.700$ after the Under 21 recommendation. We estimate $\alpha,\beta_2,\beta_1$ by fitting Equation~\eqref{eq:qb} and evaluating predictions in posterior-probability units.\footnote{Endpoint reports must be mapped to interior beliefs before estimating qB with the standard Grether least-squares implementation. For the main qB parameter estimates, we use $p^\dagger=\min\{\max\{p,0.005\},0.995\}$ for both prior and posterior reports. Thus, a reported 0 is treated as 0.005 and a reported 1 is treated as 0.995, while non-endpoint reports are unchanged.} This gives $\hat{\beta}_1=0.51$, $\hat{\beta}_2=0.78$, and $\hat{\alpha}=0.14$ using all 60,252 pairs of prior and posterior beliefs in the analysis sample. 

When $\alpha=0$, qB reduces to the model of \citet*{grether1980bayes}, which is one of the most widely applied non-Bayesian updating rules in the behavioral literature.\footnote{See \cite*{benjamin2019errors} and \cite*{ortoleva2022alternatives} for more careful discussions of the Grether rule.} The bias terms, $\beta_1$ and $\beta_2$, can have two interpretations. In the first interpretation, the DM is in fact Bayesian, but they only have noisy estimates of recommendation informativeness and prior probabilities, and the bias terms capture directions of noise. Under the second interpretation, which is more behavioral, the DM knows the DGP, but fails to apply Bayes' rule appropriately. Typically, $\beta_1<1$ captures underweighting of recommendation information or underreaction to news, while $\beta_2<1$ captures base-rate neglect.\footnote{There is a growing literature that considers the possibility that the DM has an incorrect mental model of the DGP. In this simple setting, a static distortion of the recommendation process can be represented by the $\beta_1$ term, while richer misspecification of that process would require a different advice term. Our robustness checks below therefore also consider advice terms derived from prior-conditioned Bayesian benchmarks. Richer dynamic effects, like those studied in \cite{esponda_vespa_yuksel_2024}, would require a further extension.} In the context of AI recommendations, underweighting or overweighting priors and recommendations could be driven by feelings or attitudes about AI or uncertainty about AI. Following \citet*{agarwal2023combining}, we say that a participant exhibits \emph{automation bias} if they over-weight the AI recommendation relative to their own information ($\beta_2<\beta_1$) and \emph{automation neglect} if they under-weight it ($\beta_1<\beta_2$). The right panel of Figure~\ref{fig:qb-predictions} provides an example of automation neglect, which decreases updating relative to the Bayesian benchmark. Finally, $\alpha>0$ yields a reduced-form pull toward believing that state $s_1$ is more likely, such as a state-direction bias or motivated belief distortion \citep*{kovach2020twisting}.

Figure~\ref{fig:main-qb} illustrates the predictions of qB under these parameter estimates overlaid with the actual aggregate average beliefs from our experiment. As illustrated in this figure and Figure~\ref{fig:qb-predictions}, qB struggles to satisfy the testable properties. The model satisfies monotonicity when $\beta_2>0$ because the posterior is increasing in the prior. However, qB violates reactionary updating for any nondegenerate specification with $\beta_2>0$: as priors converge to $0$ after $R_1$ or to $1$ after $R_2$, the posterior converges to the prior, so the update goes to zero exactly when the recommendation is most contradictory. This is the opposite of the reactionary-updating property. qB also violates threshold updating because its posterior is smooth in the prior and therefore does not have an interval of confirming priors on which the posterior is equal to the prior. Finally, with the estimated parameters, the prior distortion and state bias imply consistency violations at confirming extreme priors: sufficiently high priors are pulled down after $R_1$, and sufficiently low priors are pulled up after $R_2$.

\subsection{Contraction Rule}

In the Contraction Rule of \citet*{ke2024learning}, the DM first selects a ``representative" belief $\rho_R$ for each recommendation $R$ from $I(R)$, the set of probability distributions that are consistent with the recommendation. This is then mixed with their prior with the weight $\epsilon_{R}$. Hence, the posterior is given by
\begin{equation}\label{eq:cr}
    \mu_R=\epsilon_{R}\,\mu+(1-\epsilon_{R})\,\rho_R.
\end{equation}
The parameter $\epsilon_{R}$ is the weight on the DM's prior. The corresponding recommendation or AI weight is $1-\epsilon_R$.

\begin{figure}
    \centering

\begin{subfigure}[t]{0.49\textwidth}
\includegraphics[width=3.15in]{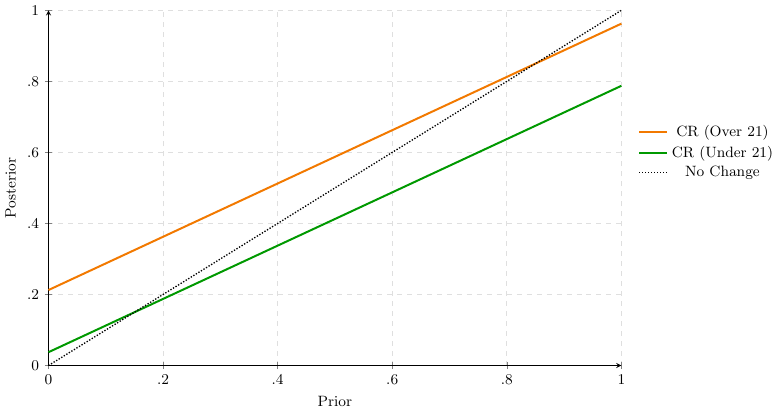}
\caption{$\epsilon_{R_i}=0.75$, $\rho_{R_1}=0.73, \rho_{R_2}=0.20$}
\end{subfigure}
\begin{subfigure}[t]{0.49\textwidth}
\includegraphics[width=3.15in]{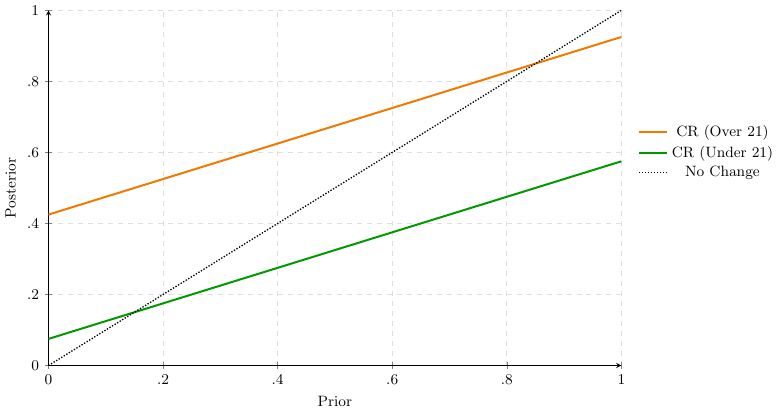}
\caption{$\epsilon_{R_i}=0.5$, $\rho_{R_1}=0.73, \rho_{R_2}=0.20$}
\end{subfigure}
\caption{Predictions of CR.}
\label{fig:cr-predictions}
\end{figure}

Because the parameters $\rho_{R}$ and $\epsilon_{R}$ are allowed to vary across recommendations, in our setting CR has four free parameters, with two parameters $(\rho_{R_i},\epsilon_{R_i})$ for each recommendation 
$R_i$. Figure~\ref{fig:cr-predictions} illustrates the predictions of CR for two symmetric values of $\epsilon_{R_i}$ ($0.75$ and $.5$) and values of $\rho_{R_i}$ that correspond to the actual probabilities that people are over 21 for each recommendation in our experiment. Estimating CR on the aggregate data gives $\hat{\epsilon}_{\text{Over}}=0.70$, $\hat{\epsilon}_{\text{Under}}=0.77$, $\hat{\rho}_{\text{Over}}(s_1)=0.85$, and $\hat{\rho}_{\text{Under}}(s_1)=0.14$. Figure~\ref{fig:main-cr} illustrates these estimates overlaid with the actual aggregate average beliefs from our experiment.

CR satisfies monotonicity whenever $\epsilon_R>0$, since the posterior is linear and increasing in the prior. It can also satisfy reactionary updating on the contradicting side of each recommendation: after $R_1$, the signed update is $(1-\epsilon_{R_1})(\rho_{R_1}(s_1)-\mu(s_1))$, which is larger when the prior is lower, and after $R_2$ the analogous signed update is larger when the prior is higher. The same linear contraction, however, leads CR to violate two of the testable properties. First, CR violates consistency at confirming extreme priors unless the representative belief is at the boundary of the state space. With the estimated parameters, priors above $\rho_{R_1}(s_1)=0.85$ are pulled down after $R_1$, and priors below $\rho_{R_2}(s_1)=0.14$ are pulled up after $R_2$. Second, CR violates threshold updating because the posterior equals the prior only at the representative belief (or everywhere in the degenerate case $\epsilon_R=1$), not on an interval of confirming priors. This failure to allow partial non-updating is the main reason CR misses the threshold property.

\subsection{Weighted Inertial Updating}

In weighted Inertial Updating (wIU) of \cite*{dominiak2021minimum}, the DM first picks the belief $\pi_R$ from the information set $I(R)$ that is closest to their prior $\mu$. In the objective version of wIU, we assume that $I(R_1)=\{\pi\in\Delta(S)\mid \pi(s_1)\in [\frac{1}{2}, 1]\}$ and $I(R_2)=\{\pi\in\Delta(S)\mid \pi(s_1)\in [0, \frac{1}{2}]\}$. For example, the recommendation of Over 21 is consistent with all possible beliefs of Over 21 from fully uncertain ($\frac{1}{2}$) to fully certain of Over 21 ($1$), so a DM whose prior assigns probability at least $\frac{1}{2}$ to Over 21 already has a prior in $I(R_1)$, while a DM below $\frac{1}{2}$ is moved to the boundary $\frac{1}{2}$ in the first step.\footnote{In the subjective version of the wIU, we relax this and estimate a subjective information constraint.} The DM then mixes that belief $\pi_R$ with their prior to form their posterior. 

Formally, the posterior $\mu_R$ is given by the formula:
\begin{equation}\label{eq:wiu}
\mu_R=w\,\mu+(1-w)\pi_R,    
\end{equation}
where $w\in [0, 1]$ and $\pi_R$ solves the minimization problem:
\[\pi_R=\arg\min_{\pi\in I(R)} d_\mu(\pi),\]
where $d_\mu(\pi)=\sum_{s\in S} -\mu(s)\sigma\big(\frac{\pi(s)}{\mu(s)}\big)$ with a strictly increasing and concave $\sigma$. The divergence $d_\mu(\pi)$ is known as the $f$-divergence and nests many popular divergences in information theory. For example, when $\sigma=\ln$, the divergence $d_\mu(\pi)$ is the Kullback-Leibler divergence and when $\sigma(x)=x^\alpha$ with $x\in (0, 1)$, $d_\mu(\pi)$ is an $\alpha$-divergence.\footnote{In \cite*{dominiak2021minimum}, they consider a more general class of divergences and also axiomatically characterize $f$-divergences. While the former is more general, the predictions of wIU when $w=0$ and the pseudo-Bayesian model of \cite{zhao2022pseudo} coincide in this experimental environment.}

Unlike CR, the prior weight in wIU does not vary by recommendation. We call $1-w$ the ``AI weight'' because it is the weight placed on the closest belief satisfying the AI recommendation.

\begin{figure}
    \centering

\begin{subfigure}[t]{0.49\textwidth}
\includegraphics[width=3.15in]{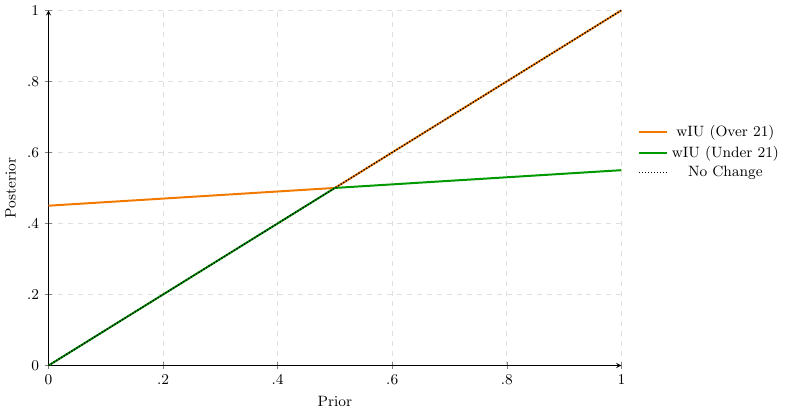}
\caption{AI weight $1-w=0.9$}
\end{subfigure}
\begin{subfigure}[t]{0.49\textwidth}
\includegraphics[width=3.15in]{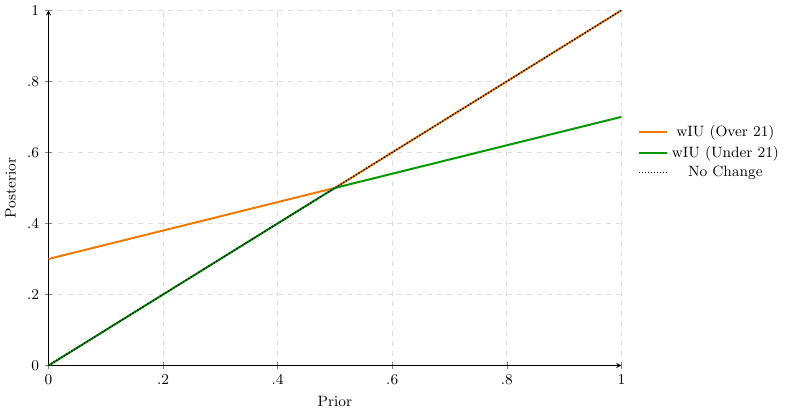}
\caption{AI weight $1-w=0.6$}
\end{subfigure}
\caption{Predictions of wIU.}
\label{fig:wiu-predictions}
\end{figure}

Figure~\ref{fig:wiu-predictions} illustrates the predictions for two values of the AI weight: one that puts high weight on the AI recommendation ($0.9=1-0.1$) and another that puts lower weight on the AI recommendation ($0.6=1-0.4$). The estimated parameter $w$ of the wIU from the aggregate data is $0.44$, which is shown in Figure~\ref{fig:main-wiu}.

We consider the objective and subjective versions of this model. In the objective version of this model (which we simply call wIU), the predictions of wIU in our setting can be summarized as follows. Suppose $I(R)=\{\pi\in \Delta(S)\mid \pi(s)\in [\frac{1}{2}, 1]\}$ for some $s$, then
\[\mu_R(s)=
\begin{cases}
w\,\mu(s)+(1-w)\,\frac{1}{2} &\text{ when } \mu(s)< \frac{1}{2}\\
\mu(s) &\text{ when } \mu(s)\ge \frac{1}{2}.\\
\end{cases}\]
\noindent In words, confirming recommendations do not change beliefs, but when recommendations are disconfirming, the DM takes the minimum belief consistent with the recommendation (here $0.5$) and mixes it with their prior. Thus, as illustrated in Figure~\ref{fig:wiu-predictions}, the wIU posterior is a piecewise linear function with a kink at $\frac{1}{2}$.\footnote{With precise AI advice (i.e., $I(R)=\{\pi\}$), wIU predicts that the posterior will be a fixed combination of the prior and the AI prediction (i.e., $\mu_R=w\,\mu+(1-w)\,\pi$). This model then predicts that average posteriors should trace approximately straight lines as a function of the prior for a fixed level of precise AI advice. CR makes the same prediction if the weight on the prior is constant for all AI advice.} There is just one free parameter in this model, and the figure shows the predictions of wIU for two values of $w$ ($.1$ and $.4$). Our estimate of this parameter from the aggregate data in our experiment is .44, and the predictions of wIU for this parameter value are given in Figure~\ref{fig:main-wiu}.

While Equations~\eqref{eq:wiu} and~\eqref{eq:cr} look similar, there are two distinctions between wIU and CR. First, CR allows the prior weight $\epsilon_{R}$ to depend on each recommendation. Second, in CR, the representative belief $\rho_R$ after recommendation $R$ is independent of the DM's prior, while in the wIU, $\pi_R$ is the closest belief to the DM's prior and it depends on the prior. Hence, the graph for the CR posterior is linear, while the one for wIU is piecewise linear, as illustrated in Figures~\ref{fig:main-cr} and~\ref{fig:main-wiu}. As a result, wIU is not a special case of CR. However, all three models (qB, wIU, and CR) coincide when the DM puts all weight on their prior.

In the subjective version of wIU, which we call \emph{weighted subjective Inertial Updating} (wsIU), we allow participants to differ in how much information they think the Over 21 recommendation conveys. However, we restrict the subjective interpretation not to contradict the recommendation. That is, the set of beliefs $\tilde{I}(R)$ must be a subset of the objective belief set $I(R)$ and the DM interprets $R_1$ as the information set $\tilde{I}(R_1)=\{\pi\in\Delta(S)\mid \pi(s_1)\in[\phi_1, 1]\}$ and $R_2$ as the information set $\tilde{I}(R_2)=\{\pi\in\Delta(S)\mid \pi(s_1)\in[0, \phi_2]\}$. For example, the Over 21 recommendation may be interpreted as $\pi(s_1)\in [0.6, 1]$ rather than $\pi(s_1)\in [0.5, 1]$. The predictions of the wsIU can be summarized as follows. Suppose $\tilde{I}(R_1)=\{\pi\in \Delta(S)\mid \pi(s_1)\in [\phi, 1]\}$. Then
\[\mu_{R_1}(s_1)=
\begin{cases}
w\,\mu(s_1)+(1-w)\,\phi &\text{ when } \mu(s_1)< \phi\\
\mu(s_1) &\text{ when } \mu(s_1)\ge \phi.\\
\end{cases}\]

This generalization of wIU has three free parameters: $w, \phi_1, \phi_2$. While $\phi_1$ and $\phi_2$ work as representative beliefs in CR, the wsIU still features a kink in belief updating. That is, when the prior is extreme and confirming, the DM does not update their belief in the wsIU. Estimating wsIU on the aggregate data gives $\hat{\phi}_{\text{Over}}=0.68$, $\hat{\phi}_{\text{Under}}=0.41$, and $\hat{w}=0.59$.

Importantly, both the objective and subjective versions of the wIU satisfy the four testable properties whenever there is positive but incomplete weight on the AI recommendation ($0<1-w<1$). Consistency follows because the closest belief in $I(R)$ is either the prior itself, when the prior already satisfies the recommendation, or the relevant boundary of the information set, which lies in the direction of the recommendation. Monotonicity follows from the piecewise-linear form: the slope is $w$ outside the information set and $1$ inside it. Reactionary updating follows because the signed update is $(1-w)(\phi_1-\mu(s_1))$ after $R_1$ when $\mu(s_1)<\phi_1$ and $(1-w)(\mu(s_1)-\phi_2)$ after $R_2$ when $\mu(s_1)>\phi_2$, so updating increases with contradiction. Finally, wIU and wsIU satisfy threshold updating directly: in the objective version the threshold is $\frac{1}{2}$, while in wsIU the subjective thresholds $\phi_1$ and $\phi_2$ are exactly the non-updating thresholds. In both versions, the posterior remains equal to the prior for all more confirming priors.

\subsection{Model Discussion}

\hspace*{.5cm} \emph{AI Trust:} All of these models have parameters that can be interpreted as model-implied weights on AI recommendations (relative to weight on the prior). Specifically, higher $\beta_1/\beta_2$ in qB, higher $1-\epsilon_{R}$ in CR, and higher $1-w$ in wIU and wsIU capture higher AI weight. While these objects are not directly comparable in quantitative terms across models, we estimate $\beta_1/\beta_2=0.51/0.78=0.65$ for qB, $1-\epsilon_{R_1}=0.30$ and $1-\epsilon_{R_2}=0.23$ for CR, $1-w=.56$ for wIU, and $1-w=.41$ for wsIU.\footnote{Appendix Figure~\ref{fig:hist-ai-weight} shows that there is variation in the estimated AI weight for the wsIU model at the individual level.}

\emph{Revisiting Non-Updating:}  While threshold updating is consistent with wIU, it is natural to consider whether the aggregate level of non-updating we observe (78.6\%) is ``reasonable'' under wIU. To obtain a simple intuition, consider a scenario in which priors and AI recommendations are completely random. In this case, the prior and posterior would be the same under wIU for approximately 50\% of observations. However, since priors and AI recommendations are positively correlated, we should observe a higher percentage of non-updating under wIU. Recall that the average accuracy for priors and AI recommendations is 61\% and 76\%, respectively. Given these accuracy levels and assuming non-negative correlation, the maximum and minimum percentage of non-updating observations under wIU are 85\% and 55.7\%, respectively.\footnote{Note that $0.557\approx 0.76\times 0.61+0.24\times 0.39$ and $0.85=0.61+(1-0.76)$. Hence, the minimum (maximum) level can be achieved by assuming zero (maximum) correlation between priors and AI recommendations, given the accuracy levels.} Hence, the percentage of non-updating observations is well within the wIU prediction, i.e., $78.6\% \in (55.7\%, 85\%)$.

\section{Model Assessments}\label{sec:comp}

\subsection{Comparing Models}\label{sec:fit} 

We compare the three models (qB, CR, and wIU) along two dimensions: fit and parsimony. In practice, there is often a tradeoff between fitness and parsimony, i.e., less restrictive models tend to have lower fit error. Hence, we distinguish between fit that comes from flexibility and fit that comes from capturing the features of the data that motivate the model comparison. We also quantify fitness and parsimony when considering tradeoffs. 

We evaluate these tradeoffs using out-of-sample performance, the completeness measure of \cite*{fudenberg2021complete}, and the restrictiveness measure of \cite*{fudenberg2023flexible}. Out-of-sample performance measures predictive fit on held-out data, but it does not separately assess fit and parsimony. Completeness measures the fraction of predictable variation in the data that the model captures, while restrictiveness quantifies parsimony by measuring how well a model fits predefined synthetic data. These normalized measures provide a Pareto frontier of completeness and restrictiveness.

Our preregistered way to evaluate model fit is via out-of-sample performance. In order to do so, we created 100 random splits of the dataset. For each split, 70\% of the observations were used to estimate model parameters. The remaining 30\% of observations were used to test the predictions of the estimated model.

The first column of Table~\ref{tab:agg} reports average mean squared error across the test splits for the four models. The non-Bayesian models, CR, wIU, and wsIU, have similar average MSEs and all have lower MSE than qB (paired tests against qB: all $p<0.001$). Columns two and three report the fraction of splits for which a particular model had the smallest MSE. CR and wIU have lower MSE than qB for every split, with wIU having the lowest MSE among the three models in $76\%$ of the splits. When the subjective version of wIU is included, it has the lowest MSE in all splits (paired tests against CR and wIU: both $p<0.001$).\footnote{Appendix Table~\ref{tab:benchmark-variants} reports analogous out-of-sample results for additional benchmark variants, including the Bayesian benchmark, a model that combines the prior and the Bayesian benchmark, and qB without signal-dependence neglect.}

\begin{figure}[ht!]
\centering
\begin{subfigure}[t]{0.48\textwidth}
\centering
\begin{tikzpicture}
\begin{axis}[
    width=0.98\textwidth,
    height=0.72\textwidth,
    xlabel={Completeness (\%)},
    ylabel={Restrictiveness (\%)},
    xmin=0, xmax=105,
    ymin=0, ymax=65,
    grid=both,
    tick label style={font=\footnotesize},
    label style={font=\footnotesize},
    legend=false
]
\addplot[only marks, black, mark=*, mark size=2.4pt] coordinates {(42.0,3.7)};
\addplot[only marks, black, mark=square*, mark size=2.4pt] coordinates {(90.9,2.3)};
\addplot[only marks, black, mark=triangle*, mark size=2.7pt] coordinates {(91.4,56.5)};
\addplot[only marks, black, mark=diamond*, mark size=2.7pt] coordinates {(95.3,7.9)};
\addplot[black, dashed, thick] coordinates {(91.4,56.5) (95.3,7.9)};
\node[anchor=south west, font=\footnotesize] at (axis cs:42.0,3.7) {qB};
\node[anchor=south east, font=\footnotesize] at (axis cs:90.9,2.3) {CR};
\node[anchor=north west, font=\footnotesize] at (axis cs:91.4,56.5) {wIU};
\node[anchor=south east, font=\footnotesize] at (axis cs:95.3,7.9) {wsIU};
\end{axis}
\end{tikzpicture}
\caption{Monotonic eligible class.}
\end{subfigure}
\hfill
\begin{subfigure}[t]{0.48\textwidth}
\centering
\begin{tikzpicture}
\begin{axis}[
    width=0.98\textwidth,
    height=0.72\textwidth,
    xlabel={Completeness (\%)},
    ylabel={Restrictiveness (\%)},
    xmin=0, xmax=105,
    ymin=0, ymax=65,
    grid=both,
    tick label style={font=\footnotesize},
    label style={font=\footnotesize},
    legend=false
]
\addplot[only marks, black, mark=*, mark size=2.4pt] coordinates {(42.0,31.7)};
\addplot[only marks, black, mark=square*, mark size=2.4pt] coordinates {(90.9,24.6)};
\addplot[only marks, black, mark=triangle*, mark size=2.7pt] coordinates {(91.4,43.2)};
\addplot[only marks, black, mark=diamond*, mark size=2.7pt] coordinates {(95.3,24.7)};
\addplot[black, dashed, thick] coordinates {(91.4,43.2) (95.3,24.7)};
\node[anchor=south west, font=\footnotesize] at (axis cs:42.0,31.7) {qB};
\node[anchor=north east, font=\footnotesize] at (axis cs:90.9,24.6) {CR};
\node[anchor=north west, font=\footnotesize] at (axis cs:91.4,43.2) {wIU};
\node[anchor=south east, font=\footnotesize] at (axis cs:95.3,24.7) {wsIU};
\end{axis}
\end{tikzpicture}
\caption{Recommendation-consistent eligible class.}
\end{subfigure}
\caption{Completeness--restrictiveness frontier. Dashed lines connect the nondominated models under each restrictiveness benchmark.}
\label{fig:completeness-restrictiveness-frontier}
\end{figure}
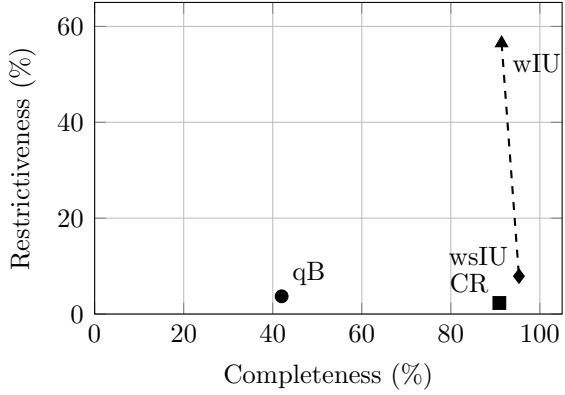
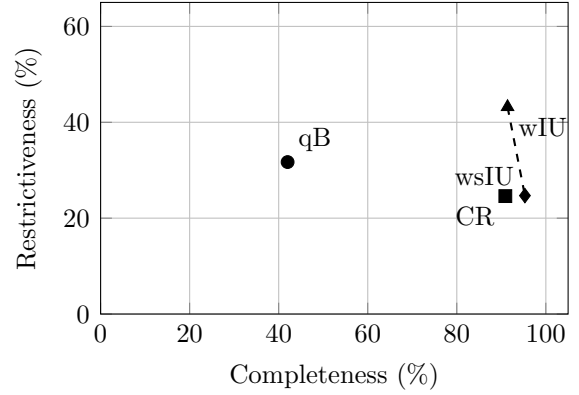

Finally, column five reports the completeness measure of \cite*{fudenberg2021complete}, while columns six and seven report two versions of the restrictiveness measure of \cite*{fudenberg2023flexible}.\footnote{Let $\overline{\mathcal{F}}$ be the set of all models (all deterministic updating rules for our case). Let $\mathcal{F}\subseteq \overline{\mathcal{F}}$ be the set of all ``eligible" models which satisfy some regularity properties. We report two eligible classes: recommendation-consistent weakly monotonic updating rules and recommendation-consistent updating rules without imposing monotonicity. The restrictiveness of model $\mathcal{F}_{\Theta}\subset \mathcal{F}$ is $r(\mathcal{F}_{\Theta}, \mathcal{F})=\frac{\mathbb{E}_{\lambda_{\mathcal{F}}}[d(\mathcal{F}_{\Theta}, f)]}{\mathbb{E}_{\lambda_{\mathcal{F}}}[d(f_{base}, f)]}$ where the expectation is calculated by randomly drawing synthetic data $f$ from $\mathcal{F}$ using uniform distribution $\lambda_{\mathcal{F}}$ and $d(\mathcal{F}_{\Theta}, f):=\inf_{f_\theta\in \mathcal{F}_{\Theta}} d(f_\theta, f)$. Here, $f_{base}$ is a base model that is nested by $\mathcal{F}_{\theta}$. For our case, $f_{base}$ is a rule with no updating, because no updating is the only model nested by our three models. The completeness of model $\mathcal{F}_{\Theta}$ is $\kappa(\mathcal{F}_{\Theta})=\frac{e_{P}(f_{base})-\min_{f_{\theta}\in \mathcal{F}_{\Theta}} e_{P}(f_\theta)}{e_{P}(f_{base})-e_{P}(f^*)}$ where $e_{P}(f)$ measures the distance between model $f$ and observed data, and $f^*=\arg\min_{f\in \overline{\mathcal{F}}}$. We use mean squared error (MSE) for $d(f_\theta, f)$ and $e_{P}(f)$.} The non-Bayesian models have completeness measures above 90\%, compared with 42\% for qB. Figure~\ref{fig:completeness-restrictiveness-frontier} plots the resulting completeness--restrictiveness tradeoff. With monotonicity imposed, wIU is the most restrictive, followed by wsIU, qB, and CR. Without imposing monotonicity, wIU remains the most restrictive, followed by qB, wsIU, and CR.

\begin{table}[ht!]
\centering
\caption{Aggregate model fit and flexibility comparisons.}
\label{tab:agg}
\footnotesize
\setlength{\tabcolsep}{3pt}
\begin{tabular*}{\textwidth}{@{\extracolsep{\fill}}lcccccc@{}}
\toprule
Model & Test MSE & \multicolumn{2}{c}{Test wins} & Completeness & \multicolumn{2}{c}{Restrictiveness} \\
\cmidrule(lr){3-4}\cmidrule(lr){6-7}
 & & Three models & Four models & & Monotone & No monotonicity \\
\midrule
qB & 0.0402 &   0\% & 0\% & 42.0\% &  3.7\% & 31.7\% \\
CR & 0.0336 &  24\% & 0\% & 90.9\% &  2.3\% & 24.6\% \\
wIU & 0.0336 & \textbf{76\%} & 0\% & 91.4\% & \textbf{56.5\%} & \textbf{43.2\%} \\
wsIU & \textbf{0.0330} & --- & \textbf{100\%} & \textbf{95.3\%} &  7.9\% & 24.7\% \\
\bottomrule

\end{tabular*}
\end{table}

\subsection{Updating Types}\label{sec:updating-types}

The aggregate model comparisons mask heterogeneity in updating behavior. This is already visible in the property tests at the individual level, where 44.3\% of participants satisfy all four stricter properties. We therefore also compare models using data for each participant. We use essentially the same procedure for measuring out-of-sample error as with the aggregate data. For each individual, model parameters are estimated on 70\% of the observations, and the remaining 30\% of observations are used to test the model predictions. Table~\ref{tab:individual-fit} reports the average and median mean squared error across participants.

\begin{table}[ht!]
\centering
\caption{Model fit for each participant.}
\label{tab:individual-fit}
\footnotesize
\begin{tabular}{lccc}
\toprule
\textbf{Model} & \textbf{Params}  &  \textbf{Avg. MSE} &  \textbf{Median MSE} \\
\midrule
qB & 3
  & 0.0256 & 0.0119  \\ [0.3em]
CR & 4
   & 0.0159 & 0.0105 \\ [0.3em]
wIU & 1
   & 0.0221 & 0.0134 \\ [0.3em]
   \textbf{wsIU} & 3
   & \textbf{0.0158} & \textbf{0.0098} \\ 
\bottomrule
\end{tabular}
\end{table}

With our estimates for each individual, we are able to classify individuals. To do this, we calculate the fraction of individuals for which a particular model yields the lowest average mean squared error across splits. Table~\ref{tab:individual-classification} reports these for three different sets of models. Notably, qB explains the largest fraction of participants in the three-model comparison, while wsIU explains the largest fraction when it is included.

\begin{table}[ht!]
\centering
\caption{Model classification by lowest MSE.}
\label{tab:individual-classification}
\footnotesize
\begin{tabular}{lccc}
\toprule
\textbf{Model} & \textbf{Three models} & \textbf{Four models} & \textbf{qB vs. wsIU} \\
\midrule
qB 
   & 37.7\% & 32.1\% & 37.7\% \\ [0.3em]
CR 
   & 34.5\% & 13.0\% &  \\ [0.3em]
wIU 
    & 27.9\% & 22.5\% & \\ [0.3em]
   \textbf{wsIU} &   
  & \textbf{32.4\%}  & \textbf{62.3\%} \\  [0.3em]
\bottomrule
\end{tabular}
\end{table}

The classification by lowest MSE is transparent, but it treats the type assignment as known. As an additional population summary, we estimate a hierarchical model with two types in the spirit of \citet*{bruhin_fehr_duda_epper_2010}. Participants are assumed to be either qB or wsIU types, and the population share of each type is estimated from the fit of the two models for each participant. This latent type approach estimates that 32.5\% of participants are qB types and 67.5\% are wsIU types, with a 95\% bootstrap confidence interval of [27.2\%, 38.2\%] for the qB share. Thus, the conclusion is similar to the simple classification: wsIU describes a majority of participants, while qB describes 32.5\%.

\begin{table}[ht!]
\centering
\caption{Population shares of qB and wsIU updating types.}
\label{tab:hierarchical-type-shares}
\footnotesize
\begin{tabular}{lcccc}
\toprule
Model & Lowest MSE share & Hierarchical share & 95\% CI & Modal posterior type \\
\midrule
qB & 37.7\% & 32.5\% & [27.2\%, 38.2\%] & 32.1\% \\
wsIU & 62.3\% & 67.5\% & [61.8\%, 72.8\%] & 67.9\% \\
\bottomrule

\end{tabular}
\begin{minipage}{0.95\textwidth}
\footnotesize
\emph{Notes:} Lowest MSE shares classify each participant by which model has lower average out-of-sample MSE across splits. Hierarchical shares come from a finite mixture model with two types. For participant $i$ and model $m\in\{\text{qB},\text{wsIU}\}$, we form the profile Gaussian log likelihood $\ell_{im}=-(160/2)\log(\text{MSE}_{im})$, where $\text{MSE}_{im}$ is the participant's average out-of-sample MSE, bounded below by $10^{-10}$ before taking logs. The qB share $\pi$ is estimated by maximizing $\sum_i \log\left(\pi e^{\ell_{i,\text{qB}}}+(1-\pi)e^{\ell_{i,\text{wsIU}}}\right)$. Confidence intervals are participant bootstrap intervals from 5,000 resamples. The modal posterior type classifies each participant using the posterior type probability from the hierarchical model.
\end{minipage}
\end{table}

We next ask whether updating types vary by treatment. Appendix Table~\ref{tab:type-treatment} shows that the type composition is similar in the INFO and NOINFO treatments. In the hierarchical specification, the qB share is 34.9\% in the INFO treatment and 30.0\% in the NOINFO treatment, a difference of -4.9 percentage points ($p=0.412$). The classification by lowest MSE gives a treatment difference of -2.3 percentage points ($p=0.662$). Thus, providing aggregate information about AI accuracy does not appear to change the population mix of qB and wsIU updating types.

\subsection{Beliefs About AI and wsIU Parameters}\label{sec:wsiu-parameter-heterogeneity}

The wsIU estimates also allow us to summarize heterogeneity in how participants interpret and use the AI recommendation. We use two summaries of the three wsIU parameters. The first is the \emph{subjective recommendation threshold}, defined as $(\phi_1+1-\phi_2)/2$, which summarizes how restrictive the participant's subjective recommendation thresholds are. The second is \emph{AI weight}, defined as $1-w$, which summarizes how much the participant moves toward the subjective recommendation threshold when the prior contradicts the recommendation.

Table~\ref{tab:wsiu-survey-parameters} relates these wsIU summaries to the survey measures collected after the experiment. The relationship is concentrated in AI weight. Participants who believe the AI is more accurate place more weight on the AI recommendation, while participants who are more confident in their own prior place less weight on the recommendation. The coefficient on belief about AI accuracy is 0.0034 and the coefficient on confidence in own prior is -0.0028 (both $p<0.001$). By contrast, these survey measures have relationships closer to zero with the subjective recommendation threshold, and both coefficients have $p>0.25$. Thus, survey beliefs appear to predict how much participants move toward the recommendation more than how they interpret the recommendation's subjective threshold.

\begin{table}[ht!]
\centering
\caption{Survey correlates of wsIU parameters.}
\label{tab:wsiu-survey-parameters}
\footnotesize
\begin{tabular}{lcc}
\toprule
& Subjective threshold & AI weight \\
\midrule
NOINFO treatment & -0.0075 & 0.0368 \\[-0.6ex]
 & (0.0124) & (0.0338) \\[0.6ex]
Woman & 0.0047 & -0.0467 \\[-0.6ex]
 & (0.0127) & (0.0345) \\[0.6ex]
Older (above median) & 0.0266 & -0.0256 \\[-0.6ex]
 & (0.0126) & (0.0343) \\[0.6ex]
Belief about AI accuracy (0--100\%) & 0.0003 & 0.0034 \\[-0.6ex]
 & (0.0003) & (0.0008) \\[0.6ex]
Confidence in own prior (0--100) & -0.0003 & -0.0028 \\[-0.6ex]
 & (0.0003) & (0.0008) \\[0.6ex]
\addlinespace
Mean dep. var. & 0.725 & 0.343 \\
N & 377 & 377 \\
\bottomrule

\end{tabular}
\begin{minipage}{0.95\textwidth}
\footnotesize
\emph{Notes:} Entries are OLS coefficients with standard errors in parentheses. The unit of observation is a participant. The subjective recommendation threshold is $(\phi_1+1-\phi_2)/2$, where $\phi_1$ and $\phi_2$ are the participant's estimated wsIU thresholds for the Over 21 and Under 21 recommendations. AI weight is $1-w$, where $w$ is the estimated weight on the prior in wsIU.
\end{minipage}
\end{table}

\section{Discussion}\label{sec:lit}

\subsection{Bayesianism}\label{sec:bayesianism} 

Our baseline qB model imposes signal-dependence neglect because this restriction yields sharp predictions while accommodating several familiar departures from Bayesian updating, including underreaction or overreaction to information, base-rate neglect, and motivated beliefs. In an environment with an unknown DGP, however, it is not obvious what should count as the appropriate Bayesian benchmark. Indeed, when the DGP is unrestricted and information has no qualitative content, Bayesian updating may have essentially no testable implications \citep*{shmaya2016experiments,molavi2021empirical, dominiak2021minimum}.

To illustrate this issue, consider the following generalization of qB, which relaxes signal-dependence neglect:
\[
\mu(s_1|X_k,R)=w\,\mu(s_1|X_k)+(1-w)\,\frac{(\Pr(R|s_1,X_k))^{\beta_1}(\mu(s_1|X_k))^{\beta_2}\,e^\alpha}
{(\Pr(R|s_1,X_k))^{\beta_1}(\mu(s_1|X_k))^{\beta_2}+(\Pr(R|s_2,X_k))^{\beta_1}(\mu(s_2|X_k))^{\beta_2}},
\]
where $X_k$ is the visual information that a subject/DM perceives from picture $k$, and $\Pr(R|s_1,X_k)$ and $\Pr(R|s_2,X_k)$ are the conditional recommendation probabilities. This reduces to our baseline qB when $w=0$ and $\Pr(R|s_1,X_k)=\delta(R|s_1)$, and reduces to the standard Bayesian model introduced in Section 4.1 when $w=\alpha=0$ and $\beta_1=\beta_2=1$. Once the conditional recommendation probabilities are unrestricted, any interior prior–posterior pair can be rationalized by an appropriate choice of conditional recommendation probabilities, even under standard Bayesian updating. Thus, relaxing signal-dependence neglect makes the model effectively untestable for interior priors. Nevertheless, it retains a sharp endpoint implication: a prior of zero must produce a posterior of zero, and a prior of one must produce a posterior of one, regardless of the recommendation.

This endpoint implication conflicts with our second behavioral pattern. Participants often move substantially away from extreme priors when the AI recommendation contradicts them. Therefore, relaxing signal-dependence neglect does not improve the Bayesian account of our findings: it removes meaningful restrictions for interior priors while preserving endpoint predictions that are inconsistent with the data.

A related possibility is that the subjects are Bayesian but learn about the AI’s DGP over the course of the experiment. Such learning could generate changing perceived recommendation probabilities, rather than fixed distortions summarized by $\beta_1$ and $\beta_2$. However, Bayesian learning still implies that a prior of zero remains zero after any recommendation, and therefore cannot explain substantial updating from extreme contradictory priors. Moreover, we find little evidence that learning is the main driver of behavior. Prior and posterior accuracy do not display meaningful dynamics over rounds, and updating is similar in the INFO and NOINFO treatments despite the substantial difference in information provided about the AI’s DGP.

Finally, there is a separate Bayesian interpretation of the updating behavior of the subjects who are consistent with wIU. 
\cite{dominiak2021minimum} prove that Inertial Updating (i.e., wIU with $w=0$) is behaviorally equivalent to Bayesian updating when the divergence $d_\mu(\pi)$ is an $f$-divergence and the information set $I(R)$ satisfies an appropriate linear moment condition. Under these conditions, one can construct a signal structure from the information sets $I(R)$ such that Inertial Updating under an unknown DGP generates the same posterior as Bayesian updating under the constructed, known DGP. Accordingly, wIU can be interpreted as placing positive weight on the prior and the remaining weight on this Bayesian-equivalent posterior. This interpretation does not, however, reduce wIU in our experiment to standard Bayesian updating: the information sets associated with our recommendations violate the required linear moment condition.

\subsection{Human-AI Interaction}\label{sec:AIAdvice}

Our work contributes to a growing literature on human-AI interaction that spans economics, psychology, management, computer science, and human-computer interaction. Our work is most closely related to papers that consider the impacts of AI advice on beliefs \citep*{agarwal2023combining,caplin2024abc} and AI advice that comes in the form of qualitative recommendations \citep*{hoong2025improving, NBERw33949}. 

\citet*{agarwal2023combining} study how professional radiologists combine their own information with AI predictions, and they find that AI predictions do not improve radiologist performance on average. They explain the failure to benefit from AI assistance as arising from errors in belief updating: radiologists underweight AI predictions and neglect the statistical dependence between their own information and the AI prediction. Our experiment is complementary to theirs in two respects. First, the AI advice in their experiment is precise, such as a probability of pneumonia, whereas the AI recommendation in our experiment is qualitative. Second, their welfare analysis uses the qB model, while we compare qB to CR and wIU, models designed for general or black box information. 

\begin{figure}
\centering
\begin{subfigure}[t]{0.49\textwidth}
\includegraphics[width=\textwidth]{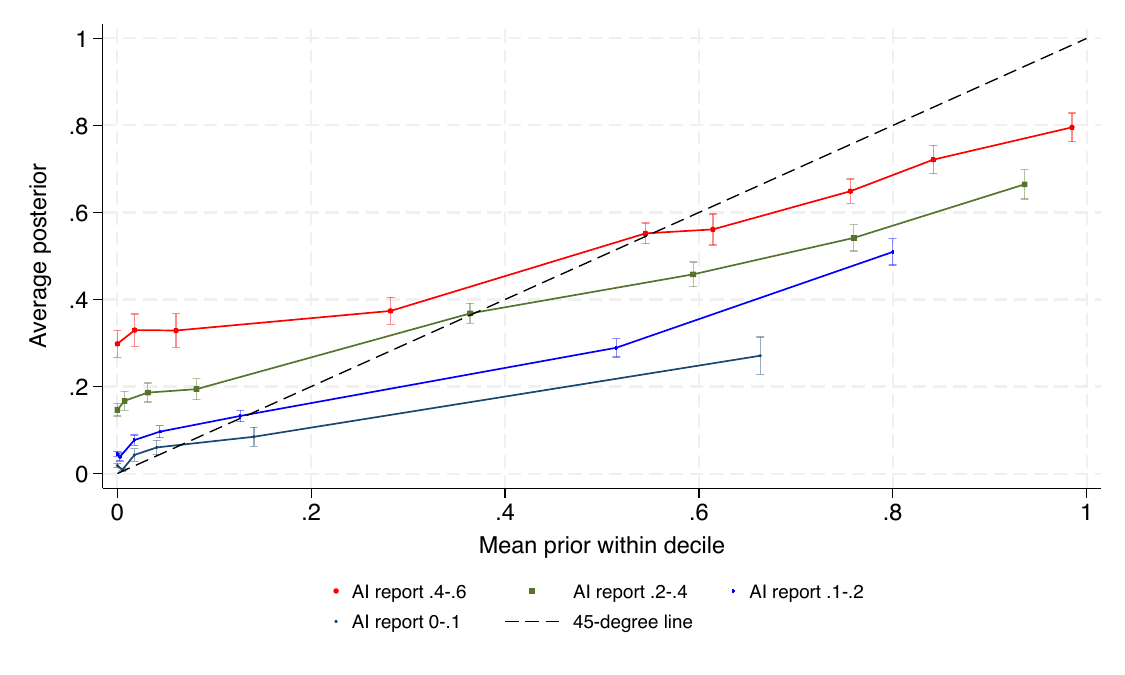}

\end{subfigure}

\caption{Average posterior at binned priors for different bins of AI advice for radiologist reports in \citet*{agarwal2023combining}.}

\label{fig:agarwal}
\end{figure}

Their data, described in \cite*{moehring2025dataset}, also allow us to compare our results with evidence on probabilistic AI advice and professional radiologists. In both experiments, AI advice shifts elicited beliefs in line with the advice. However, participants in both experiments also place weight on their own initial assessments even though the AI is often more accurate than human decision makers. In line with this, both papers find evidence of automation neglect when they estimate the parameters of qB. In addition, when we re-analyze their data we find suggestive evidence in line with the latter two behavioral patterns in updating. As shown in Figure~\ref{fig:agarwal}, for ``more certain'' AI advice (a confidence score above .66 or below .33), updates are larger when recommendations contradict extreme priors and smaller for intermediate priors. However, with precise AI advice the definition of ``contradiction'' is more subtle.

Our results also relate to \citet*{caplin2024abc}, who use the Bouncer task to study who benefits from AI assistance. They show that AI increases payoffs more for low-ability participants, but that conditional on ability, the value of AI assistance depends on whether participants are calibrated about their own ability. Our results suggest a belief-updating channel for this finding. If participants who are more confident in their own assessments place more weight on their priors, then miscalibration can reduce the value of AI even when the AI is informative. Consistent with this interpretation, our wsIU estimates suggest that AI weight is increasing in beliefs about AI accuracy and decreasing in confidence in own prior. Thus, overconfidence may matter twice: first by affecting the initial prior, and second by affecting the weight placed on that prior after observing the AI recommendation.

\subsection{Human vs. AI Recommendations}\label{sec:humansvsAI}

We chose to use AI recommendations to study belief updating with unknown DGPs because of the three features mentioned previously: humans may treat it as a black box, it is possible to estimate the DGP, and the DGP is stable. However, a potentially interesting avenue for follow-up work could be to compare updating from AI recommendations to updating from human recommendations. 

There is a large existing literature on how humans respond to human recommendations. For example, several papers study social learning environments where the DM observes another DM's actions, but does not know their DGPs \citep*{de2022non,chen2026sequential, chan2024prior}. Studies using the Judge-Advisor System introduced by \cite*{sniezek1995cueing} find that DMs tend to underweight human advice relative to their own initial opinions \citep*{bonaccio2006advice,sniezek2001trust}, which is consistent with what we observe in our qB estimates.\footnote{Our design shares the basic judge-advisor structure, but the traditional tasks in that literature involve continuous point estimates, such as distances, ages, or counts, which do not allow direct comparisons of formal updating rules.} 

However, gender may mediate the comparison between AI and human advice. Evidence on algorithmic judgment and AI adoption suggests that men adopt generative AI more often than women \citep*{stephany2026women}. In line with this, Figure~\ref{fig:update-gender} shows that men in our experiment update more than women on average when the AI recommendation contradicts an extreme prior. On the other hand, there is evidence that men are less likely to take advice from humans than women \citep*{see2011detrimental}.

\begin{figure}
    \centering

\begin{subfigure}[t]{0.49\textwidth}
\includegraphics[width=\textwidth]{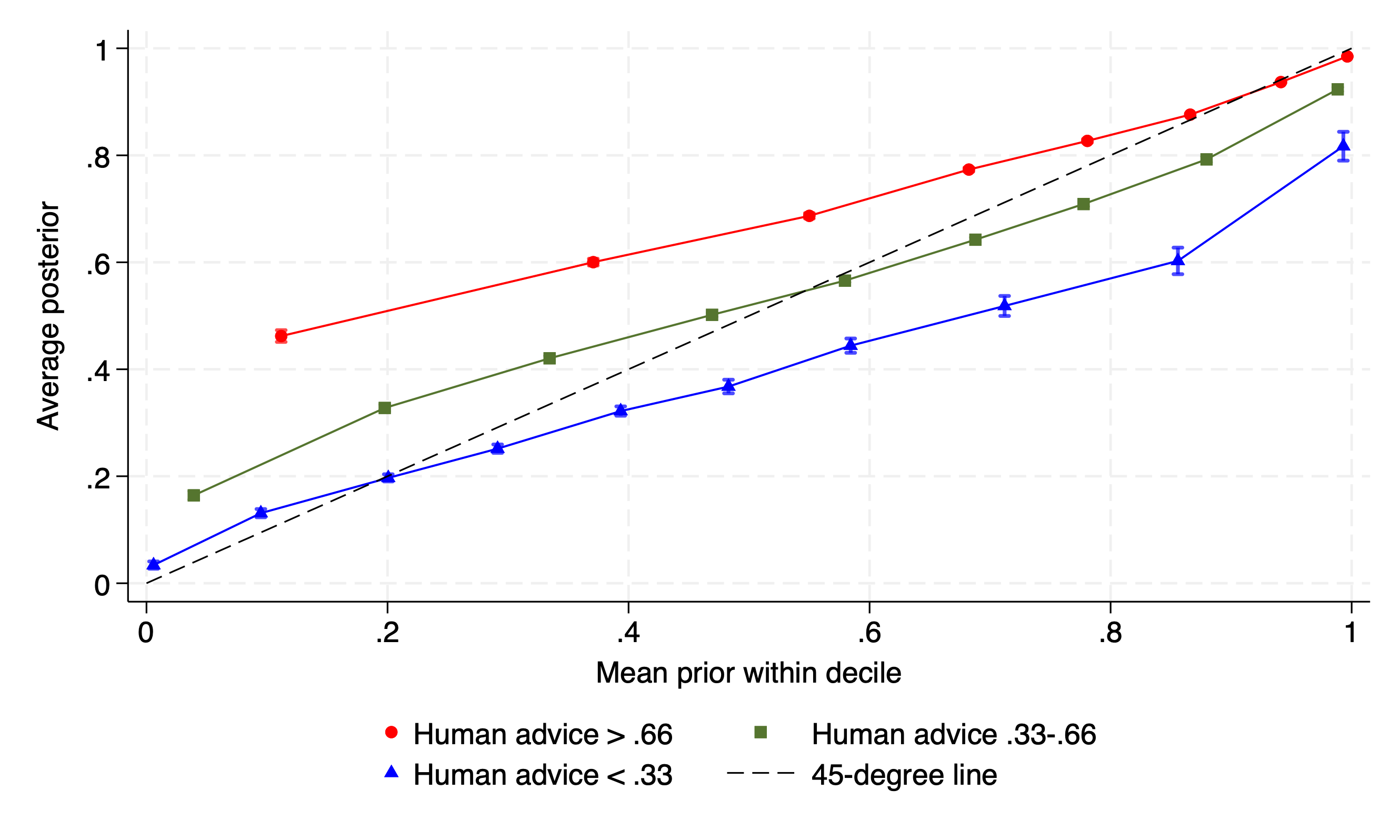}
\caption{Human-labeled advice.}
\end{subfigure}
\begin{subfigure}[t]{0.49\textwidth}
\includegraphics[width=\textwidth]{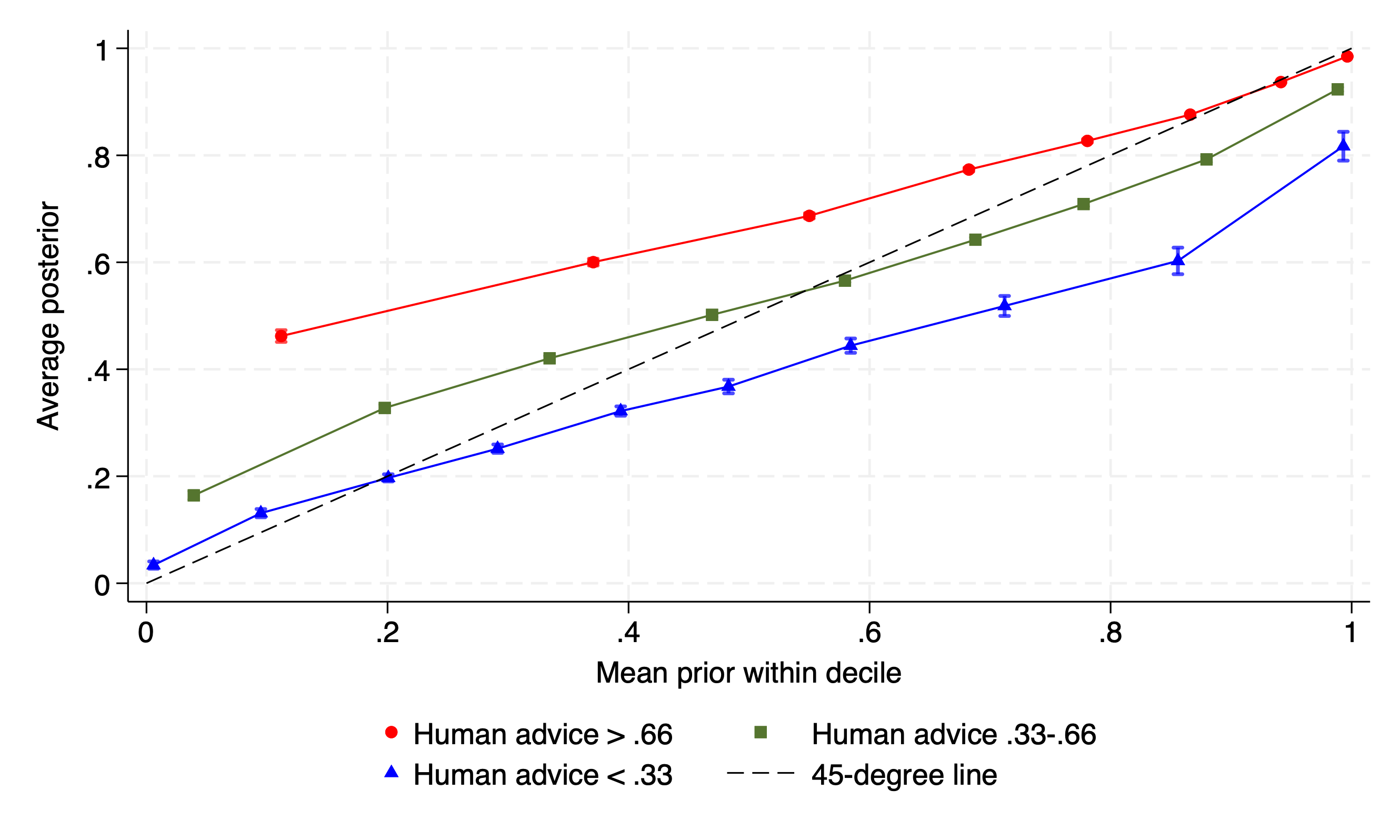}
\caption{AI-labeled advice.}
\end{subfigure}
\caption{Updating on advice labeled as human or AI from \cite*{vodrahalli2022humans}.}
\label{fig:vod}
\end{figure}

Also, in our setting human and AI recommendations might differ because human and AI judgments are related but not highly correlated. Defining the participant's binary judgment as whether the prior belief is above 50\%, the overall correlation between the participant's binary judgment and the AI recommendation is 0.233, with agreement in 62.5\% of observations.\footnote{Exact 50\% priors are coded as not Over 21 for this calculation. Excluding them gives an overall correlation of 0.234.} At the participant level, the median correlation is 0.266 and the 95th percentile is 0.434. Only five participants have correlations above 0.5, with a maximum of 0.762. Thus, even the participants most aligned with AI are not simply reproducing the AI recommendation in binary form.

This comparison raises two issues for future designs that compare human and AI recommendations. First, the low correlation between human judgments and AI recommendations may make human recommendations systematically different from AI recommendations. Second, participants may have correct or incorrect beliefs about those differences, and those beliefs may affect how they use the recommendations. Deception about the source of advice, or selecting human advisors whose recommendations mimic the AI, can hold the recommendation fixed across perceived sources. But doing so changes the object being studied: a human advisor selected to look like the AI is not a typical human advisor, which may distort participants' beliefs about the advice-generating process.

To overcome the first challenge, \citet*{vodrahalli2022humans} use deception around the stated source of advice, and their data provide suggestive evidence that individuals may not update differently when otherwise-equivalent advice is labeled as coming from a human or AI. In their four layperson tasks, participants first reported a continuous belief about the correct state, received precise advice experimentally labeled as either human or AI, and then reported a posterior belief. For these layperson tasks the underlying advice was held fixed across the human- and AI-labeled conditions: the comparison therefore identifies the effect of the perceived advice source rather than differences in the actual production of advice. Figure~\ref{fig:vod} plots the relationship between prior and posterior beliefs by advice strength, separately for advice labeled as human or AI. The figure shows that human- and AI-labeled advice produce similar updating patterns.

\subsection{Behavioral Microfoundations}

\begin{figure}[ht!]
\begin{center}
\includegraphics[width=2.65in]{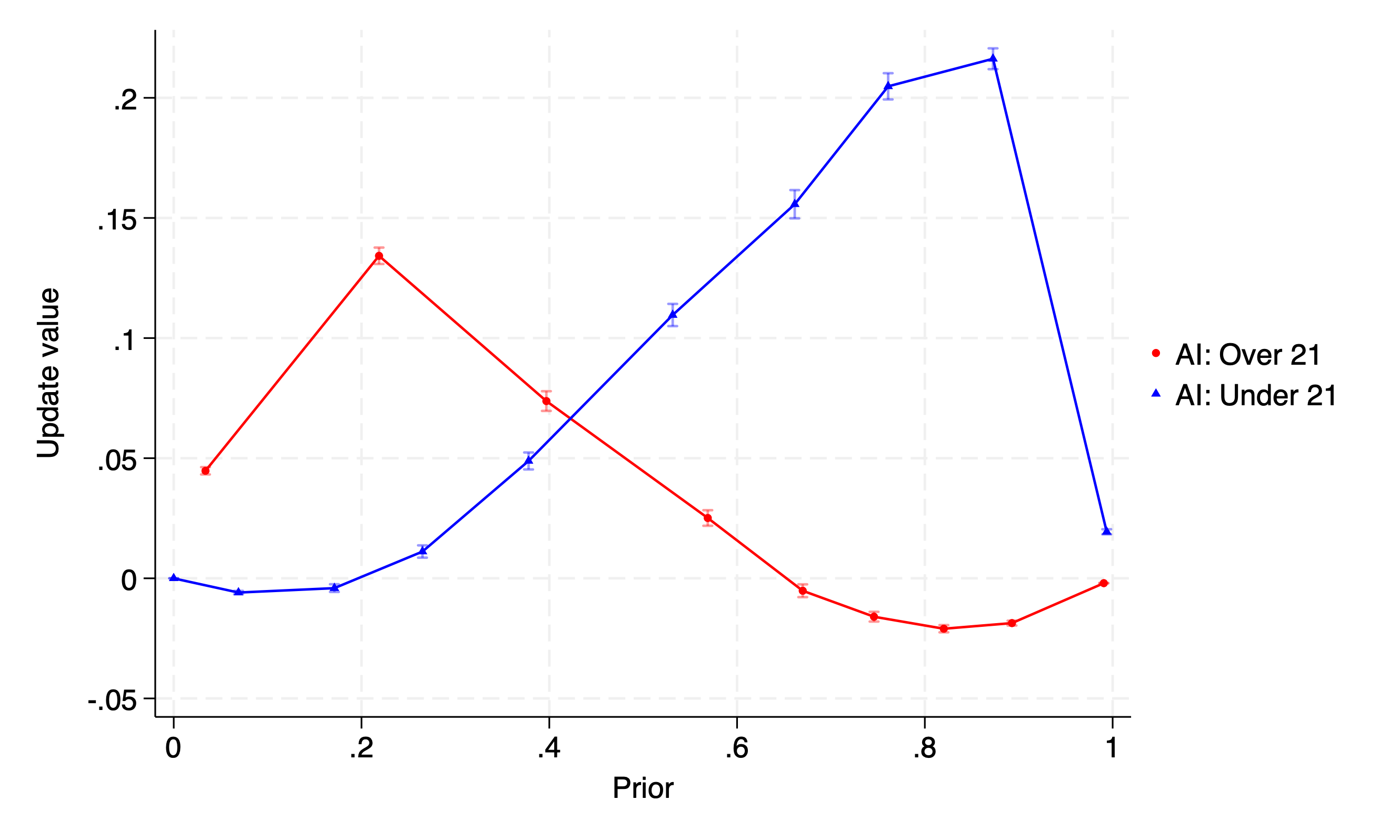}
\end{center}
\caption{Ex post updating value under signal-dependence neglect, with bars showing one standard error.}
\label{fig:update-value}
\end{figure}

We find that CR and wIU provide parsimonious descriptions of the data, but as noted previously, they are not designed to identify a unique behavioral microfoundation. Figure~\ref{fig:update-value} shows one reason why the asymmetric updating pattern may arise. The ex-post value of moving from the prior to the Bayesian posterior under signal-dependence neglect is largest when the AI recommendation contradicts an extreme prior and is close to zero, or even negative, when the recommendation confirms an extreme prior.

Three behavioral forces might provide possible microfoundations for the updating we observe in our experiment. First, inertia with qualitative information maps directly into wIU: DMs may choose posteriors close to the prior while satisfying the signal \citep{dominiak2021minimum,dominiak2023inertial}. Second, attention to contradiction can explain larger movements from extreme priors; limited attention and cognitive imprecision can distort perceived information \citep{ba2025over}, and Figure~\ref{fig:update-value} shows that updating has highest value where advice is most surprising. This connects to cue weighting and surprise \citep{lynch1989effects,friston2008hierarchical}. Third, interpretive uncertainty can attenuate updating near one half or lead DMs to reassess their prior after contradiction \citep{enke2023cognitive,ortoleva2012modeling,de2022non,dominiak2023inertial,chan2024prior}. Coarse average reliability beliefs and contingent thinking may also explain why INFO has limited impact \citep{augenblick2025overinference,aina2023contingent}.

\section{Conclusion}\label{sec:conclusion}

We experimentally study how individuals revise beliefs after receiving qualitative AI recommendations. In each round, participants first report a prior belief that a person in an image was over 21, then receive an AI recommendation, and then report a posterior belief. The recommendation is qualitative: it is either Over 21 or Under 21. This design generates 60,252 pairs of prior and posterior beliefs and allows us to study belief updating when information is payoff relevant, but the DGP behind the information is unknown or only summarized statistically.

We document three behavioral patterns. First, participants update close to zero when an extreme prior is confirmed by the AI. Second, participants update more when an extreme prior is contradicted by the AI. Third, participants update less when the prior is close to one half. These patterns motivate four testable properties: consistency with the recommendation, monotonicity in the prior, reactionary updating, and threshold updating. In the aggregate data, the first three properties are not rejected, and the hinge regressions imply nondegenerate non-updating thresholds of 0.620 after an Over 21 recommendation and 0.310 after an Under 21 recommendation. The INFO treatment does not change these patterns, even though it provides the state-contingent recommendation rates needed to construct a Bayesian posterior under signal-dependence neglect.

The evidence at the individual level is more heterogeneous. At the observation level, 98.1\% of updates are weakly in the direction of the AI recommendation. At the participant level, however, 44.3\% satisfy all four criteria and another 55.2\% fail exactly one criterion. Most participants satisfy the weak consistency, monotonicity, and reactionary updating criteria, while threshold updating varies more across participants.

These patterns differ from the predictions of qB updating. A qB model updates the prior through a distorted likelihood ratio, and therefore tends to predict smaller absolute movements near extreme priors and larger movements near intermediate priors. The data exhibit the opposite pattern for contradicting recommendations. CR and wIU fit better because they treat the recommendation as qualitative information. CR and wIU both have lower aggregate out-of-sample MSE than qB, while wIU has the highest restrictiveness among the aggregate models. When the subjective version of wIU is included, it has the lowest aggregate MSE and the highest completeness. At the individual level, wsIU has lower average MSE than qB and wIU and is statistically close to CR, which reinforces the conclusion that updating behavior is heterogeneous rather than governed by one rule for everyone.

Importantly, we do not claim that these are the only three models that could explain updating from qualitative AI recommendations. Rather, the comparison shows that benchmark models based on likelihood ratios miss the three patterns in beliefs,\footnote{See Appendix~\ref{app:alternative-model-benchmarks} for more comparison models.} while models that treat recommendations as general information capture those patterns more directly. Also, our experiment does not test the full axiomatic foundations of wIU or CR, which require richer variation in information sets and sequential recommendations. Instead, it tests their main implications for beliefs in a binary environment with qualitative recommendations: monotonicity, movement toward beliefs that are consistent with the recommendation, linear contraction under CR, and threshold inertia under wIU.

Our results have implications for AI interface design. If decision makers respond to qualitative AI recommendations using inertial or contraction rules, then providing a global accuracy rate may not be enough to induce Bayesian use of the recommendation. This does not imply that accuracy information is irrelevant. It may matter in settings where decision makers' beliefs about AI accuracy are miscalibrated. But our evidence suggests that when decision makers already have roughly accurate average beliefs about AI accuracy, aggregate accuracy information alone may leave the shape of updating largely unchanged. Decision makers may need information about local informativeness, the dependence between their own information and the AI prediction, or the features that generated the recommendation.

The results also suggest that recommendation format may matter, consistent with \citet*{hoong2025improving}. A firm designing an AI assistant can provide a qualitative recommendation, a calibrated probability, an AI confidence category, an explanation, or a comparison between the decision maker's judgment and the model's historical performance on similar cases. These formats need not be behaviorally equivalent. A qualitative recommendation may induce contraction or threshold updating, while a calibrated probability or local explanation may move decision makers closer to integration based on likelihood ratios. Which format is optimal may depend on the decision environment, the decision maker's prior information, and the cost of erroneous reliance on the AI.

At a more aggregate level, the findings speak to models of non-Bayesian opinion dynamics. \cite*{cerreia2024dynamic} study dynamic opinion aggregation when DMs use nonlinear, non-Bayesian aggregators. Our experiment studies a single DM interacting with an AI assistant, not a network. Still, if many DMs process qualitative AI recommendations using inertial or contraction-style rules, shared AI advice need not produce Bayesian convergence. DMs with different priors may respond differently to the same recommendation because the recommendation is confirming for some and contradicting for others. This implication should be treated as a theoretical possibility rather than a direct empirical conclusion from our experiment.

Several comparative statics are left for future work. The first is the precision and wording of the recommendation. Future experiments could vary whether the AI says ``more likely,'' makes an explicit recommendation, provides a calibrated probability, reports a confidence interval, or gives local evidence. They could also vary the threshold used to convert the AI model probability into a qualitative recommendation, such as requiring a higher probability before the AI gives an Over 21 or Under 21 recommendation. This would help distinguish contraction toward a qualitative threshold from more general pull toward one half. A related design could ask participants, before seeing the image, what they think ``more likely'' means mathematically, connecting the experiment to work on how individuals construct boundaries for contradictory evidence, such as \cite*{sadler2021practical}.

The second comparative static is feedback. Our experiment does not provide explicit feedback, and we do not observe dynamic learning in the aggregate. Future work could provide feedback about the true state, feedback about AI accuracy, or feedback about both the participant's own prior and the AI recommendation. Such designs would test whether participants learn the AI's reliability, learn their own reliability, or continue to rely on inertial heuristics after feedback.

The third comparative static is the decision context. Future work could study higher stakes, other tasks, including ones that require more cognitive effort or domain expertise, or more complex decisions, and compare advice from human experts, traditional machine learning systems, and large language models. LLM recommendations may be more anthropomorphic, more persuasive, and more capable of providing explanations. These features could change reliance on the recommendation, perceived opacity of the DGP, or the updating rule itself.

Our experiment is deliberately stylized to isolate a basic force in the interaction between humans and AI: when decision makers receive qualitative advice from an opaque algorithm, they may not integrate it as a signal with a known likelihood ratio. Instead, they may rely on an updating rule that is inertial, weighted by confidence, and sensitive to whether the advice confirms or contradicts the prior. As AI systems become more common in economic decision-making, understanding these updating rules is necessary for predicting when AI assistance improves decisions, when its measured effect is close to zero, and when it may reinforce existing beliefs.

\pagebreak
\appendix

{\footnotesize
\bibliographystyle{ecta}
\bibliography{bibliography}
}
\normalsize

\newpage

\section{Preregistration Report}\label{app:preregistration}

This appendix compares the analyses in the paper to AsPredicted \#228821. The preregistration was filed before data collection and stated that the main question was how participants update beliefs after imprecise AI advice. It also listed two secondary questions: whether updating differs with information about the AI and whether there are systematic differences across participants. The dependent variables in the preregistration were the reported likelihoods that the person in an image was over 21 before and after AI advice. These match the prior and posterior beliefs analyzed in the paper.

The implemented design follows the preregistration. The preregistered Baseline and AI Info treatments correspond to the NOINFO and INFO treatments in the paper. Participants completed repeated age classification tasks, received a qualitative AI prediction, and could revise their belief. The preregistration stated that participants who failed attention checks would not be allowed to complete the experiment, which is how the study was implemented. The final sample contains 188 participants in INFO and 189 in NOINFO, below the preregistered target of 200--300 participants per treatment. We also exclude 68 rounds in which participants timed out from the analysis sample. This exclusion at the round level was not listed in the AsPredicted exclusion rule field, but it affects only 0.11\% of rounds.

The preregistered primary analysis was to compare model performance on held-out data at the aggregate and individual levels for correlation-neglect models, Grether style models with belief biases, Inertial Updating, Inertial Updating with prior weighting, and several robustness variants. The paper implements this plan using Bayesian benchmarks with and without signal-dependence neglect, qB specifications based on the Grether rule, wIU, and wsIU. It also adds CR as an additional model motivated by general information. The main criterion for model comparison is MSE of posterior beliefs on held-out data. The preregistration also listed accuracy, distance from truth, MSE, and cross-entropy as performance measures. We report binary accuracy descriptively, but use MSE of posterior beliefs as the main criterion for model fit because the paper's model comparison is about belief updating rather than state prediction. The paper also adds grouped cross-validation, model diagnostics for updates and exact non-updating, latent prior robustness for qB, and completeness--restrictiveness comparisons.

The preregistration specified heterogeneity analyses by confidence in AI, confidence in oneself, calibration, age, gender, and response time. The paper implements these in several places: Table~\ref{tab:property-tests-splits} reports property tests by participant and image splits, Section~\ref{sec:comp} compares updating types and wsIU parameters across participants, Table~\ref{tab:wsiu-survey-parameters} relates wsIU parameters to beliefs about AI accuracy and confidence in own prior, and Appendix~\ref{app:accRT} reports response-time analyses. Participant ability and MSE of prior beliefs are used as empirical proxies for calibration. The preregistered secondary analyses on aggregate updating patterns, survey beliefs about AI accuracy, human and AI performance, dynamics over rounds, and difficult images are reported in Sections~\ref{sec:results}--\ref{sec:comp} and Appendices~\ref{app:summary-tables},~\ref{app:splits}, and~\ref{app:accRT}.

Several analyses in the current paper were not specified in the preregistration. These include the four behavioral property tests, the non-updating threshold estimates, two-way clustered inference by participant and image, local empirical Bayes benchmarks, checks of prior sufficiency, power calculations for treatment effects, the finite mixture summary of qB and wsIU types, the comparison to external datasets on AI advice, and the descriptive check for dynamic discouragement after contradicting advice at extreme priors. These additions are motivated by the observed updating patterns and by the theoretical comparison between likelihood ratio and general information models.

\section{Additional Analyses}

\subsection{Additional Model Benchmarks}\label{app:alternative-model-benchmarks}

The Bayesian and qB benchmark variants are reported in Table~\ref{tab:benchmark-variants}. The qB comparison clarifies the role of signal-dependence neglect. In the original split at the observation level, qB without signal-dependence neglect has lower MSE than qB with signal-dependence neglect (0.0390 rather than 0.0402). We therefore do not interpret signal-dependence neglect as necessary for fitting our data. However, even the qB variants with lower MSE remain behind CR, wIU, and wsIU in Table~\ref{tab:agg}, and the diagnostic comparisons below show why: qB smooths updates but does not capture exact non-updating or the asymmetry between confirming and contradicting recommendations as well as the inertial models.

\begin{table}[ht!]
\centering
\caption{Bayesian and qB benchmark variants.}
\label{tab:benchmark-variants}
\footnotesize
\setlength{\tabcolsep}{6pt}
\begin{tabular*}{0.70\textwidth}{@{\extracolsep{\fill}}lc@{}}
\toprule
Model & Test MSE \\
\midrule
Bayesian benchmark without signal-dependence neglect & 0.0530 \\
Bayesian benchmark with signal-dependence neglect & 0.0548 \\
Epstein, Noor, and Sandroni (ENS) updating rule & 0.0408 \\
qB with signal-dependence neglect & 0.0402 \\
qB without signal-dependence neglect & 0.0390 \\
Weighted qB & 0.0389 \\
\bottomrule

\end{tabular*}
\begin{minipage}{0.95\textwidth}
\footnotesize
\emph{Notes:} Entries are means across the same 100 train-test splits at the observation level used in Table~\ref{tab:agg}. The Bayesian benchmark applies Bayes' rule using an estimated recommendation process that conditions on the prior. The Bayesian benchmark under signal-dependence neglect applies Bayes' rule using only the empirical state-contingent recommendation rates. ENS is a convex combination of the prior and the Bayesian benchmark with one parameter. The qB rows compare the baseline qB specification, which imposes signal-dependence neglect, to the same qB specification using the Bayesian advice term instead. Weighted qB is a convex combination of the prior and the qB prediction with one parameter.
\end{minipage}
\end{table}

\subsection{Robustness of Aggregate Property Tests}\label{app:property-robustness}

Table~\ref{tab:exclusion-robustness} reports robustness checks that exclude participants or observations based on pre-specified data-quality and behavior measures.

\begin{table}[H]
\centering
\caption{Robustness of aggregate property tests to exclusion rules.}
\label{tab:exclusion-robustness}
\footnotesize
\setlength{\tabcolsep}{3pt}
\begin{tabular*}{\textwidth}{@{\extracolsep{\fill}}lrrrrrrr@{}}
\toprule
Sample & $N$ & Cons. & Mono. & React. & Thresh. & Over thresh. & Under thresh. \\
\midrule
Main sample & 60,252 & \ensuremath{\checkmark} & \ensuremath{\checkmark} & \ensuremath{\checkmark} & \ensuremath{\checkmark} & 0.62 & 0.31 \\
Exclude non-updaters & 55,134 & \ensuremath{\checkmark} & \ensuremath{\checkmark} & \ensuremath{\checkmark} & \ensuremath{\checkmark} & 0.58 & 0.32 \\
Exclude no-update rate $\geq$ 95\% & 40,427 & \ensuremath{\checkmark} & \ensuremath{\checkmark} & \ensuremath{\checkmark} & \ensuremath{\checkmark} & 0.59 & 0.33 \\
Exclude fastest 5\% & 58,174 & \ensuremath{\checkmark} & \ensuremath{\checkmark} & \ensuremath{\checkmark} & \ensuremath{\checkmark} & 0.62 & 0.32 \\
Exclude poor INFO recall proxy & 55,455 & \ensuremath{\checkmark} & \ensuremath{\checkmark} & \ensuremath{\checkmark} & \ensuremath{\checkmark} & 0.63 & 0.30 \\
Exclude degenerate priors & 58,973 & \ensuremath{\checkmark} & \ensuremath{\checkmark} & \ensuremath{\checkmark} & \ensuremath{\checkmark} & 0.59 & 0.33 \\
\bottomrule

\end{tabular*}
\begin{minipage}{0.95\textwidth}
\footnotesize
\emph{Notes:} Checkmarks are defined as in Table~\ref{tab:property-tests-splits}. Non-updaters are participants whose posterior belief equals their prior belief in every analyzed task. The exclusion based on extreme rates of no updating removes participants whose rate of no updating is at least 95\%. The fastest sample exclusion removes participants in the bottom 5\% of median response times for prior beliefs. The poor-recall proxy removes INFO participants whose stated overall belief about AI accuracy is more than 25 percentage points from the empirical AI accuracy. Degenerate prior distributions are participants whose prior standard deviation is below 0.05 or whose priors fall in one decile more than 90\% of the time.
\end{minipage}
\end{table}

\subsection{Response to Obvious AI Mistakes}\label{app:dynamic-wrong-ai}

A possibility is that participants may respond dynamically when the AI makes an ``obvious mistake'' in the eyes of the participant.  There are potentially two ways to identify that the AI has made a clear mistake from the perspective of the participants: all subjects agree on the state and the AI disagrees, or a subject holds an extreme prior and does not move much in response to the AI. To capture the latter, we identify rounds in which a participant had a prior at most 0.10 or at least 0.90, a recommendation that contradicts the prior, an absolute update of at most 0.05, and at least one previous nonzero update by the same participant.

These events occur in 3,122 observations, or 5.2\% of the analysis sample, and are observed for 264 participants. In the raw data, updating is lower on the next task after such an event. Participants update on the next task in 10.7\% of cases after such an event, compared with 22.0\% of other cases. The mean absolute update is 0.049 after such an event and 0.079 otherwise, while the mean signed update toward the recommendation is 0.036 after such an event and 0.071 otherwise. This pattern is consistent with dynamic discouragement or disengagement after observing advice that contradicts an extreme prior and leaves beliefs nearly unchanged.

The evidence is only suggestive, however, because these events are mechanically concentrated among participants and histories with low updating. After residualizing outcomes with respect to participant fixed effects, round fixed effects, and current controls for recommendation by prior decile, the one-round association is smaller: the coefficient is $-0.007$ for any update ($p=0.184$), $-0.004$ for absolute update ($p=0.154$), and $-0.007$ for signed update toward the recommendation ($p=0.021$), using two-way clustered standard errors by participant and image. Thus, the raw pattern is consistent with short-run discouragement, but the controlled evidence is mixed and does not change the main pooled updating results.

\subsection{Additional Summary Tables}\label{app:summary-tables}

\begin{table}[H]
\centering
\caption{Summary statistics and treatment balance.}
\label{tab:summary-treatment-balance}
\footnotesize
\setlength{\tabcolsep}{4pt}
\emph{Panel A: Participant summary and treatment balance}\par\vspace{0.25em}
\begin{tabular*}{\textwidth}{@{\extracolsep{\fill}}p{0.38\textwidth}cccc@{}}
\toprule
 & All & INFO & NOINFO & $p$-value \\
\midrule
$N$ & 377 & 188 & 189 &  \\
Age, mean (sd) & 37.7 (13.3) & 39.0 (13.9) & 36.5 (12.5) & 0.071 \\
Age, median & 35.0 & 37.0 & 34.0 &  \\
Woman & 59.4\% & 61.2\% & 57.7\% & 0.490 \\
White & 61.5\% & 63.3\% & 59.8\% & 0.485 \\
Prior binarized accuracy & 61.8\% & 62.1\% & 61.4\% & 0.320 \\
Posterior binarized accuracy & 66.2\% & 66.3\% & 66.1\% & 0.822 \\
Belief about AI accuracy, mean (sd) & 67.6 (20.6) & 68.2 (19.0) & 67.0 (22.1) & 0.544 \\
Confidence in own prior, mean (sd) & 73.9 (20.8) & 72.5 (21.7) & 75.4 (19.9) & 0.178 \\
Confidence in posterior, mean (sd) & 74.3 (20.9) & 73.0 (22.1) & 75.6 (19.6) & 0.238 \\
Prior response time, median seconds & 3.8 (1.6) & 3.8 (1.7) & 3.7 (1.6) & 0.511 \\
Posterior response time, median seconds & 6.9 (2.5) & 7.0 (2.6) & 6.8 (2.5) & 0.506 \\
No-update rate & 78.6\% & 78.6\% & 78.6\% & 0.984 \\
Above-median Prolific approvals & 49.9\% & 52.7\% & 47.1\% & 0.280 \\
\bottomrule

\end{tabular*}

\vspace{0.9em}
\emph{Panel B: Image and task-instance summary}\par\vspace{0.25em}
\begin{tabular*}{0.74\textwidth}{@{\extracolsep{\fill}}lc@{}}
\toprule
Measure & Value \\
\midrule
Images & 160 \\
Over 21 images & 80 (50.0\%) \\
Under 21 images & 80 (50.0\%) \\
AI recommends Over 21 & 56.2\% \\
AI binary accuracy & 76.2\% \\
AI model confidence, mean (sd) & 0.76 (0.13) \\
White image & 78.8\% \\
Woman image & 50.0\% \\
\bottomrule

\end{tabular*}
\begin{minipage}{0.95\textwidth}
\footnotesize
\emph{Notes:} Panel A reports means across participants, with standard deviations in parentheses where applicable. Response time rows report median response times by participant. The no-update rate is the participant fraction of rounds in which the posterior belief equals the prior belief. Above-median Prolific approvals is a proxy for platform experience. The current cleaned data file does not retain separate fatigue point or previous paid classification experience variables, so those measures are not reported. The $p$-values are two-sided Welch tests of equality across treatments. In Panel B, the unit of observation is an image, so the AI binary accuracy row is at the image level rather than observation weighted after round exclusions. AI confidence is $\max\{p,1-p\}$, where $p$ is the AI model probability that the person in the image is over 21.
\end{minipage}
\end{table}

\subsection{Updating Types by Treatment}\label{app:type-treatment}

\begin{table}[H]
\centering
\caption{Updating types by treatment.}
\label{tab:type-treatment}
\footnotesize
\begin{tabular}{lcccc}
\toprule
Treatment & $N$ & qB: lowest MSE & qB: hierarchical & wsIU: hierarchical \\
\midrule
INFO & 188 & 38.8\% & 34.9\% & 65.1\% \\
NOINFO & 189 & 36.5\% & 30.0\% & 70.0\% \\
\addlinespace
NOINFO $-$ INFO &  & -2.3 pp & -4.9 pp & 4.9 pp \\
$p$-value &  & 0.662 & 0.412 & 0.412 \\
\bottomrule

\end{tabular}
\begin{minipage}{0.95\textwidth}
\footnotesize
\emph{Notes:} Lowest MSE shares classify each participant by which model has lower average out-of-sample MSE across splits. Hierarchical shares are estimated separately by treatment using the finite mixture model with two types from Table~\ref{tab:hierarchical-type-shares}. The $p$-values are two-sided permutation tests of equality across treatments. The wsIU $p$-value is the same as the hierarchical qB $p$-value because the two shares sum to one.
\end{minipage}
\end{table}

\subsection{Updating by Subgroup}\label{app:splits}

This section reports a series of robustness checks for the updating patterns documented in Section~\ref{sec:results}. We split the data along several dimensions that may plausibly affect belief revision: whether the AI recommendation is directionally correct in binary terms, AI confidence, image difficulty, round number, response time, gender, age, participant ability, beliefs about AI accuracy, and confidence in own prior. The split definitions are those used in Table~\ref{tab:property-tests-splits}: AI correctness is at the observation level, AI confidence and image difficulty are at the image level, second half is rounds 81--160, response speed is based on the median prior response time for each participant, and ability, beliefs about AI accuracy, and confidence in own prior are median splits across participants. Across these splits, the same qualitative pattern is largely preserved. Participants tend to update in the direction of the AI recommendation, but updating is largest when the recommendation contradicts an extreme prior and smallest when the recommendation confirms an extreme prior or when the prior is intermediate. These figures, therefore, suggest that the results are not driven by a particular subgroup of participants, a subset of images, fatigue, response speed, or differential beliefs about AI accuracy.

\begin{figure}[H]
\centering
\includegraphics[width=4in]{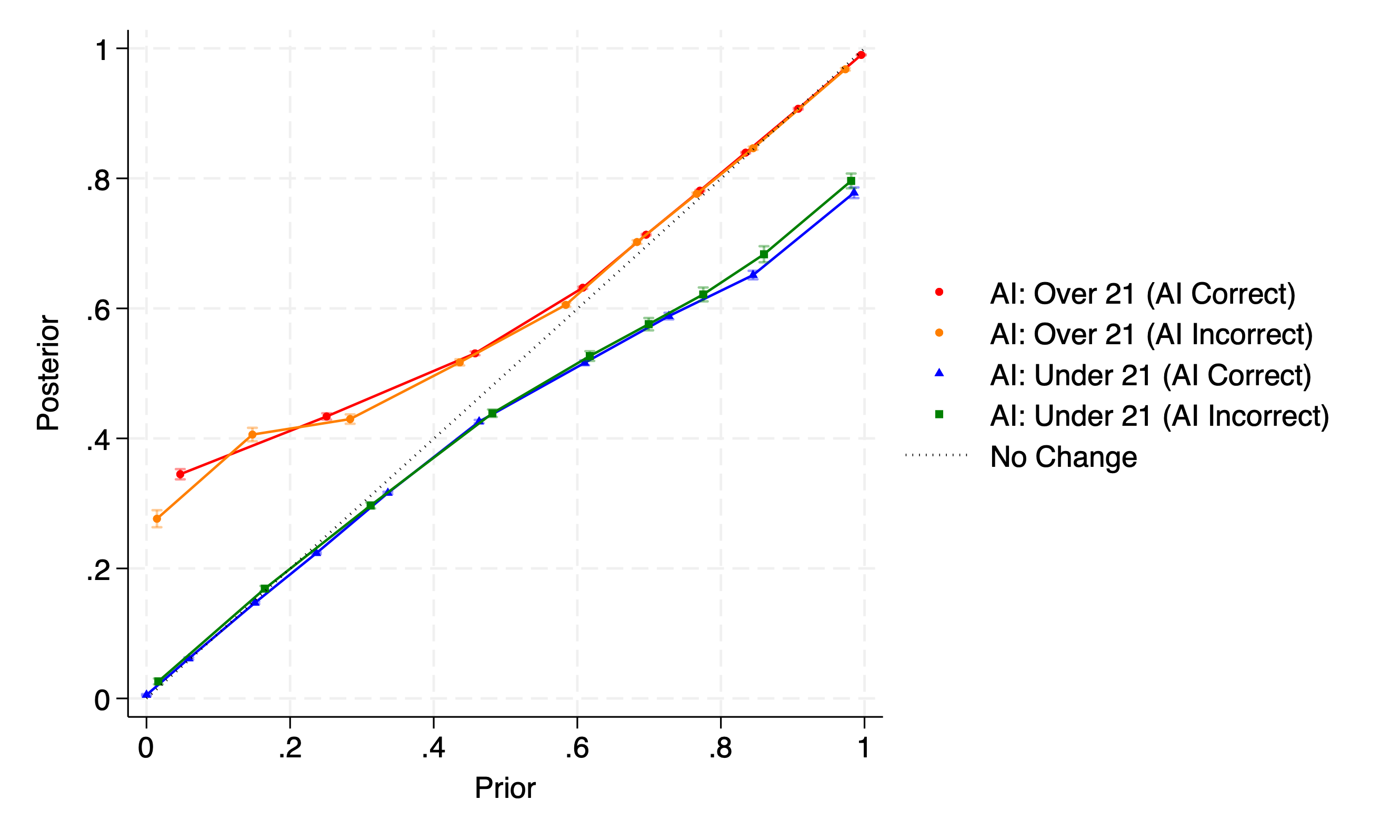}
\caption{Updating by binary directional correctness of the AI recommendation.}
\label{fig:update-ai-correct}
\end{figure}

\begin{figure}[H]
\centering
\includegraphics[width=4in]{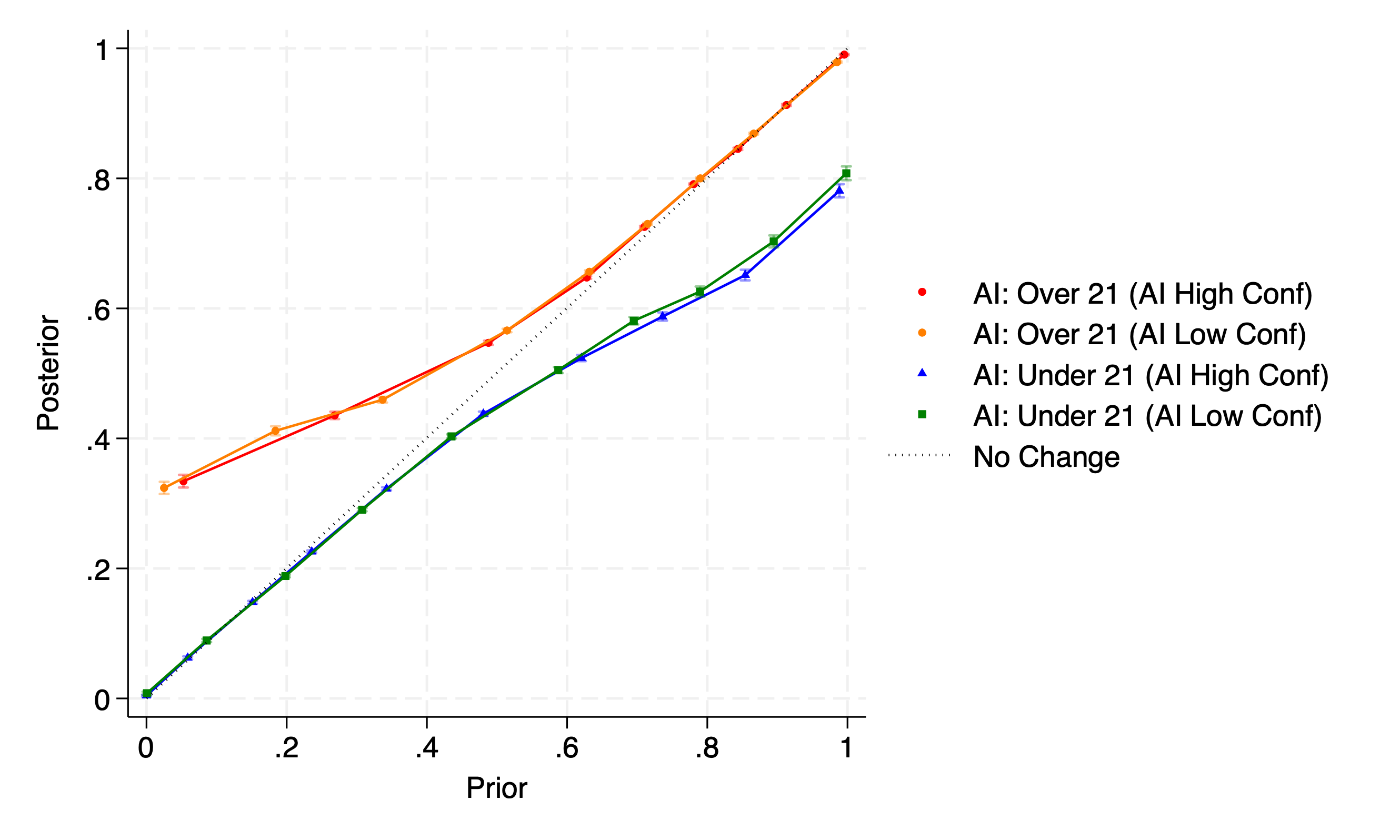}
\caption{Updating by AI confidence.}
\label{fig:update-ai-confidence}
\end{figure}

\begin{figure}[H]
\centering
\includegraphics[width=4in]{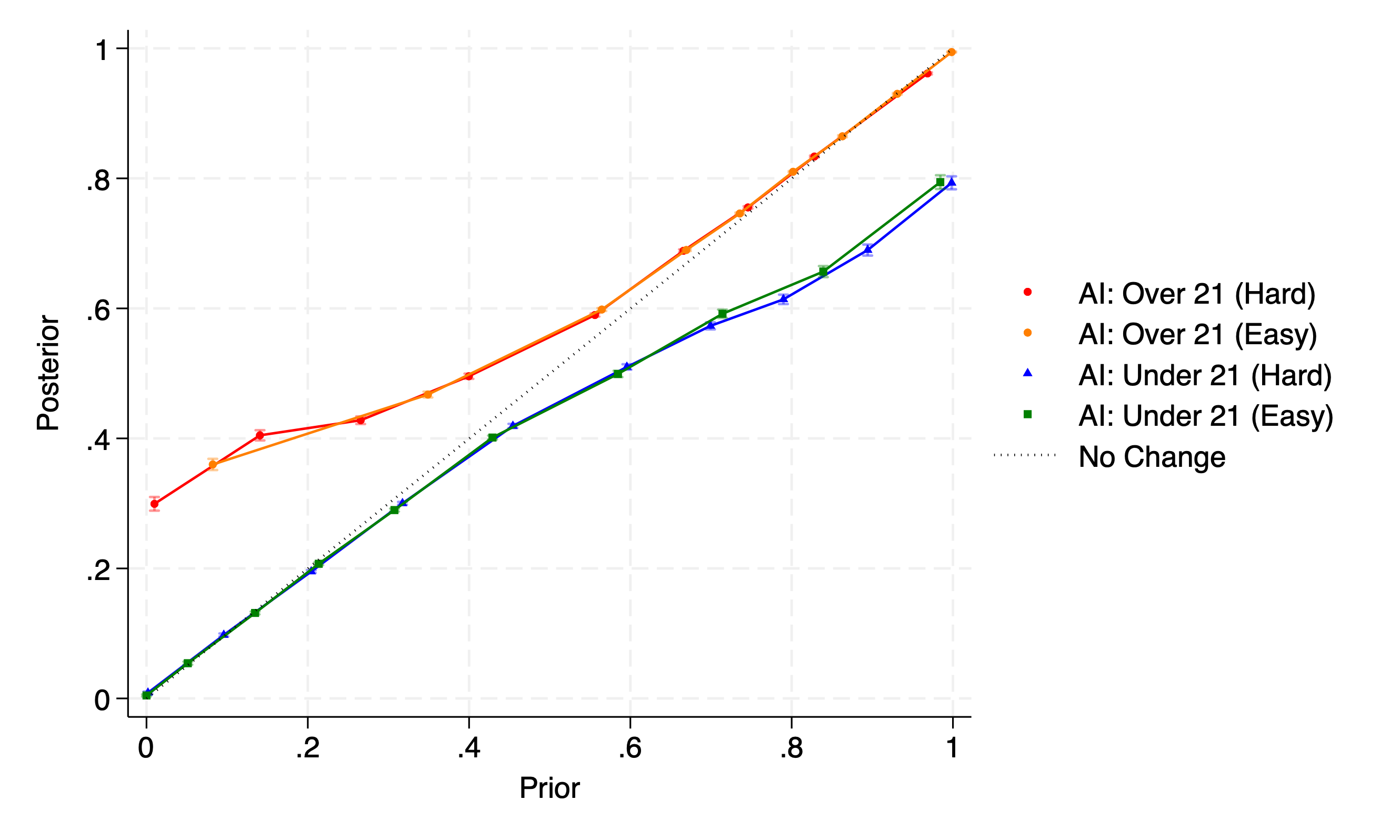}
\caption{Updating by image difficulty.}
\label{fig:update-image-difficulty}
\end{figure}

\begin{figure}[H]
\centering
\includegraphics[width=4in]{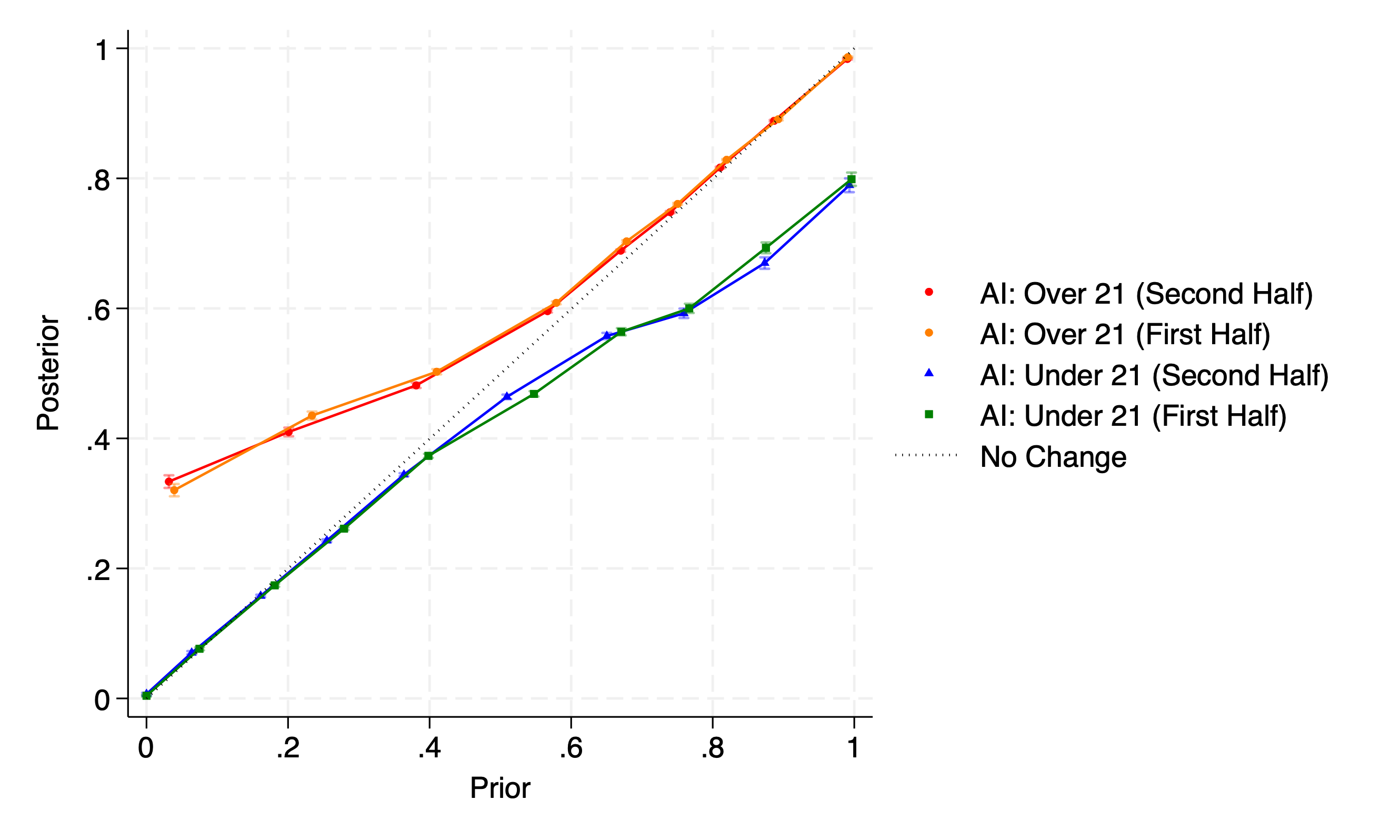}
\caption{Updating by round (rounds 1--80 or rounds 81--160).}
\label{fig:update-round}
\end{figure}

\begin{figure}[H]
\centering
\includegraphics[width=4in]{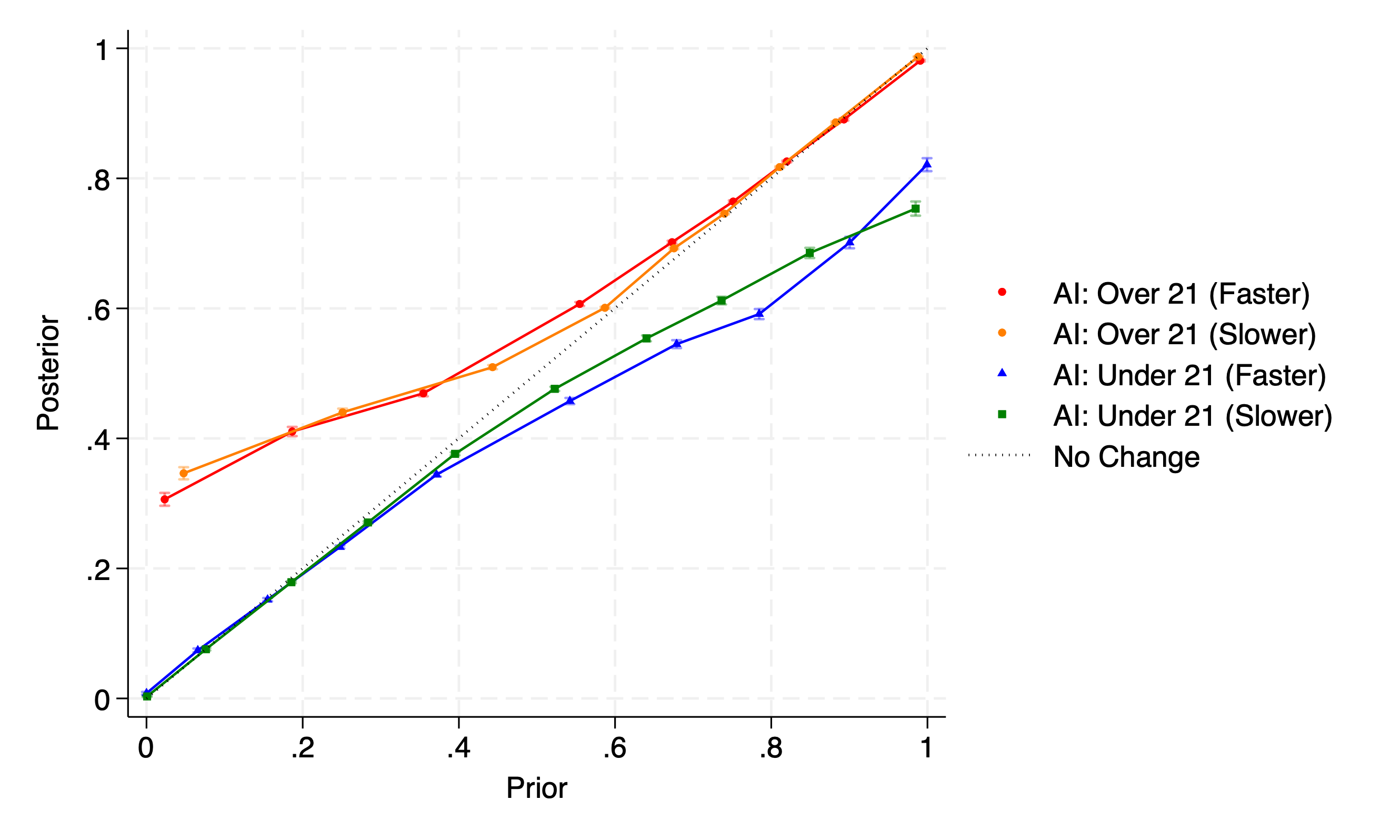}
\caption{Updating by response time (above or below the median prior response time by participant).}
\label{fig:update-response-time}
\end{figure}

\begin{figure}[H]
\centering
\includegraphics[width=4in]{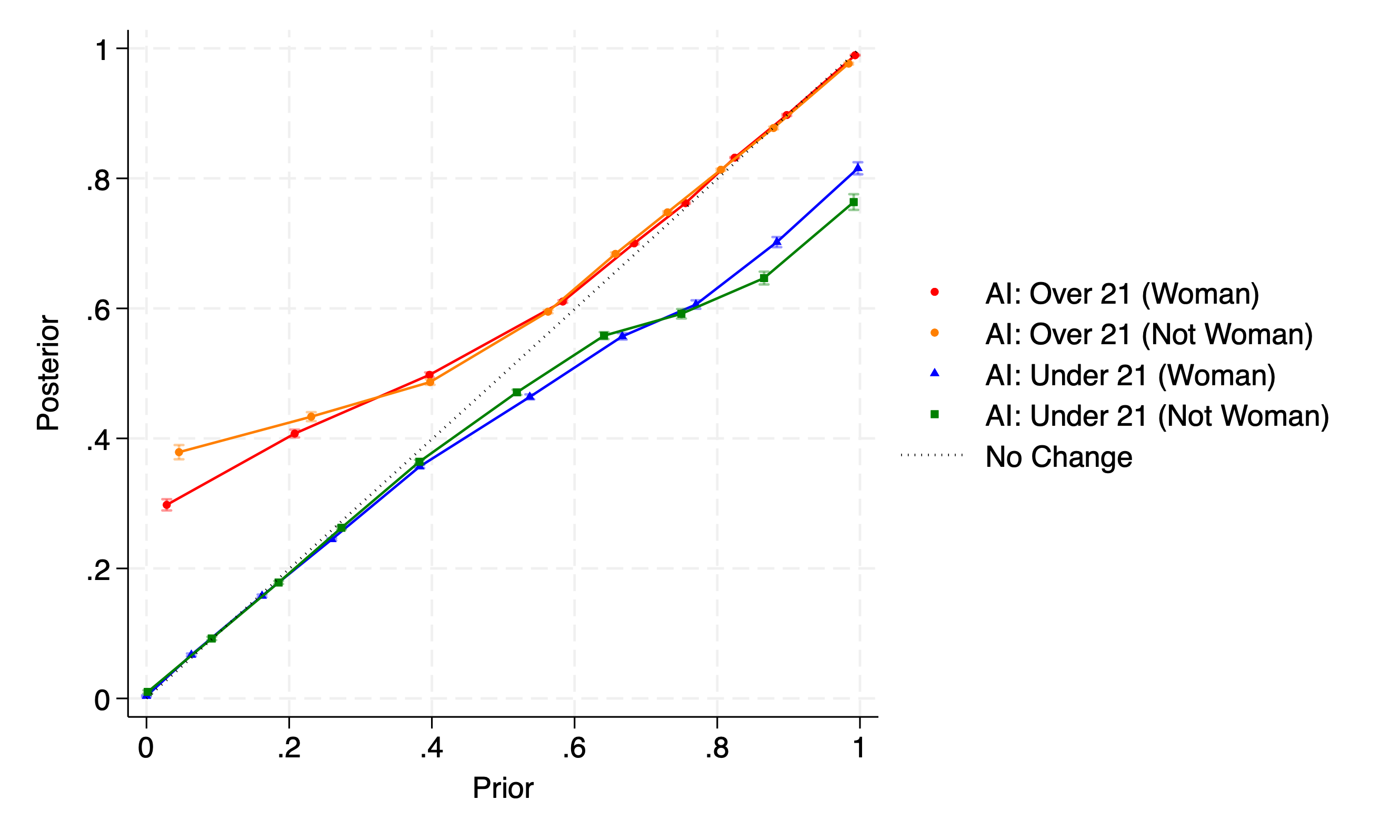}
\caption{Updating by gender (self-reported as a woman or another category).}
\label{fig:update-gender}
\end{figure}

\begin{figure}[H]
\centering
\includegraphics[width=4in]{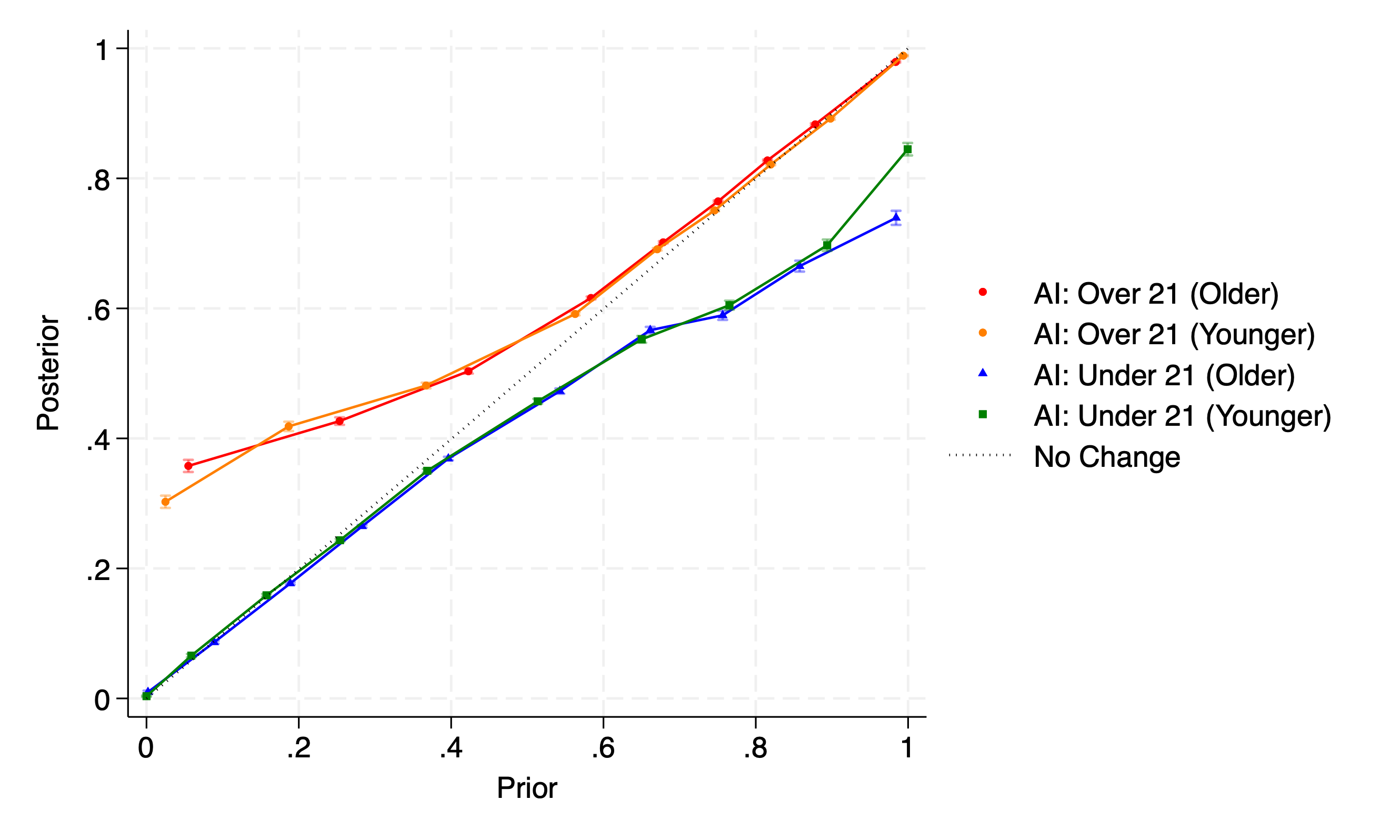}
\caption{Updating by age (above or below the median age by participant).}
\label{fig:update-age}
\end{figure}

\begin{figure}[H]
\centering
\includegraphics[width=4in]{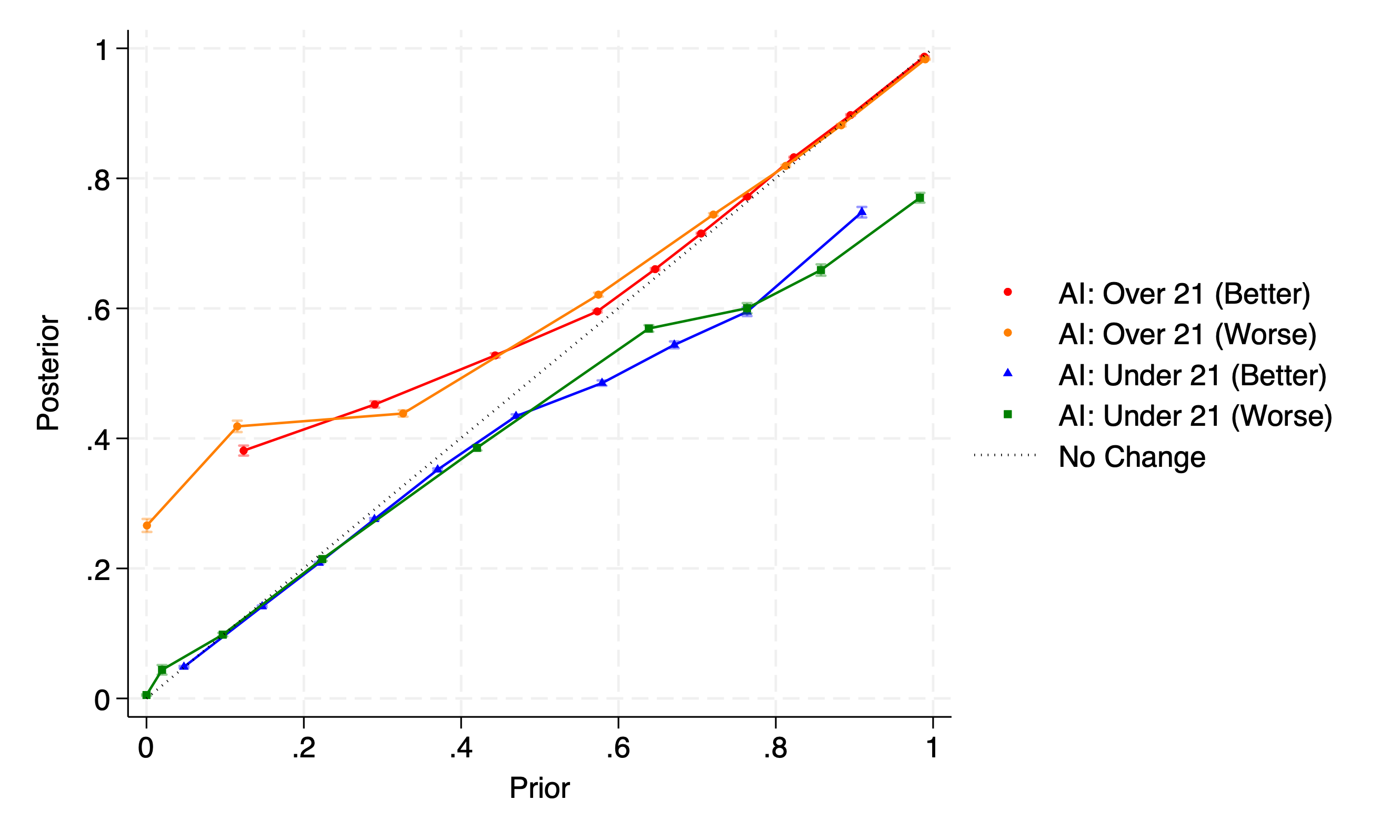}    
\caption{Updating by participant ability (lower or higher MSE of prior beliefs by participant).}
\label{fig:update-ability}
\end{figure}

\begin{figure}[H]
\centering
\includegraphics[width=4in]{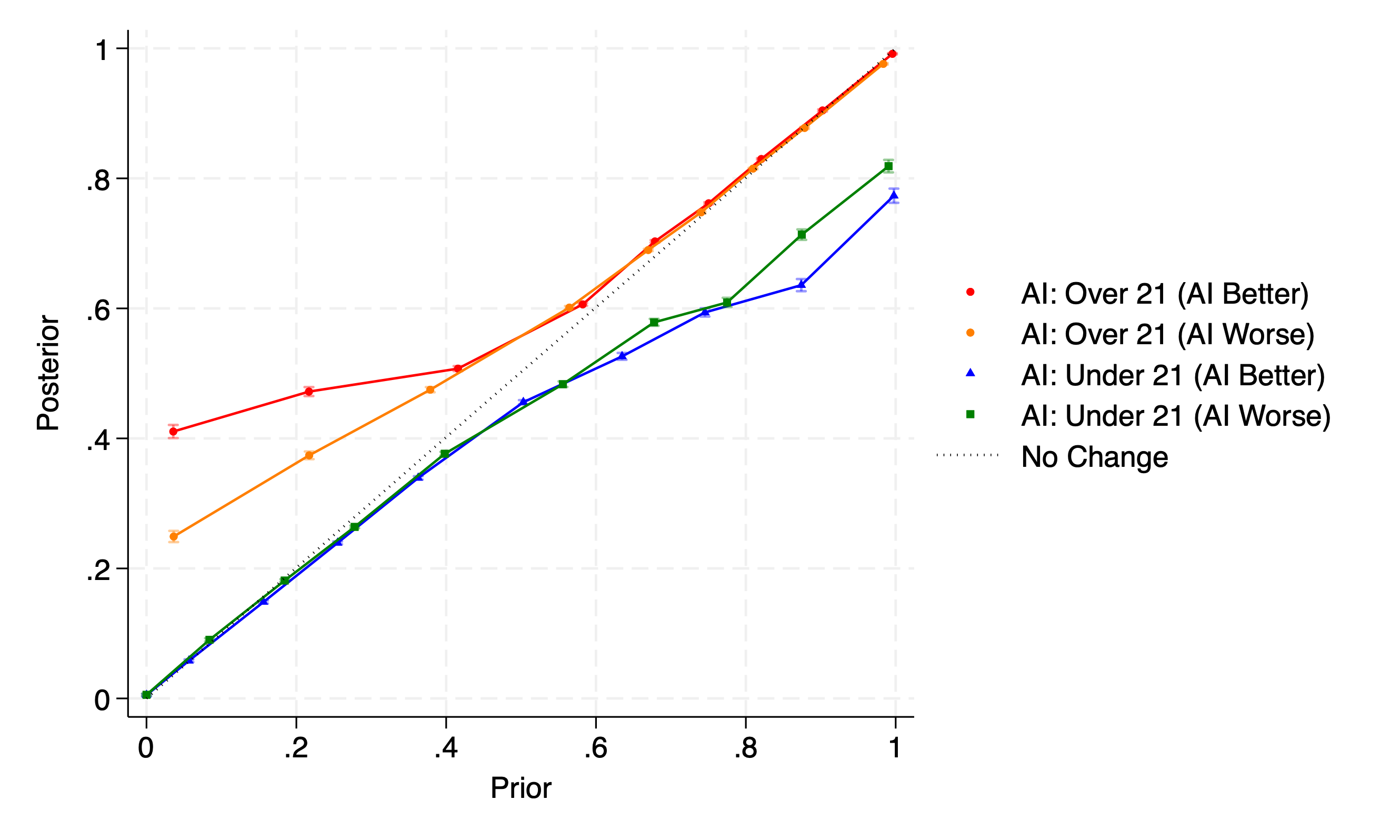}
\caption{Updating by belief about AI accuracy (above or below median belief about how often AI matched truth).}
\label{fig:update-ai-accuracy-belief}
\end{figure}

\begin{figure}[H]
\centering
\includegraphics[width=4in]{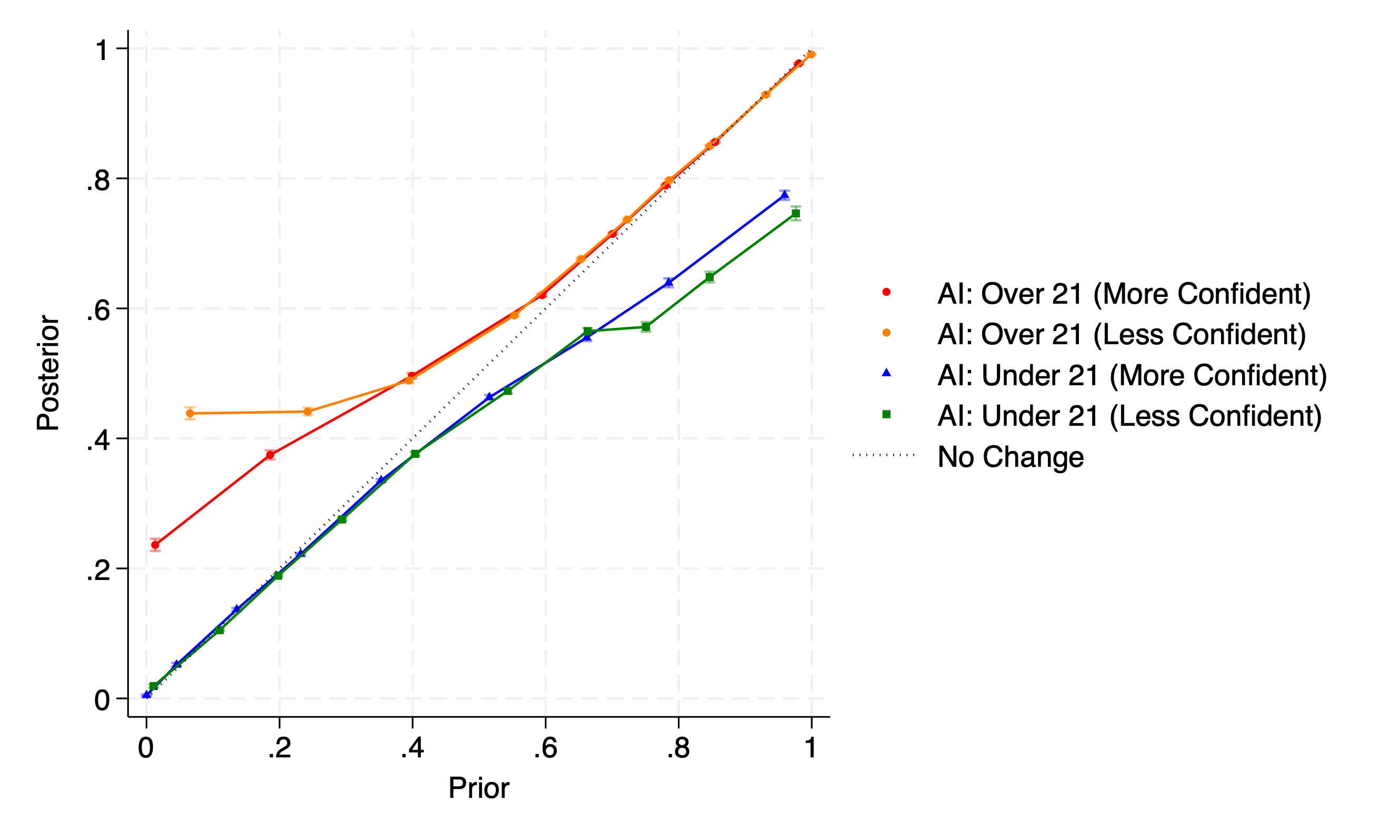}
\caption{Updating by confidence in own prior (above or below median self-reported confidence in prior beliefs).}
\label{fig:update-prior-confidence}
\end{figure}

\begin{figure}[H]
\centering
\includegraphics[width=4in]{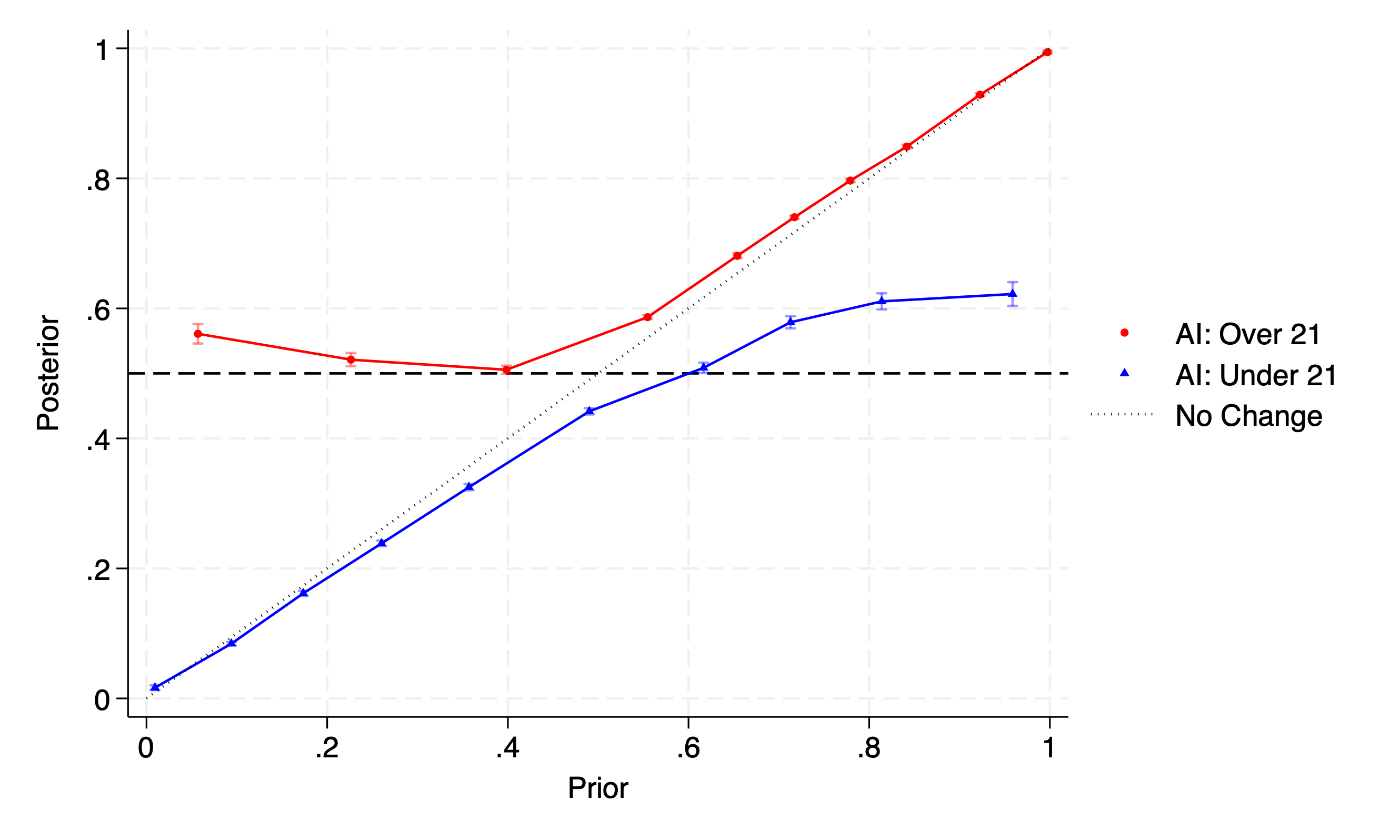}
\caption{Updating for participants with higher beliefs about AI accuracy and lower confidence in own prior.}
\label{fig:update-ai-accuracy-belief-low-prior-confidence}
\end{figure}

\subsection{Survey Responses}\label{app:survey}

\begin{figure}[H]
\centering
\includegraphics[width=4in]{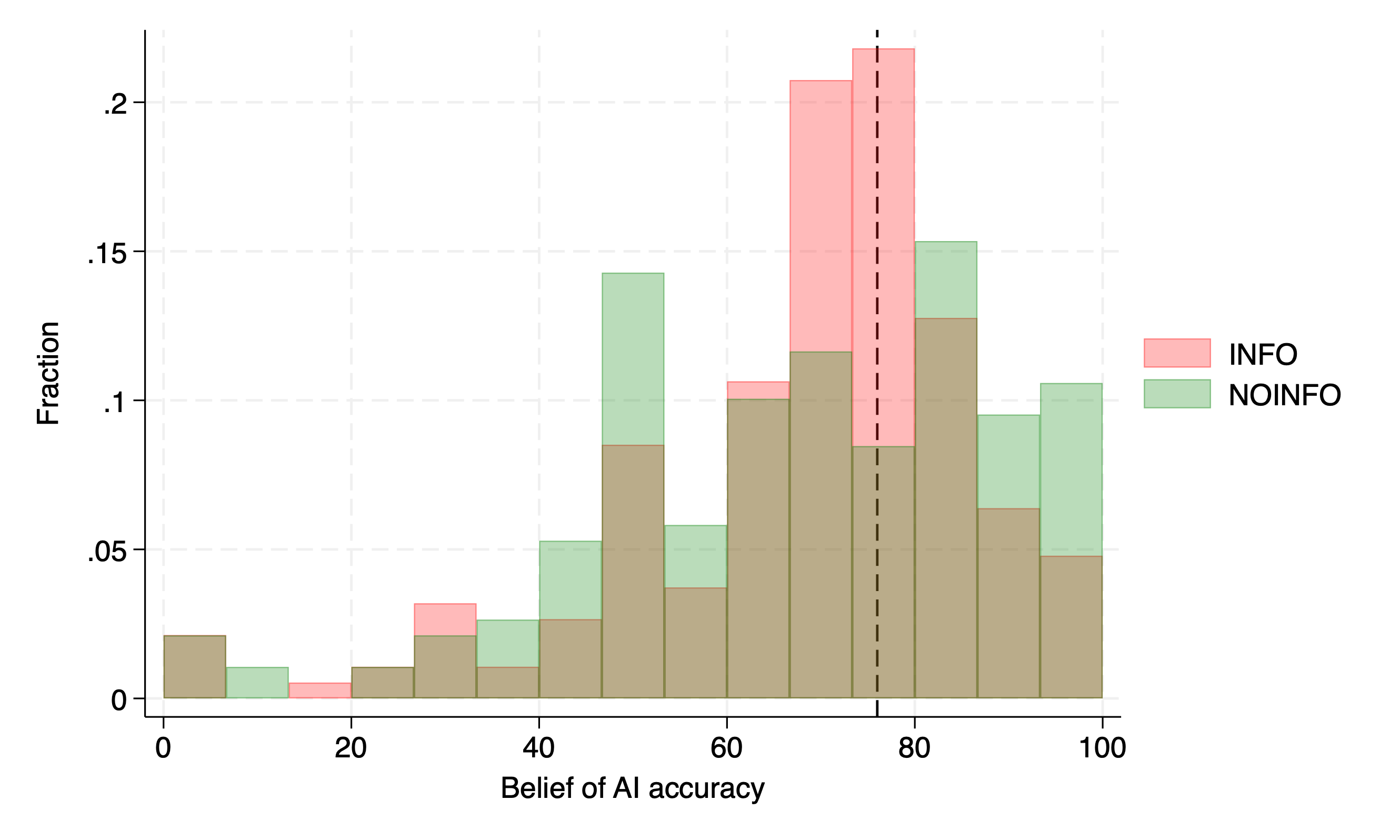}
\caption{Distribution of beliefs about AI accuracy.}
\label{fig:hist-ai-accuracy-beliefs}
\end{figure}

\begin{figure}[H]
\centering
\includegraphics[width=4in]{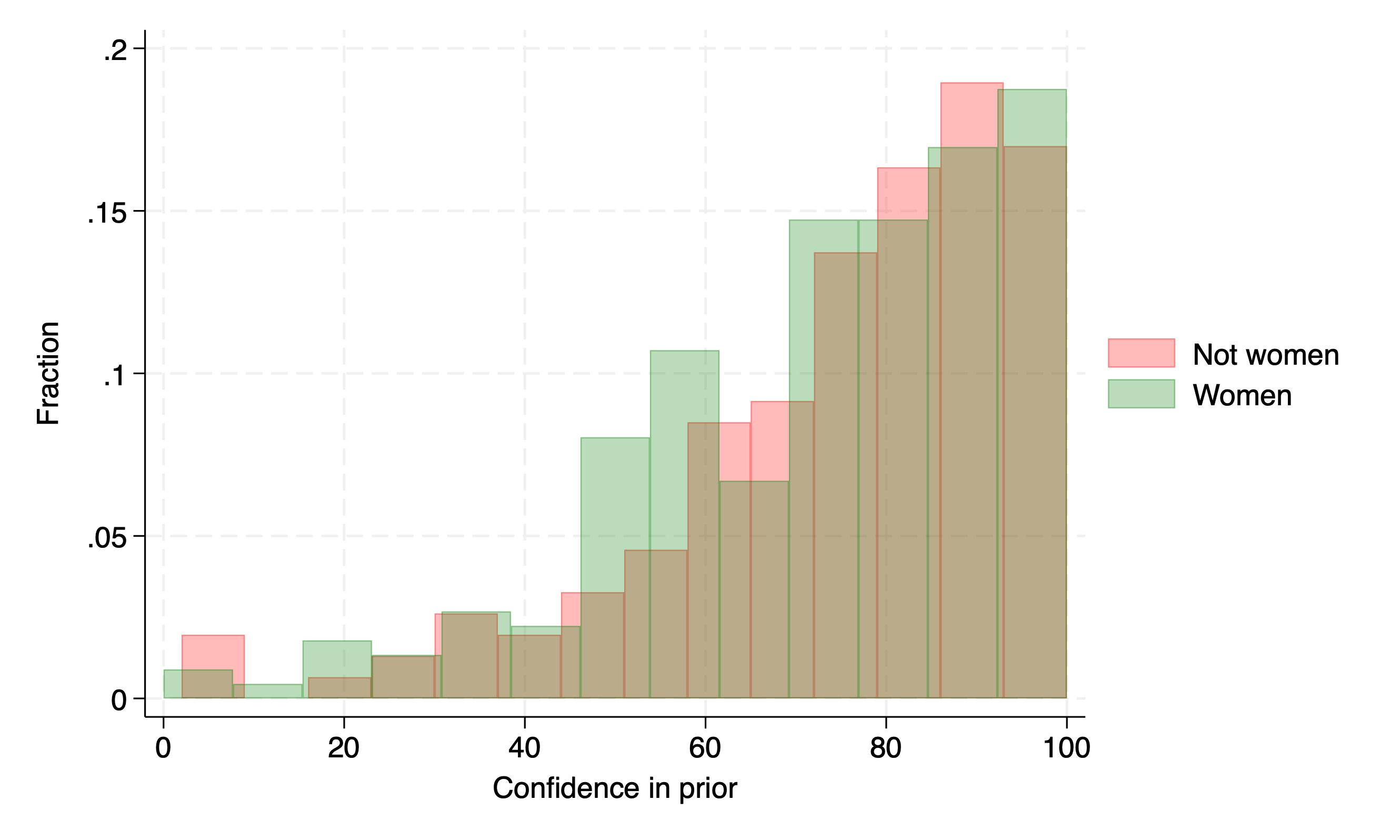}
\caption{Distribution of confidence in own prior by gender.}
\label{fig:hist-prior-confidence}
\end{figure}

Figures~\ref{fig:hist-ai-accuracy-beliefs} and~\ref{fig:hist-prior-confidence} show the distribution of the two survey measures used most often in the heterogeneity analysis: beliefs about AI accuracy and confidence in one's own prior beliefs.

\subsection{Response Time}\label{app:accRT}

\begin{figure}[H]
\centering
\includegraphics[width=4.5in]{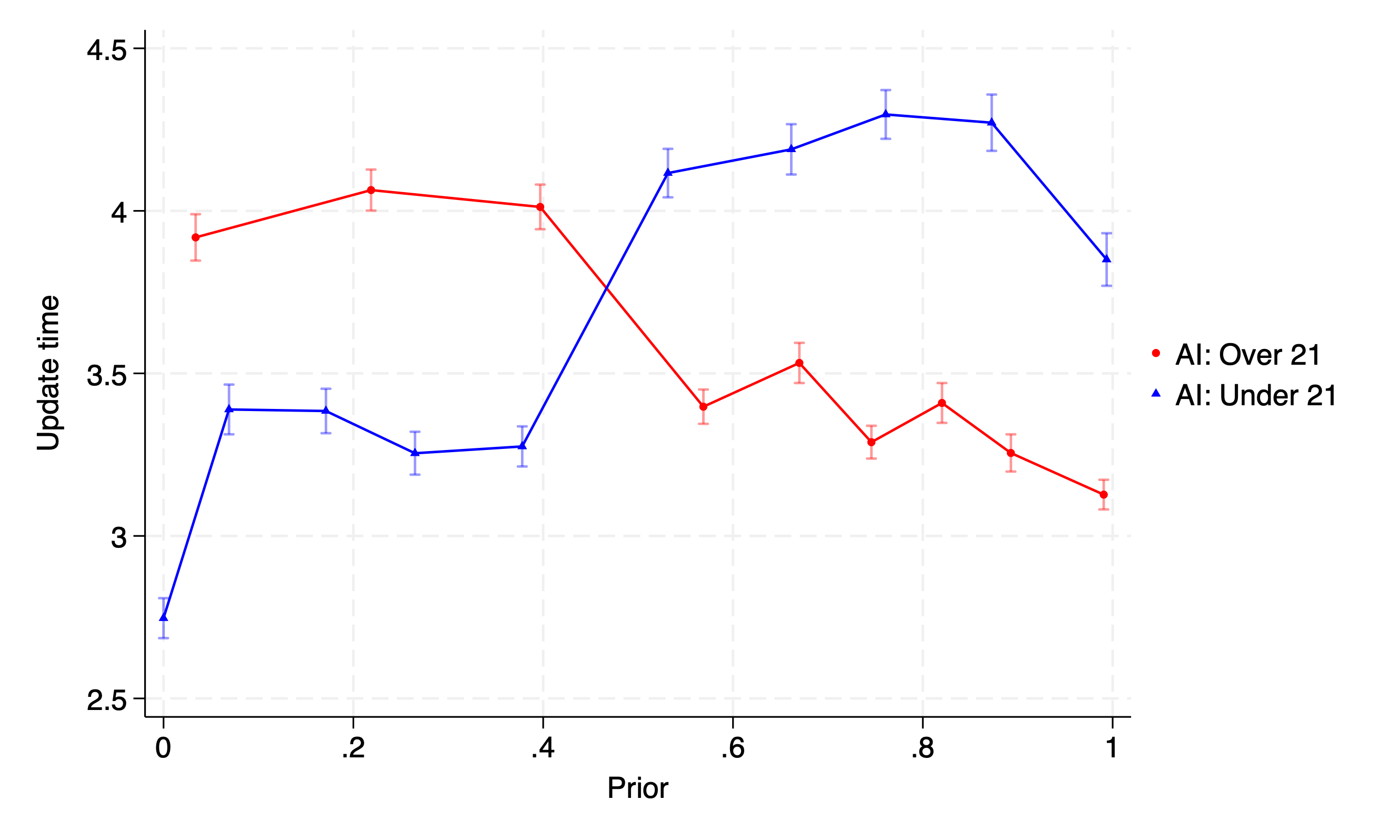}
\caption{Update time by prior belief $\mu(s_1)$.}
\label{fig:update-time}
\end{figure}

Figure~\ref{fig:update-time} summarizes the time participants took to revise beliefs after seeing the AI recommendation, and Table~\ref{tab:response-time-updating} reports a descriptive check of whether response times predict updating after flexibly controlling for prior beliefs, AI recommendations, and their interaction. Response time for the prior belief alone does not predict whether participants update or how much they update. By contrast, response time after advice is associated with updating: a longer response after advice is associated with a higher probability of any update and larger update magnitudes. This relationship should not be interpreted causally, since response time after advice is partly a consequence of deciding whether and how far to move the slider. As a check on the concern that non-updating after directionally correct initial beliefs reflects only the cost of moving the slider, participants still spent time on the decision after advice in the cases with extreme confirming priors and no movement: the median incremental response time was 2 seconds and the mean was 3.0 seconds.

\begin{table}[H]
\centering
\caption{Response times and updating.}
\label{tab:response-time-updating}
\footnotesize
\begin{tabular}{lccc}
\toprule
 & Any update & Absolute update & Signed update \\
\midrule
\multicolumn{4}{l}{\emph{Panel A: Prior-belief response time only}} \\
Log prior-belief RT & -0.009 & 0.002 & 0.007 \\
 & (0.015) & (0.005) & (0.005) \\
\(p\)-value & 0.538 & 0.758 & 0.155 \\
\addlinespace
\multicolumn{4}{l}{\emph{Panel B: Prior- and post-advice response times}} \\
Log prior-belief RT & -0.110 & -0.032 & -0.021 \\
 & (0.013) & (0.004) & (0.004) \\
\(p\)-value & \(<0.001\) & \(<0.001\) & \(<0.001\) \\
Log post-advice RT & 0.298 & 0.099 & 0.084 \\
 & (0.014) & (0.005) & (0.005) \\
\(p\)-value & \(<0.001\) & \(<0.001\) & \(<0.001\) \\
\midrule
Outcome mean & 0.214 & 0.078 & 0.070 \\
\(N\) &    60,252 &    60,252 &    60,252 \\
\bottomrule
\end{tabular}

\begin{minipage}{0.95\textwidth}
\footnotesize
\emph{Notes:} The unit of observation is a participant-image pair. All specifications include controls for recommendation by prior decile. Standard errors in parentheses are clustered by participant and image. Response time after advice is the time between submitting the prior belief and submitting the posterior belief. Any update is an indicator for the posterior belief differing from the prior belief. Absolute update is the absolute difference between posterior and prior beliefs. Signed update is movement toward the AI recommendation.
\end{minipage}
\end{table}

\subsection{Estimated AI Weight}\label{app:ai-weight}

\begin{figure}[H]
\centering
\includegraphics[width=4in]{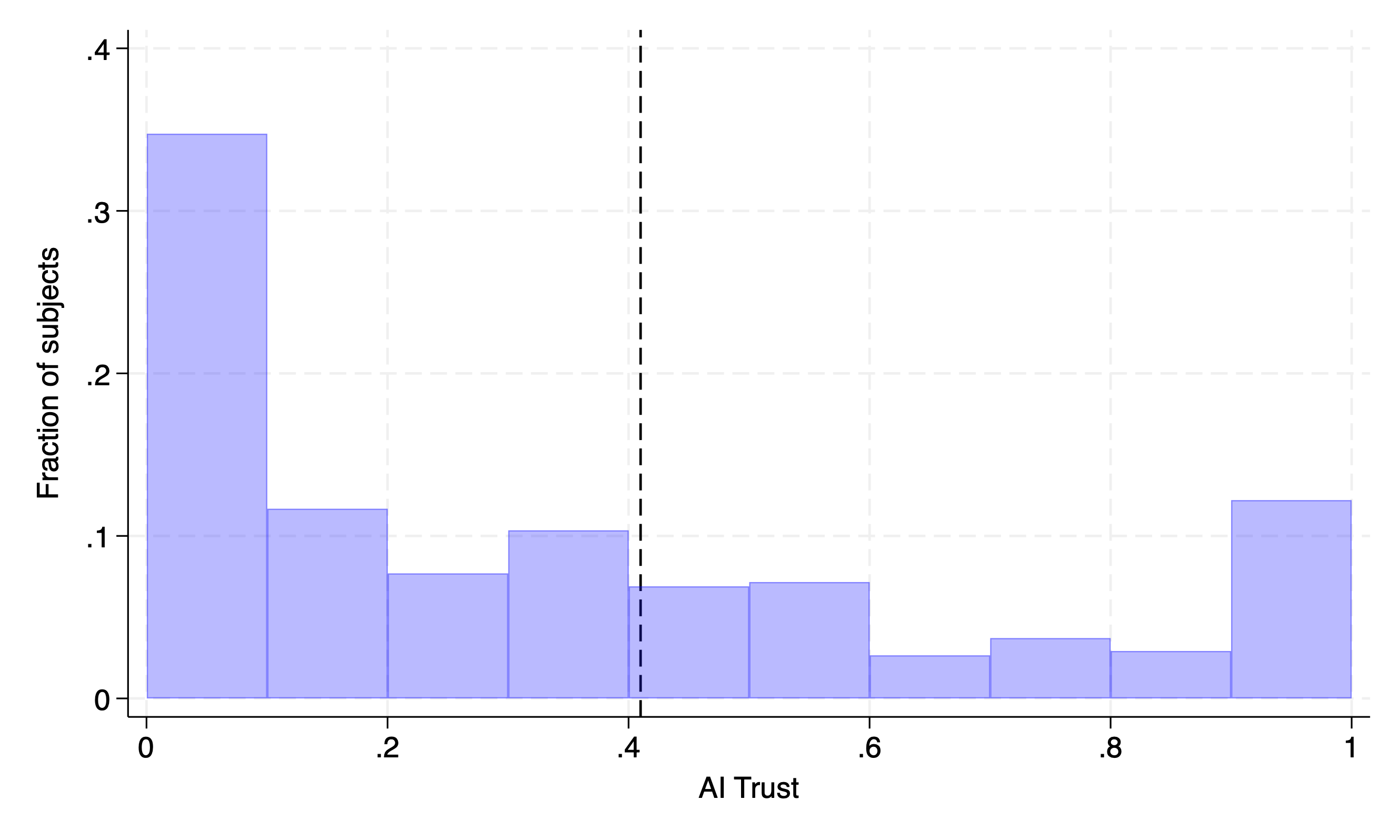}
\caption{Distribution of estimated AI weight ($1-w$).}
\label{fig:hist-ai-weight}
\end{figure}

Figure~\ref{fig:hist-ai-weight} and Table~\ref{tab:ai-weight-correlates} report distributions of estimated wsIU parameters by individual and regression results for AI weight on survey measures and demographics. 

\begin{table}[H]
\centering
\caption{Correlates of estimated AI weight.}
\label{tab:ai-weight-correlates}
\footnotesize
\begin{tabular}{l@{\hskip 0.6cm}c}
\toprule
& Coefficient \\
\midrule
NOINFO treatment             & \makebox[3.9em][r]{\(0.0368\)} \\[-0.6ex]
                             & \((0.0338)\) \\[0.6ex]

Woman                        & \makebox[3.9em][r]{\(-0.0467\)} \\[-0.6ex]
                             & \((0.0345)\) \\[0.6ex]

Older (above median)         & \makebox[3.9em][r]{\(-0.0256\)} \\[-0.6ex]
                             & \((0.0343)\) \\[0.6ex]

Confidence in own prior (0--100) & \makebox[3.9em][r]{\(-0.00283\)} \\[-0.6ex]
                             & \((0.00082)\) \\[0.6ex]

Belief about AI accuracy (0--100\%) & \makebox[3.9em][r]{\(0.00342\)} \\[-0.6ex]
                                 & \((0.00083)\) \\[0.6ex]

\bottomrule
\end{tabular}
\end{table}

\clearpage
\section{Instructions and Screenshots}\label{app:screenshots}

This appendix shows screenshots from the experiment.




\begin{figure}[H]
\centering
\includegraphics[width=0.95\textwidth,height=0.82\textheight,keepaspectratio]{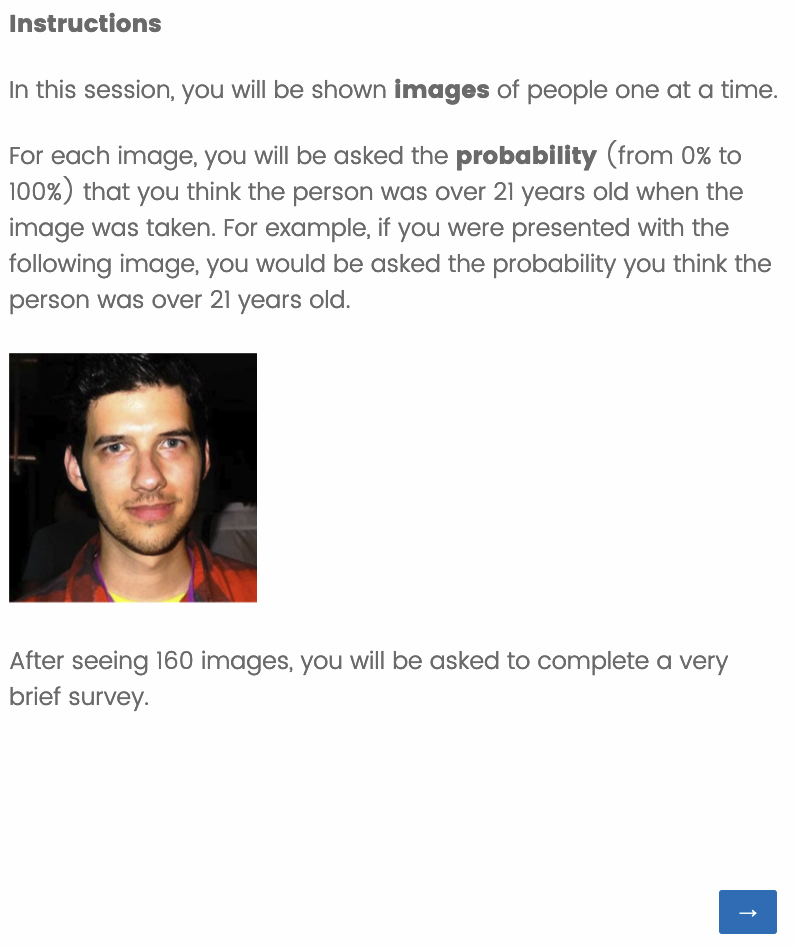}
\caption{Instruction screen.}
\label{fig:screenshot-instructions}
\end{figure}

\begin{figure}[H]
\centering
\includegraphics[width=0.95\textwidth,height=0.82\textheight,keepaspectratio]{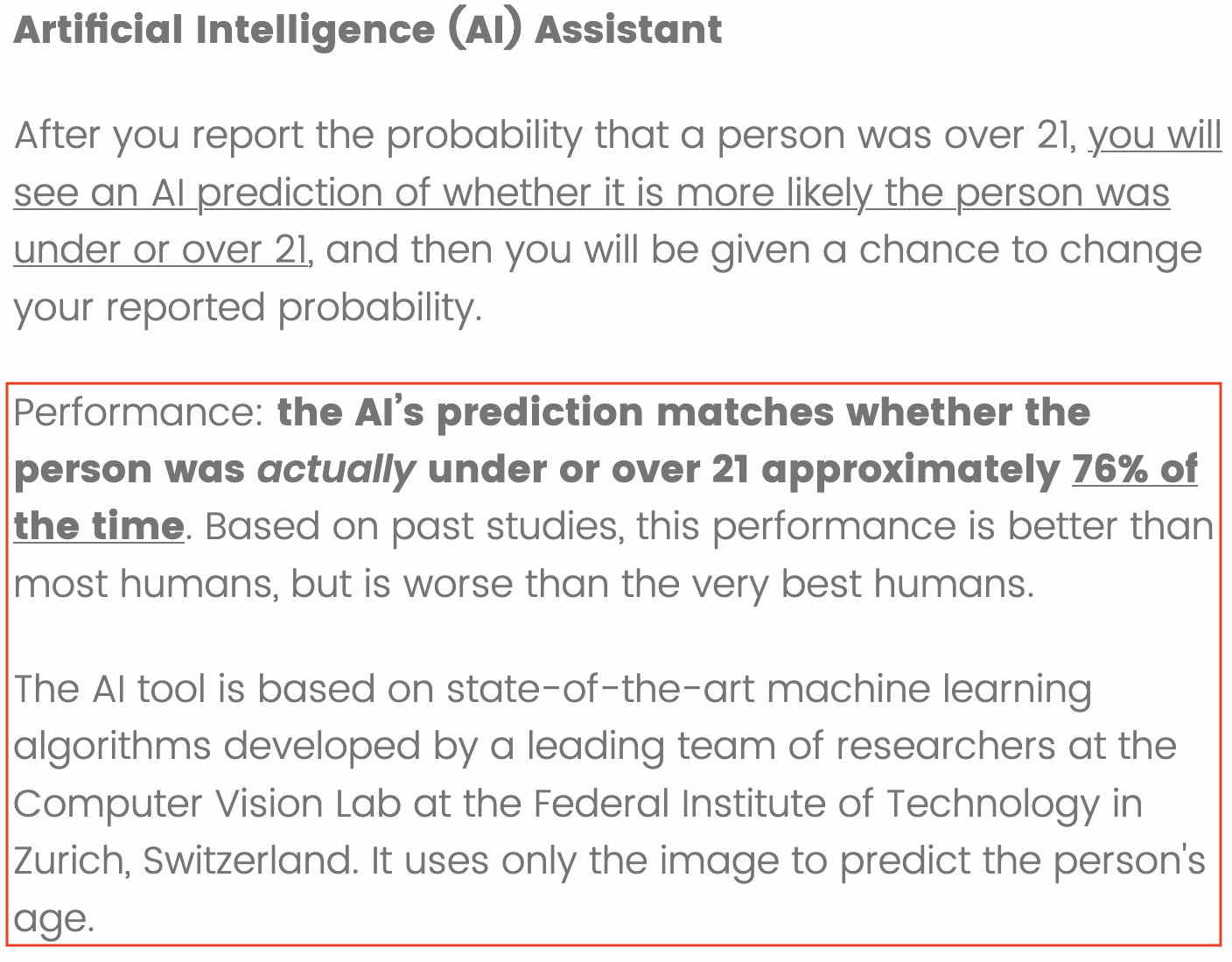}
\caption{AI accuracy information screen (text in red box for INFO treatment).}
\label{fig:screenshot-acc-ai}
\end{figure}

\begin{figure}[H]
\centering
\includegraphics[width=0.95\textwidth,height=0.82\textheight,keepaspectratio]{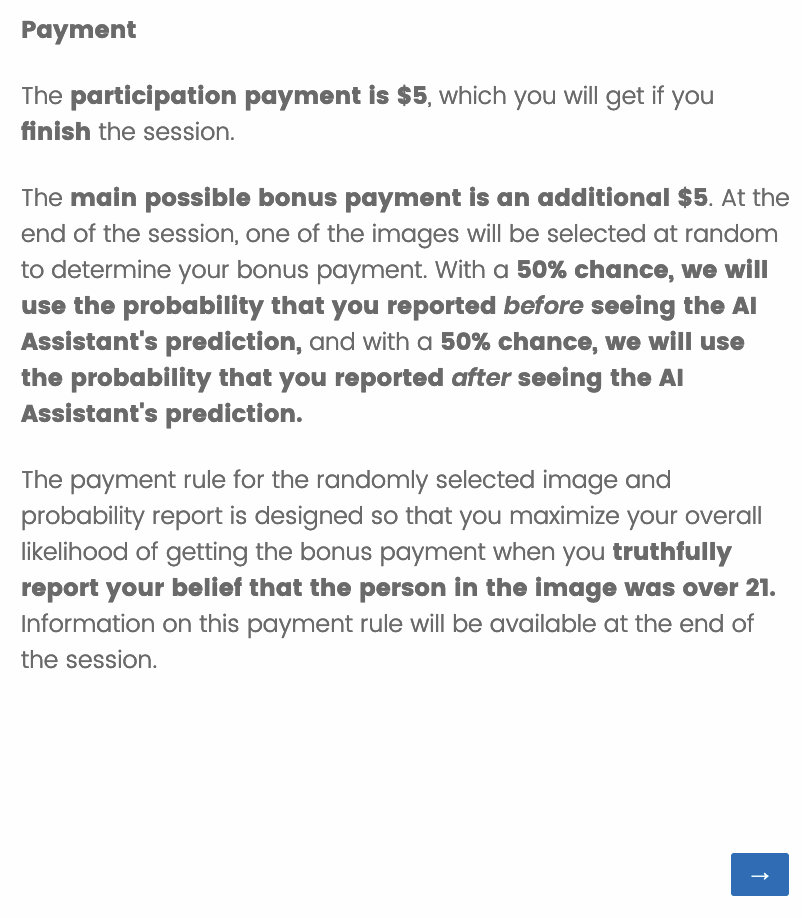}
\caption{Payment instruction screen.}
\label{fig:screenshot-payment}
\end{figure}

\begin{figure}[H]
\centering
\includegraphics[width=0.95\textwidth,height=0.82\textheight,keepaspectratio]{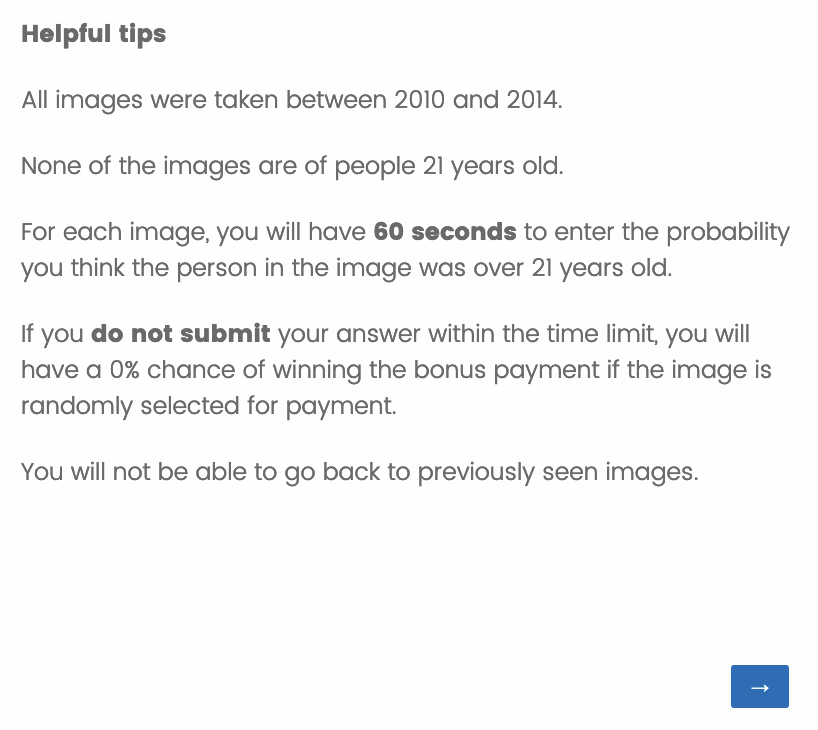}
\caption{Helpful tips screen.}
\label{fig:screenshot-helpful-tips}
\end{figure}

\begin{figure}[H]
\centering
\includegraphics[width=0.95\textwidth,height=0.82\textheight,keepaspectratio]{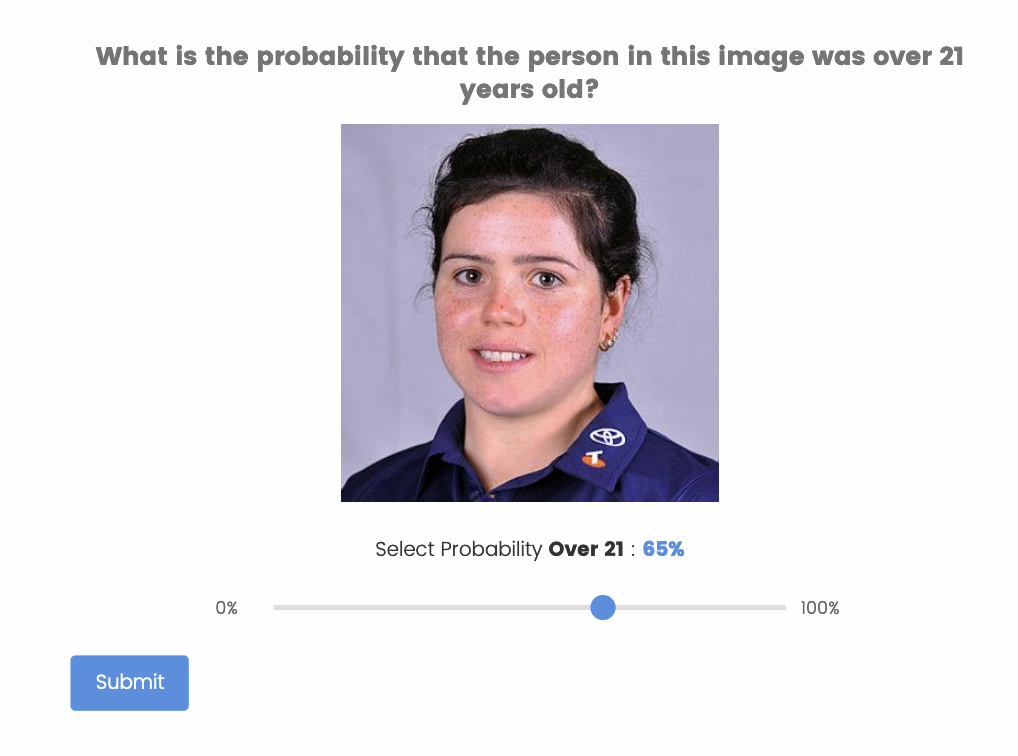}
\caption{Prior belief elicitation screen.}
\label{fig:screenshot-prob-initial}
\end{figure}

\begin{figure}[H]
\centering
\includegraphics[width=0.95\textwidth,height=0.82\textheight,keepaspectratio]{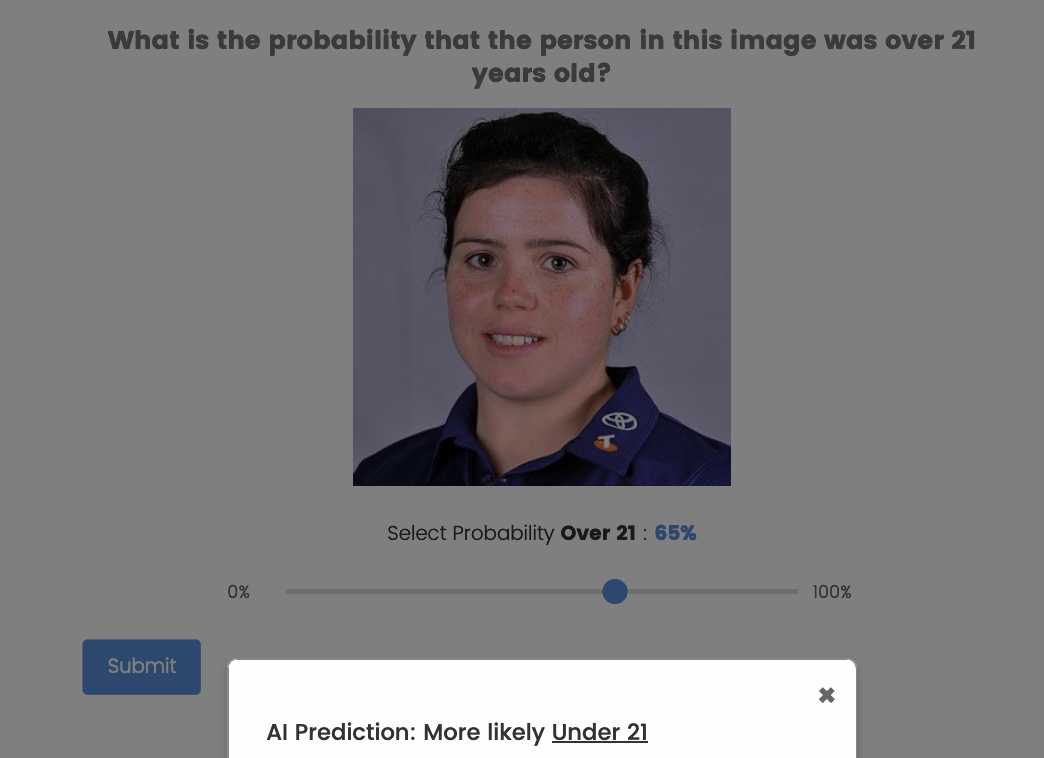}
\caption{Belief elicitation screen with pop-up.}
\label{fig:screenshot-prob-popup}
\end{figure}

\begin{figure}[H]
\centering
\includegraphics[width=0.95\textwidth,height=0.82\textheight,keepaspectratio]{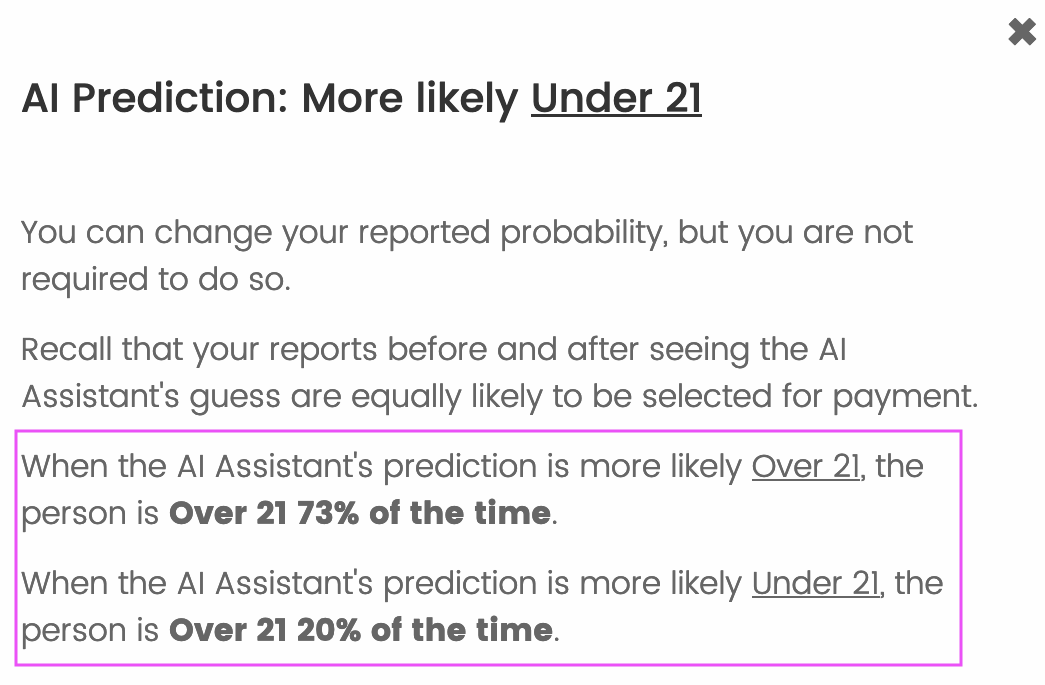}
\caption{Zoomed belief elicitation pop-up screen (text in pink box for INFO treatment).}
\label{fig:screenshot-popup-prob-zoom-box}
\end{figure}

\begin{figure}[H]
\centering
\includegraphics[width=0.95\textwidth,height=0.82\textheight,keepaspectratio]{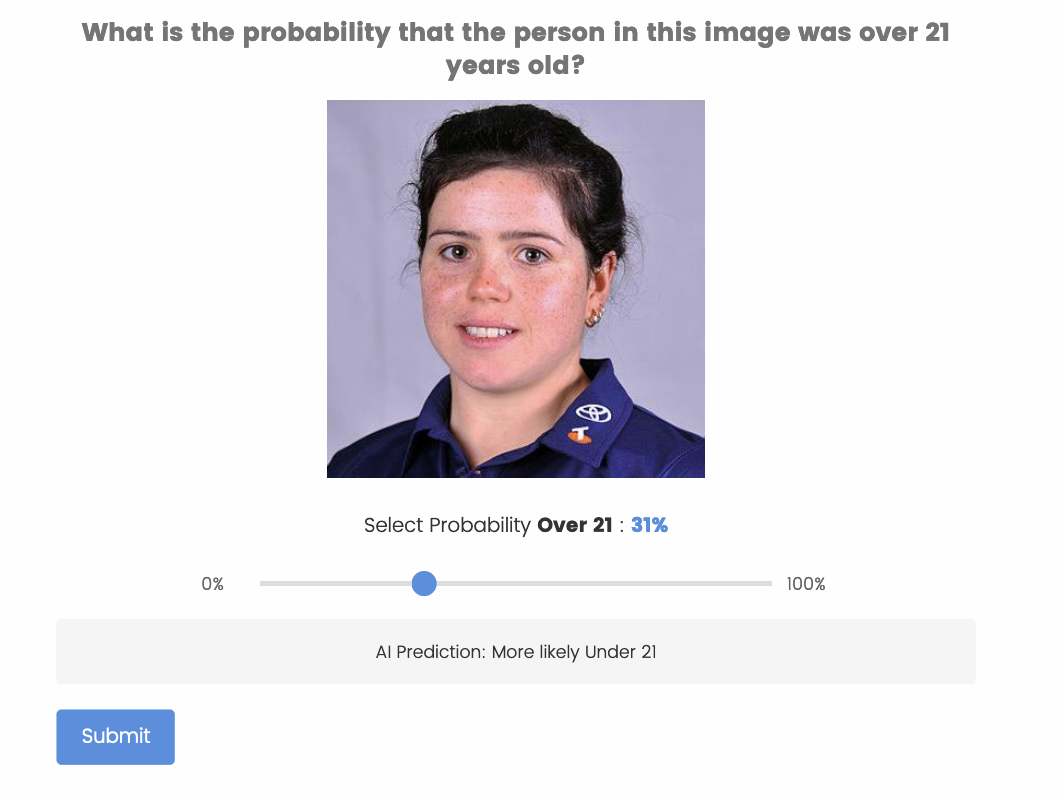}
\caption{Posterior belief elicitation screen.}
\label{fig:screenshot-prob-final}
\end{figure}

\end{document}